# Emergent trans-moiré orbitals and topology in rhombohedral graphene

Yuqin Wang[1]*, Jian Xie[1]*, Yi-Jie Wang[1]*, Jiajun Zhang[1]*, Yiting Gao[1], Zaizhe Zhang[1], Da Yi[1], Yan Xie[1], Jingjing Shi[1], Guanqin Zhao[1], Chengyu Xiong[1], Kenji Watanabe[2], Takashi Taniguchi[3], Zhi-Da Song[1,4,5†], Xiaobo Lu[1,4,5†], and Yi Chen[1,4,5,6†]

[1]International Center for Quantum Materials, School of Physics, Peking University, Beijing 100871, China

[2]Research Center for Electronic and Optical Materials, National Institute for Material Sciences, 1-1 Namiki, Tsukuba 305-0044, Japan

[3]Research Center for Materials Nanoarchitectonics, National Institute for Material Sciences, 1-1 Namiki, Tsukuba 305-0044, Japan

[4]Collaborative Innovation Center of Quantum Matter, Beijing 100871, China

[5]Beijing Key Laboratory of Quantum Devices, Peking University, Beijing 100871, China

[6]Interdisciplinary Institute of Light-Element Quantum Materials and Research Center for Light-Element Advanced Materials, Peking University, Beijing 100871, China

*These authors contributed equally to this work.

[†]E-mail: songzd@pku.edu.cn; xiaobolu@pku.edu.cn; yichen@pku.edu.cn

**Abstract:**

The fractional quantum anomalous Hall effect (FQAHE) exhibited in fractional Chern insulators[1-5] has recently been demonstrated in twisted $MoTe_2$[6-9] and rhombohedral graphene/hBN moiré superlattices[10-14], promising new routes toward topological quantum computation[15,16]. Central to realizing this promise is the understanding of the underlying microscopic mechanism. This, however, remains elusive in the case of rhombohedral graphene[10-29], with the crux being its two seemingly paradoxical conditions: a pronounced small-twist-angle ($\theta$) moiré interface, yet only when electrons are kept distant from it[10-14]. Here, by scanning tunnelling microscopic imaging with both conditions fulfilled, we capture drastic 10-meV-scale electronic structure reshaping in rhombohedral hexalayer graphene by 'trans-moiré orbitals', which emerge on the other, distant side of the moiré interface but nevertheless enforce the moiré periodicity at all measured fillings. We visualize a hierarchy of spatially and energetically distinct trans-moiré orbitals which doped electrons must sequentially occupy—the lowest-energy orbital, expectedly responsible for exotic phases at small fillings, carries a hollow-cage-like shape. Strikingly, these trans-moiré orbitals vanish at $\theta \gtrsim 1^{\circ}$, and so do quantum anomalous Hall plateaus in similar devices[30]. Simulations reveal an interaction-driven $\theta$-sensitive charge-redistribution mechanism which shapes the trans-moiré orbitals and corresponding Chern minibands. Our findings reconcile the paradoxical conditions with a natural explanation: electrons are not simply kept distant from a small-$\theta$ moiré interface; they are forced into topological trans-moiré orbitals, forged precisely under such conditions. Our microscopic diagnostics unlocks a range of possible 'synthetic' FQAHE platforms.

**Main text:**

A central theme in modern condensed matter physics is to investigate how a myriad of interacting electrons can self-organize to form exotic quantum many-body states of matter. This is exemplified in van der Waals flat-band systems[31,32] in which electrons with quenched kinetic energies are found to exhibit unusual superconductivity[33-35], correlation-induced topology[36,37], and generalized Wigner crystallisation[38-40]. Recently rhombohedral multilayer graphene provides another fertile testbed for such investigations[10-14,41-47], wherein correlated electrons arise from a pair of low-energy flat bands located on the top and bottom surfaces of the system[48-53] with energy differences set by an electrical displacement field ($D$) and band fillings ($\nu$) by electrostatic doping. Electrons in such highly tunable flat bands have first been experimentally studied in rhombohedral trilayer graphene to give rise to diverse correlated ground states including Mott insulators[42,43], superconductors[44,45], Chern insulators[46,47], and half/quarter metals[46]. In thicker (tetra-[12-14], penta-[10,12], and hexalayer[11]) rhombohedral graphene aligned to hexagonal boron nitride (hBN), electrons, when preferably injected to the surface on the distant side of the moiré interface, are found to spontaneously break time-reversal symmetry and fractionalize, forming a fractional Chern insulator that exhibits the fractional quantum anomalous Hall effect (FQAHE)[10-14].

Despite significant investigations, however, the microscopic origin of fractional and even integer quantum anomalous Hall effects of rhombohedral graphene/hBN superlattices remains contentious[10-29]. The bulk of theoretical works indicates that spatially homogeneous electrons in this system spontaneously crystallise, forming a so-called anomalous Hall crystal[17-24]. In this scenario, it is argued that the moiré potential felt by electrons polarized away from it is marginal and act merely as a pinning potential. A few works, however, suggest that the moiré influences might be underrated and are sensitive to the choice of the theoretical scheme[15,28,29]. This controversy lies at the heart of the FQAHE in rhombohedral graphene and has so far hindered the precise understanding of its origin.

To experimentally tackle this controversy, here we directly image the moiré electronic behavior of gate-tunable hBN-aligned rhombohedral hexalayer graphene (R*n*G/hBN, $n$ = 6) via scanning tunnelling microscopy/spectroscopy (STM/STS). We pick R6G because it is the thickest established member of its FQAHE family (R4G, R5G, and R6G)[10-14], hence best highlighting the dichotomy between the two aforementioned scenarios. STM imaging of R6G/hBN devices with twist angles $\theta \geq 1.22^{\circ}$ shows spatially uniform electronic states with no perceptible moiré potentials on the moiré-distant surface, in apparent agreement with predictions[17-24,28,29]. However, in R6G/hBN with $\theta \leq 0.52^{\circ}$, drastic 10-meV-scale moiré-periodic flat-band renormalization, hundreds of times stronger than single-particle estimates[17-24,28,29], emerges on the moiré-distant surface. More strikingly, the featureless electronic states now give way to emergent 'trans-moiré orbitals', which appear on the other, distant side of the moiré interface but nevertheless enforce the moiré periodicity at all measured $\nu$'s. We visualize a hierarchy of spatially and energetically distinct trans-moiré orbitals which doped electrons must sequentially occupy—the lowest-energy trans-moiré orbital, expectedly responsible for exotic phases at $\nu \leq 1$, is a distinct hexagonal-cage-like orbital. Both the trans-moiré orbitals and the flat-band modulations cease to exist beyond a threshold moiré twist angle, which agrees with the critical twist angle that the quantum anomalous Hall plateaus vanish in similar devices (both $\sim 1^{\circ}$)[30]. Self-consistent mean-field calculations allow us to identify an interaction-driven charge-redistribution mechanism at play, which reshapes the spatially homogeneous electronic states into trans-moiré orbitals and corresponding Chern minibands at small $\theta$'s.

**Atomically resolved moiré lattice relaxation in aligned R6G/hBN**

Figure 1a sketches the structure of our gate-tunable R6G/hBN devices, where the moiré interface is formed on the bottom surface of R6G, and local probe measurements are performed on its top (moiré-distant) surface. Large-scale STM topographs of R6G/hBN show spatially uniform moiré periodicity typically across hundreds of moiré unit cells (Fig. 1b and Supplementary Information Fig. 1), highlighting the quality of the fabricated R6G/hBN devices (Methods and Extended Data Fig. 1). Detailed twist-angle determination and strain analyses can be found in Supplementary Information Section 1, with results summarized in Extended Data Table 1.

A key aspect in understanding the FQAHE in R*n*G/hBN lies in a faithful atomistic description of its moiré structures at relevant small twist angles, which to our knowledge has not been visualized. Figure 1c presents an atomically resolved STM topograph of R6G/hBN D1 with a twist angle of $\theta = 0.28^{\circ}$. This is within the twist-angle range that the quantum anomalous Hall effects can occur in R6G/hBN, as shown in transport measurements of a similar-twist-angle dual-gated R6G/hBN device (Fig. 1f, D8 with $\theta = 0.17^{\circ}$)[11]. The STM topographic heights in Fig. 1c exhibit three high-symmetry sites, $C_{BN}$, $C_B$, and $C_N$ (Fig. 1c insets), identified based on stacking energy differences[54-57] and first-principle simulation results[56]. Strikingly, significant lattice relaxation is present in R6G/hBN, where the more stable $C_B$ region is seen to markedly expand into the $C_{BN}$ and $C_N$ regions, rendering the transition areas ridge-like (Fig. 1c)[54,55]. These observations are consistent with first-principle predictions of R*n*G/hBN[56] as well as topographs of bilayer or monolayer graphene aligned to hBN (Extended Data Fig. 2 and Refs. [54,55,57]). This suggests that the topographic features originate from the graphene/hBN moiré interface, to which upper graphene layers conform (as illustrated in Fig. 1b inset)[56]. Similar features are consistently observed in STM topographs of R6G/hBN across all measured (small) twist angles (Figs. 2, 3), although less sharp (non-atomically resolved) tips induce more spatial smoothening. Such lattice relaxation is found to carry profound influences on the moiré-distant electronic structure in R*n*G/hBN[13,30,58] (see Methods, Extended Data Fig. 2, and later text).

Figure 1e shows a typical STM d$I$/d$V$ spectrum on the top surface of R6G/hBN. The top-surface flat band induces a strong low-energy d$I$/d$V$ feature (Fig. 1e) with possible moiré- and correlation-induced fine structures (see below). The other flat band (located on the bottom surface) is not observed, but its position can be inferred from higher-energy remote bands (Fig. 1d, Supplementary Information Sections 2, 3).

**Absence of moiré electronic modulations in R6G/hBN with $\theta \geq 1.22^{\circ}$**

Figure 2a-d presents STM/STS measurements of R6G/hBN D6 with a twist angle of 1.40°. While its STM topograph (Fig. 2a inset) shows clear moiré-periodic corrugations (resulting from aforementioned structural buckling of R6G caused by hBN, Fig. 1b inset[56]), the local electronic structure of D6 shows no perceptible moiré modulations. This is evidenced in Fig. 2a, where spatially dependent d$I$/d$V$ curves along a high-symmetry direction across multiple moiré unit cells exhibit nearly identical flat-band features. Correspondingly, constant-current d$I$/d$V$ maps (which minimize topographic effects, Methods) exhibit spatially homogeneous electronic spectral distributions (Fig. 2b-d; see Supplementary Information Section 8 for a quantitative analysis of the homogeneity; different set points and gate voltages show similar results, Extended Data Fig. 7). Similar behavior is observed in R6G/hBN D4, D5, D7 with twist angles of 1.22°, 1.31°, 1.78°, respectively (Fig. 2e-g). The observed spatially homogeneous electronic structures apparently agree with the predicted negligible moiré potential on the moiré-distant surface[17-24,28,29].

**Emergent flat-band renormalization in small-$\theta$ R6G/hBN**

Striking results emerge when the same measurements are performed in R6G/hBN D1 with a smaller twist angle of 0.28°. Figure 3a and b shows spatially dependent STM d$I$/d$V$ spectra performed in the same fashion as Fig. 2a; in stark contrast, however, Fig. 3a features a drastic 10-meV-scale moiré-periodic modulation of the flat band, which appears on the moiré-distant surface of R6G/hBN D1 despite its hexalayer separation from the actual moiré interface. This amplitude is hundreds of times stronger than estimates of direct emanated moiré potentials vertically across R$n$G[17-24,28,29] and already approaches the order of magnitude of the moiré potential measured at the graphene/hBN interface[59].

The observed moiré renormalization strength does not vary significantly over a wide span of back gate voltages (Fig. 3a-f, Extended Data Fig. 9), indicating its robustness within this back-gate-tunable range (Fig. 3h). In fact, through a two-step fitting procedure (Supplementary Information Section 7), we find that all measurements performed in D1 under different conditions nevertheless show flat-band renormalization strengths roughly on the same order of magnitude, ranging from 6.77 meV to 27.1 meV (Extended Data Fig. 9). Qualitatively similar albeit slightly smaller renormalization is observed in R6G/hBN D2 and D3 with $\theta = 0.42^{\circ}$ and $0.52^{\circ}$, respectively (Extended Data Figs. 4-7, 9, and Supplementary Information Section 10), in stark contrast to null results from R6G/hBN D4 to D7 (with $\theta \geq 1.22^{\circ}$). Multiple STM microtips have been used and yield consistent results (Methods). The observed moiré-periodic flat-band renormalization thus defines a significant energy scale for moiré-distant electrons in small-$\theta$ R6G/hBN.

**Emergent trans-moiré orbitals in small-$\theta$ R6G/hBN**

Unlike spatially homogeneous electron filling in large-$\theta$ R6G/hBN (Fig. 2), flat-band renormalization of small-$\theta$ R6G/hBN is found to shape a hierarchy of spatially and energetically distinct trans-moiré orbitals on the moiré-distant surface, which doped carriers there must sequentially fill into (Fig. 4; a complete data set is shown in Extended Data Fig. 3 and Supplementary Movie 1). The lowest-energy trans-moiré orbital, which hosts doped electrons at small fillings, is arguably of most interest. Spectroscopic imaging allows us to visualize this lowest-energy trans-moiré orbital to form a distinct hollow-cage-like shape, exhibiting a hexagonal network connecting $C_{BN}$ and $C_N$ sites (Fig. 4a,e,i). Upon increasing the bias voltage, the higher-energy orbitals start to grow into more strongly linked shapes (Fig. 4b,f,j). At even higher energies, this gives way to pancake-like orbitals centered around the $C_B$ sites (Fig. 4c,g,k), which appear like the negative of the lowest-energy orbitals (Fig. 4a,e,i) and dominate the upper half of the flat band (Supplementary Information Section 4). In comparison, states energetically away from the flat band are homogeneously distributed in space (Fig. 4d,h,l), showing negligible set-point effects (Methods). Similar behavior is seen in the other two small-twist-angle devices, R6G/hBN D2 and D3 with $\theta = 0.42^\circ$ and $0.52^\circ$, respectively, as shown in Extended Data Figs. 4-7. To explicitly invalidate tip dependence as a possible cause of our observations, we manage to subsequently measure R6G/hBN D6 (with $\theta = 1.40^\circ$) and D3 (with $\theta = 0.52^\circ$) using the same microtip and find the trans-moiré orbitals and flat-band renormalization only to emerge when switching to the small-twist-angle device D3 (Extended Data Fig. 7).

Overall, the trans-moiré orbitals at small twist angles show robustness in real-space distributions under conditions spanning the $\nu$-$D$ diagram (covering $-4 \leq \nu \leq 4$ under a wide range of $D$ values reaching the QAHE-relevant $D$ fields, Extended Data Figs. 6, 9, Supplementary

Information Section 5). This is in agreement with the observed $v$- and $D$-insensitivity of the flat-band modulations, which shape the real-space trans-moiré orbitals in small-twist-angle R6G/hBN.

**Mechanism of trans-moiré-orbital formation**

The appearance of twist-angle-dependent trans-moiré orbitals is an unusual type of orbital formation. Direct (emanated) moiré potential experienced by the moiré-distant surface is known to be negligible given the hexalayer separation from the moiré interface[17-24,28,60]. Tip-induced charging can in principle produce a variety of moiré-periodic charging patterns[40], but they are typically accompanied by a pronounced charging peak in STM d$I$/d$V$ spectra, sensitive to the tip heights[61-63], and not present in material electronic-structure simulations without the tip. On the contrary, our STM d$I$/d$V$ spectra exhibit no obvious charging peaks (Fig. 3h, Extended Data Fig. 10), and tip-height-dependent d$I$/d$V$ maps show robust trans-moiré orbital patterns with varying STM tip heights (Extended Data Fig. 5, Methods).

Our self-consistent mean-field calculations reveal that the key to understanding the trans-moiré orbitals lies in an emergent Hartree potential from interaction-mediated charge redistributions (Extended Data Fig. 2). As shown in the bottom image of Fig. 5a, in R6G/hBN, the moiré-stacking energy from the hBN substrate induces high electron density centered around $C_B$ sites at the moiré-proximate interface. Such charge density can exert strong Coulomb repulsion vertically across R$n$G layers, leading to reduction of occupied charge density at $C_B$ sites on the moiré-distant surface by raising its electronic energy (top image in Fig. 5a). In other words, the moiré-distant electronic structure is now subject to an emergent moiré-periodic Hartree potential and hence renormalized both spatially and energetically, resulting in the observed trans-moiré orbitals. This picture well explains the experimental observations: the theoretical local density of states (LDOS) line cuts and maps (Fig. 3i-k, Fig. 5e insets, Extended Data Fig. 2, and Supplementary Information Fig. 5) reproduce the experimental observations (Figs. 3, 4).

On the contrary, at a large twist angle $\theta = 1.40^{\circ}$, simulations find vanishingly weak emergent moiré modulations in R6G/hBN, leading to nearly homogeneous electronic distributions on the moiré-distant surface (Fig. 5c), also consistent with experiment (Fig. 2).

The distinction between large- and small-twist-angle moiré lattices can be understood by considering whether the moiré reciprocal lattice vectors, $\boldsymbol{G}_{\mathbf{M}}$'s, can nest two states within the flat-band bottom (compare Fig. 5d and 5f). At a small twist angle $\theta$, the small $\boldsymbol{G}_{\mathbf{M}}$ can connect two Bloch states inside the (unfolded) flat band (i.e., two states with very similar energies), which leads to their large intermixing and hence strong moiré-periodic electronic modulations (Fig. 5a,d). On the contrary, at large $\theta$, the large $\boldsymbol{G}_{\mathbf{M}}$ indicates that at least one of the nested Bloch states must live outside the flat-band bottom (Fig. 5f). Consequently, such large-$\theta$ moiré lattices are found to induce negligibly weak moiré-periodic modulations onto the moiré-distant surface (Fig. 5c).

**Emergent Chern physics at small $\theta$'s**

Our electronic-state imaging thus highlights the presence of a strong emergent Hartree potential in small-$\theta$ R6G/hBN across various conditions. Crucially, in terms of the band structure, this potential splits the moiré-distant flat band into moiré minibands, and the lowest-energy moiré miniband can carry a Chern number of $|C| = 1$ in the parameter regime of interest ($\nu = 1$ at large negative $D$ for both stacking orientations, which corresponds to the occupation of an isospin-polarized trans-moiré Chern miniband) (Fig. 5e, Methods, and Supplementary Information Sections 5, 6). The emergent Hartree potential can thus offer a means to naturally generate a Chern insulator in rhombohedral graphene susceptible to further fractionalization[64], consistent with Ref. [28]. Our experimental imaging of trans-moiré orbitals, corresponding to real-space LDOS distributions of the emergent moiré minibands (Fig. 5e, Methods), justifies this mechanism and the employed theoretical scheme (Supplementary Information Section 6). Importantly, these emergent trans-moiré potentials are present in the large-negative-$D$ limit both experimentally (Extended Data Fig. 6) and theoretically (Supplementary Information Figure 4), thus expectedly shaping topological electronic behavior in transport-relevant parameter regime (for more extensive discussions, see Supplementary Information Section 5).

Since open-surface STM orbital imaging and dual-gated transport measurements cannot be performed on the same devices, to experimentally test whether the proposed mechanism of emergent moiré renormalization is concurrent with Chern insulator formation under transport conditions, we prepare two series of R6G/hBN devices and take advantage of sensitivity of both phenomena to twist angle $\theta$. With a series of open-surface STM devices at increasing $\theta$'s (R6G/hBN D1-D7), we plot the measured $\theta$-dependent moiré renormalization strengths and compare with transport results reported in Ref. [30] (on R6G/hBN D8 and similar dual-gated devices). We find that the 'on' and 'off' conditions of trans-moiré renormalization agree well with those of quantum anomalous Hall plateaus, suggesting an intimate linkage between STM-observed trans-moiré orbitals and transport-observed Chern-insulator formation (Fig. 5b).

Our results thus suggest that topology can be generated in R$n$G/hBN through interaction-driven trans-moiré potentials without the need of spontaneous translational symmetry breaking. This is best shown in STM imaging of the moiré-distant electronic states in small-$\theta$ R6G/hBN (Fig. 4, Extended Data Figs. 4-7, and Supplementary Information Section 10), which exhibit the same moiré periodicity over a wide range of filling numbers, instead of an electron crystal with $\nu$-sensitive real-space patterns[65-67] (Supplementary Information Section 6). We note that it is still possible that anomalous Hall crystallisation may exist in parameter regimes not covered in our experiments, especially at lower, millikelvin temperatures where the extended quantum anomalous Hall effect is observed[12,68] (for a more thorough discussion on implications of our results for R*n*G/hBN modelling, see Supplementary Information Section 6).

We anticipate qualitatively similar phenomena and mechanism to hold in R5G/hBN and R4G/hBN given that measured moiré renormalization of R6G/hBN is close in strength to the moiré potential on R3G/hBN[69]. In particular, according to our proposed mechanism, the strength of trans-moiré interactions provides us with a possible direct control knob for the FQAHE in graphene, for instance, by varying hBN thicknesses to modify gate-induced screening.

Lastly, drawing on our microscopic diagnostics of how R*n*G is transformed into an FQAHE platform, we propose a broad class of new 'synthetic' FQAHE platforms, each composed of: (1) a flat-band layer[70,71] and (2) a moiré layer (e.g., a twisted bilayer[19,72-76]). The former, no longer needing to form a moiré interface itself, hosts hundreds of promising candidates[70].

*Note added*: We note that emergent moiré renormalization induced by charge redistributions is also predicted in Refs.[77,78], consistent with our findings where they overlap. Including a Fock term in our calculations opens a larger gap at $\nu = 1$ and reverses the sign of the valley Chern number, in agreement with the exact diagonalization result of Ref. [78]; this term, however, is not necessary for reproducing our experiment (Methods), and this sign is opposite to that inferred from the phenomenological spin-orbit-coupling description of recent experiments[79,80]. Our imaging results, which delineate the structural and electronic-structure landscapes of R*n*G/hBN, motivate further dedicated theoretical investigations into the most faithful description scheme for this intriguing system (for more discussions along these lines, see Supplementary Information Section 6).

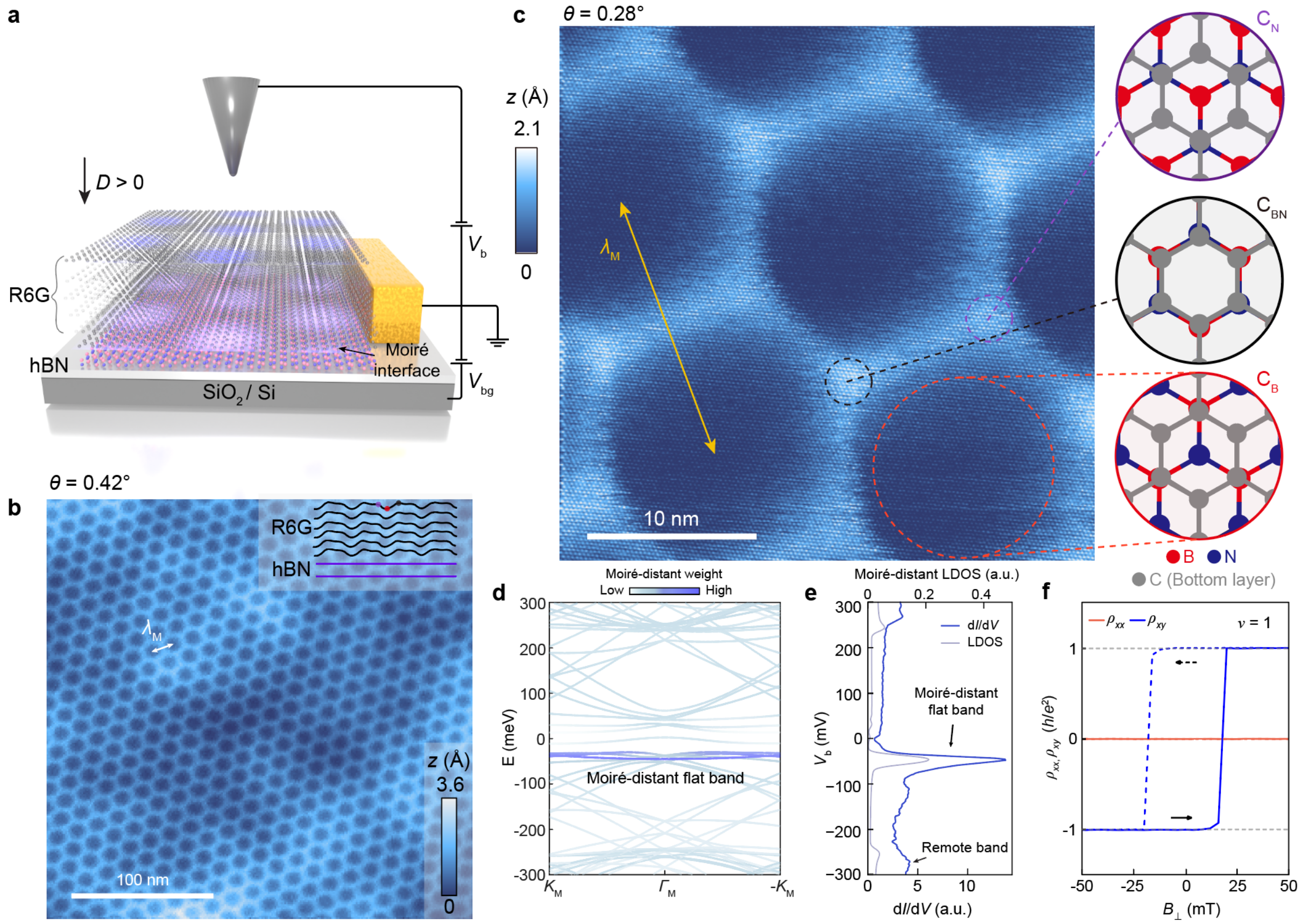

**Fig. 1. Atomically resolved moiré structure in R6G/hBN. a**, Schematic of the experimental setup where an STM tip probes the top (moiré-distant) surface of a gate-tunable R6G/hBN device. **b**, Large-scale STM topograph of moiré structure in R6G/hBN D2 showing a twist angle of $\theta = 0.42^{\circ}$ ($V_b = 0.3$ V, $I_t = 3$ pA). Inset shows a side view of R6G/hBN after moiré lattice relaxation. **c**, Atomically resolved STM topograph of R6G/hBN D1 with $\theta = 0.28^{\circ}$ (three high-symmetry stacking sites labeled in the right insets). The $C_B$ regions are seen to drastically expand into the $C_N$ and $C_{BN}$ regions due to moiré lattice relaxation ($V_b = 0.5$ V, $I_t = 5$ pA). **d**, Calculated band structure of R6G/hBN with color denoting the moiré-distant spectral weight. Parameters chosen to match with experiment in **e**. **e**, Representative STM d$I$/d$V$ spectrum (blue curve) showing the moiré-distant flat band along with higher-lying remote bands ($V_b = 0.5$ V, $I_t = 50$ pA, $V_{mod} = 3$ mV, acquired at a $C_B$ site of R6G/hBN D1). Grey curve shows calculated local density of states (LDOS) on the moiré-distant surface (derived from **d**). **f**, Magnetotransport results of a dual-gated R6G/hBN device (D8) with a small twist angle of $\theta = 0.17^{\circ}$. Magnetic-hysteresis-loop measurements at filling factor $\nu = 1$ shows quantized Hall resistivity $\rho_{xy}$ and vanishing longitudinal resistivity $\rho_{xx}$ at zero magnetic field.

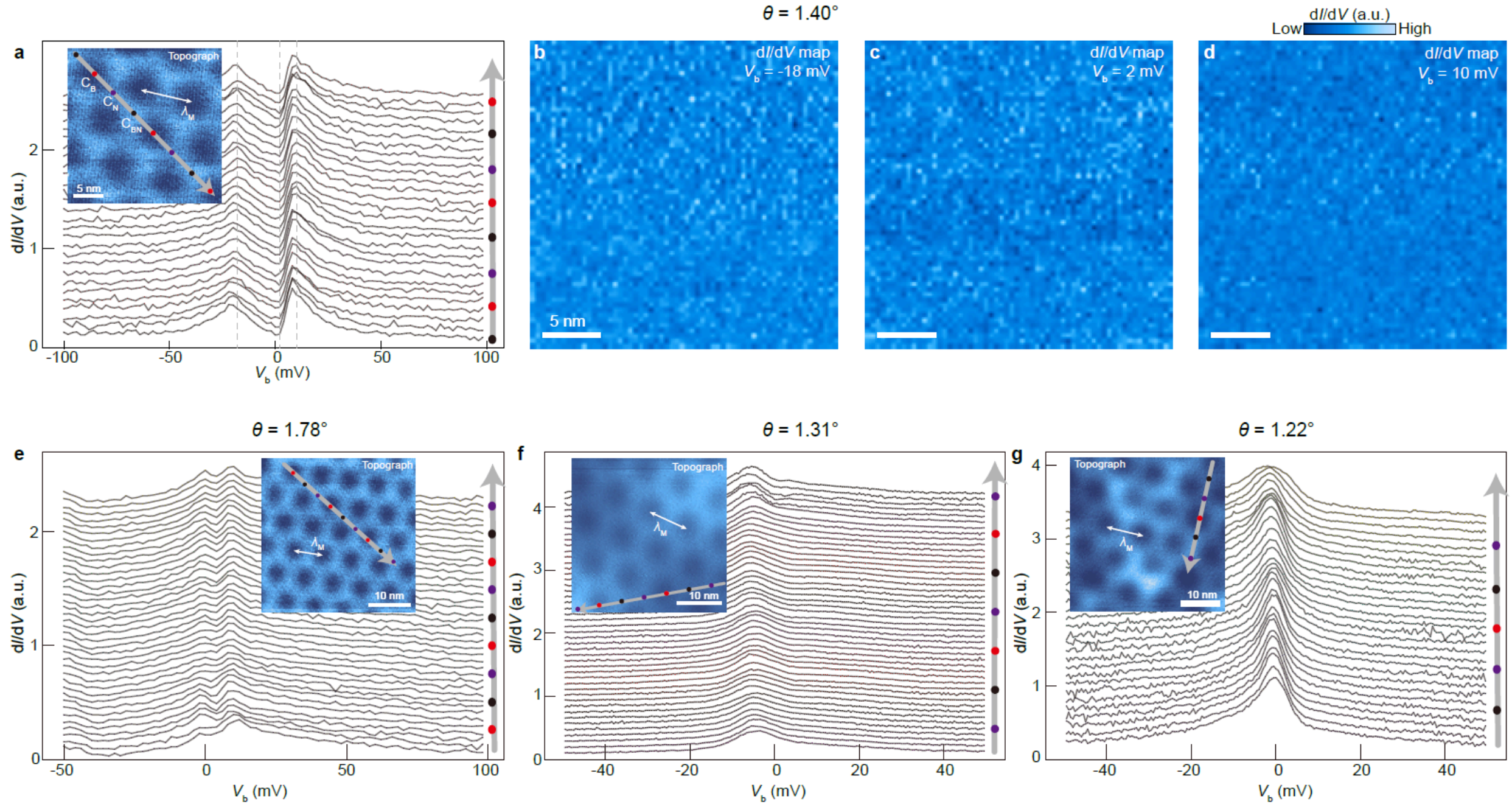

**Fig. 2. Absence of moiré-distant electronic modulations in R6G/hBN devices with $\theta \geq 1.22^{\circ}$.** **a,** Spatially dependent STM d$I$/d$V$ spectra measured along a high-symmetry direction across multiple moiré cells in D6 with $\theta = 1.40^{\circ}$ (grey arrow in the inset) showing no noticeable moiré-periodic electronic modulations ($V_b$ = -0.5 V, $I_t$ = 40 pA, $V_{mod}$ = 3 mV, $V_{bg}$ = 44 V). Inset: STM topograph showing the moiré superlattice of D6 (topographic contrast from structural buckling of R6G caused by hBN, see text) ($V_b$ = -0.1 V, $I_t$ = 5 pA). High-symmetry sites are indicated by purple ($C_N$), red ($C_B$) and black ($C_{BN}$) dots. **b-d,** Constant-current d$I$/d$V$ maps in the same area as the inset of **a** showing spatially homogeneous flat-band spectral distributions ($V_b$ = -0.1 V, $I_t$ = 10 pA, $V_{mod}$ = 3 mV, $V_{bg}$ = 44 V). **e-g,** same as **a**, but for D7 with $\theta = 1.78^{\circ}$ (**e**) ($V_b$ = -0.5 V, $I_t$ = 30 pA, $V_{mod}$ = 3 mV, $V_{bg}$ = 60 V), D5 with $\theta = 1.31^{\circ}$(**f**) ($V_b$ = 0.3 V, $I_t$ = 60 pA, $V_{mod}$ = 3 mV, $V_{bg}$ = 14 V), D4 with $\theta = 1.22^{\circ}$ (**g**) ($V_b$ = -0.3 V, $I_t$ = 80 pA, $V_{mod}$ = 2 mV, $V_{bg}$ = 16 V), also showing no noticeable moiré-periodic electronic modulations. Set points of the insets: **e** ($V_b$ = 0.5 V, $I_t$ = 3 pA), **f** ($V_b$ = -0.1 V, $I_t$ = 3 pA), **g** ($V_b$ = -0.3 V, $I_t$ = 3 pA). Measurements performed at $T$ = 4.0 K (**a-d**) & $T$ = 2.2 K (**e-g**).

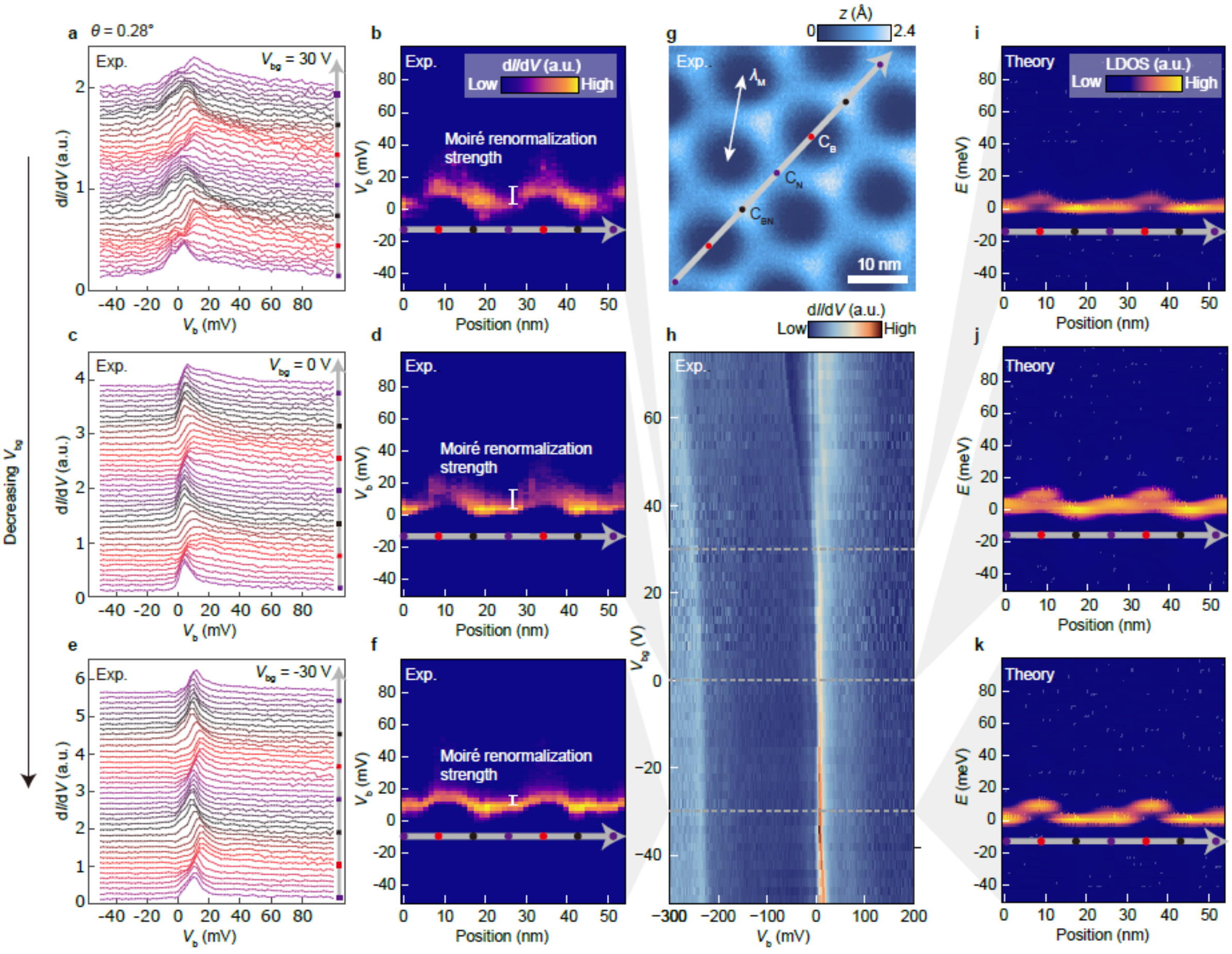

**Fig. 3. Emergent moiré-periodic flat-band renormalization in R6G/hBN D1 with $\theta = 0.28^{\circ}$.** **a**, **c & e**, Spatially dependent d$I$/d$V$ spectra (along the grey single arrow in **g**) showing strong moiré-periodic flat-band modulations, taken at $V_{bg}$ = 30 V (**a**), 0V (**c**) and -30V (**e**) ($V_b$ = -0.4 V, $I_t$ = 30 pA, $V_{mod}$ = 3 mV). **b**, **d & f**, Same data as in **a**, **c** & **e**, respectively, but plotted in color and rotated by 90° to better visualize the flat-band renormalization strength (white bar in each panel). **g**, STM topograph of R6G/hBN D1 with similar annotations to Fig. 2a inset ($V_b$ = -0.1 V, $I_t$ = 3 pA). **h**, Gate-dependent d$I$/d$V$ spectra acquired at a $C_{BN}$ site of R6G/hBN D1 ($V_b$ = -0.4 V, $I_t$ = 30 pA, $V_{mod}$ = 3 mV). **i, j & k**, Calculated local density of states (LDOS) along the same high-symmetry line shown in **g** with the same parameters as in **b**, **d & f**, respectively. Estimated ($\nu$, $D$) parameters are summarized in Extended Data Fig. 9. Measurements performed at $T$ = 3.4 K.

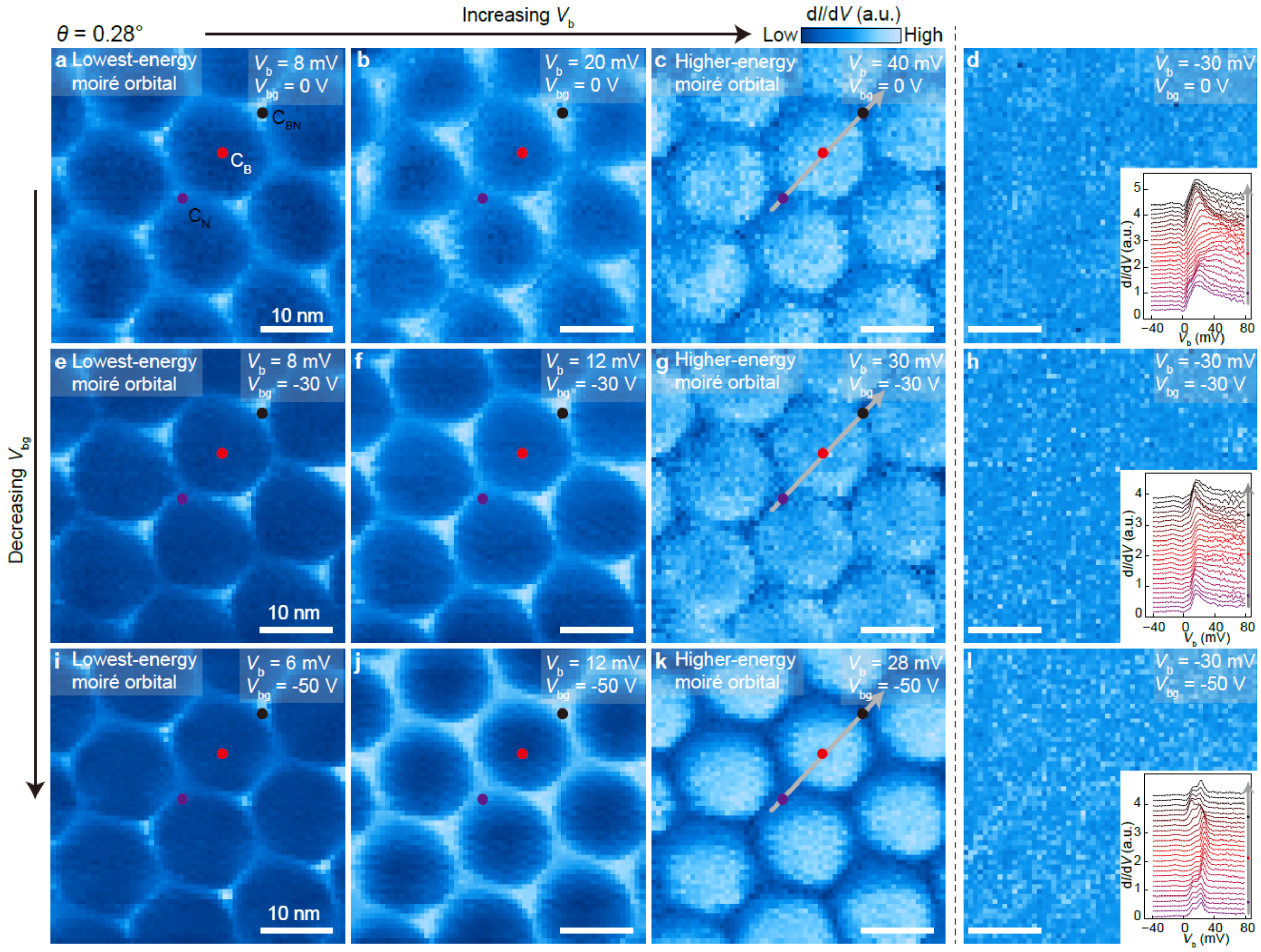

**Fig. 4. Emergent trans-moiré orbitals in R6G/hBN D1 with $\theta = 0.28^{\circ}$, all taken in the same area. a-c,** Constant-current d$I$/d$V$ maps at $V_{bg}$ = 0 V showing emergent trans-moiré orbitals: low-energy hollow-cage-like orbital forming a hexagonal network connecting $C_{BN}$ and $C_N$ sites (**a**); higher-lying orbitals showing more strongly linked shapes (**b**); pancake-like orbitals dominating the upper half of moiré-distant flat band that appear like the negative of **a** (**c**). **d,** d$I$/d$V$ map at a bias voltage outside the flat band showing spatially homogeneous spectral distributions, indicating negligible setpoint effects. Inset: Spatially dependent d$I$/d$V$ spectra acquired along a high-symmetry direction (grey arrow in **c**) ($V_b$ = -0.1 V, $I_t$ = 30 pA, $V_{mod}$ = 3 mV). **e-h,** Same as **a-d** but at $V_{bg}$ = -30 V. **i-l,** Same as **a-d** but at $V_{bg}$ = -50 V. All d$I$/d$V$ maps are taken in the same area using the same setpoint: $V_b$ = -0.1 V, $I_t$ = 30 pA, $V_{mod}$ = 3 mV. The same high-symmetry sites are marked in each map (purple dot: $C_N$, red: $C_B$, black: $C_{BN}$). Estimated ($\nu$, $D$) parameters are summarized in Extended Data Fig. 9. All scale bars: 10 nm. Measurements performed at $T$ = 3.4 K.

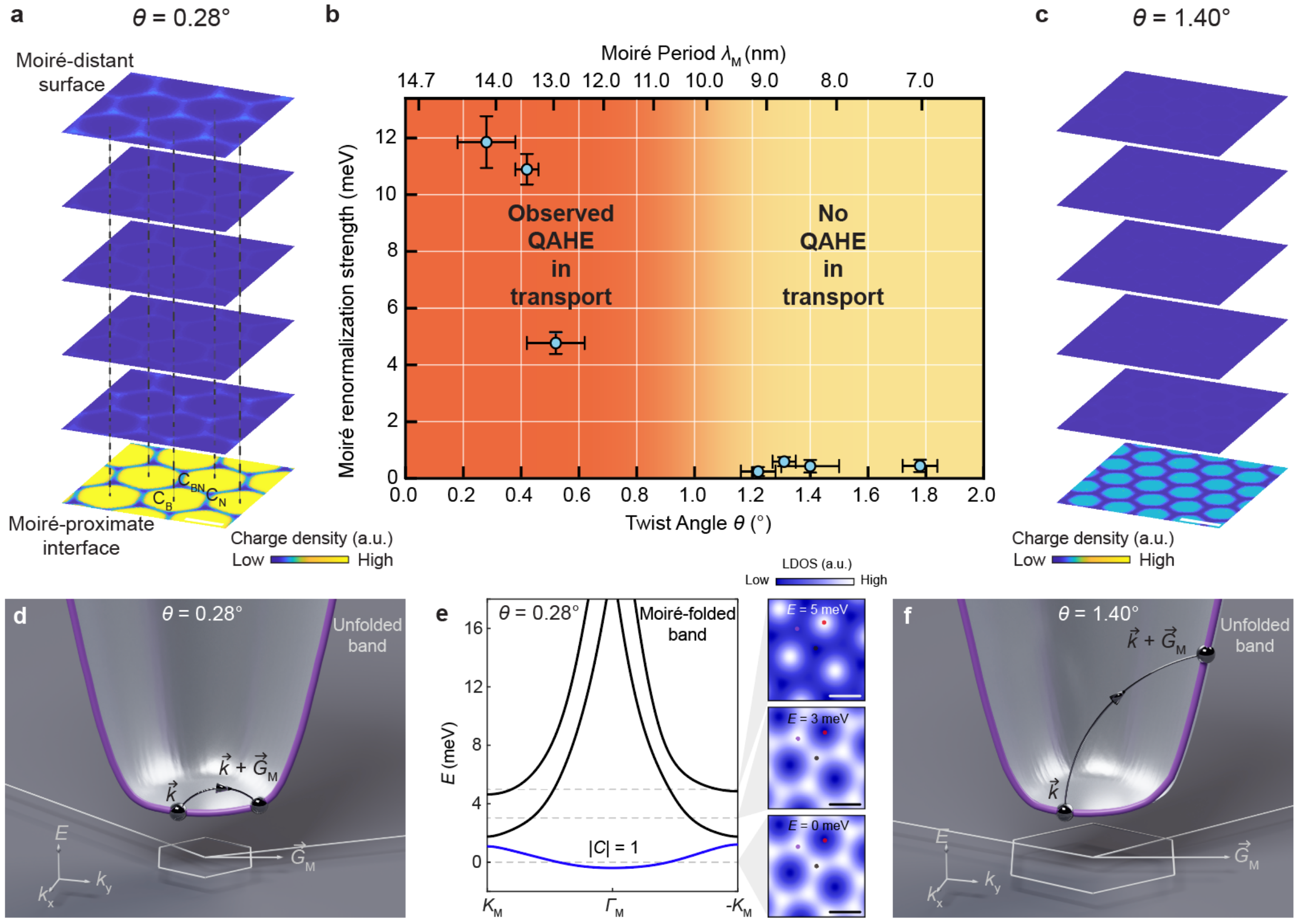

**Fig. 5. Mechanism of $\theta$-sensitive trans-moiré-orbital and Chern-miniband formation. a, c**, Simulated layer-resolved charge density variations in R6G/hBN with $\theta = 0.28^{\circ}$ **(a)** and $\theta = 1.40^{\circ}$ **(c)** showing drastically different moiré renormalization. All layers plotted with the same color bar after subtracting each layer's minimal value. Scale bars: 10 nm. **b,** Summary of experimentally measured moiré renormalization strengths as a function of $\theta$ (all measured at $\nu \sim 1$) overlaid with transport results of dual-gated R6G/hBN[30]. The onset condition of observed moiré renormalization agrees well with that of the quantum anomalous Hall effect (QAHE) in transport (both with a critical angle of $\sim 1^{\circ}$)[30]. **d, f,** Reciprocal-space schematics illustrating the difference between R6G/hBN with $\theta = 0.28^{\circ}$ **(d)** and $\theta = 1.40^{\circ}$ **(f)**. At small $\theta$, the small moiré reciprocal lattice wavevector $\boldsymbol{G}_{\mathbf{M}}$ can connect two Bloch states, $\boldsymbol{k}$ and $\boldsymbol{k} + \boldsymbol{G}_{\mathbf{M}}$, inside the unfolded flat-band bottom, leading to strong moiré renormalization (see text). **e,** Representative emergent Chern miniband with Chern number $|C| = 1$ in R6G/hBN with $\theta = 0.28^{\circ}$ ($\nu = 1$, $V_D = 15$ meV, $\xi = 1$). Insets: three representative trans-moiré orbitals including the hollow-cage-like orbital. Scale bar: 10 nm.

**Acknowledgments**

This research was supported by National Science Foundation of China (Grant No. 92365114, 12521006, 12274006, 12404044 and 12141401), National Key R&D Program of China (Grant No. 2022YFA1405100), National Science and Technology Major Project (Grant No. 2023ZD0301351010), and Beijing National Laboratory for Condensed Matter Physics (Grant No. 2025BNLCMPKF001). Z.-D. S. and Y.-J. W. were supported by National Natural Science Foundation of China (General Program No. 12274005), National Key Research and Development Program of China (No. 2021YFA1401900), and Innovation Program for Quantum Science and Technology (No. 2021ZD0302403). This work is also supported by Peking Nanofab.

**Author contributions**

Z.S., X.L., and Y.C. initiated the project. Y.W., J.Z., and Y.C. designed the experiment. J.X., and J.Z. fabricated the devices with help from Z.Z., D.Y., and Y.X. under the supervision of X.L. Y.W., J.Z., Y.G., J.S., G.Z., and C.X. performed the STM/STS measurements under the supervision of Y.C. Y.-J.W. performed theoretical modelling under the supervision of Z.-D.S. T.T. and K.W. contributed the hBN substrates. Y.W., Y.-J.W., J.Z., and Y.C. wrote the paper with input from all co-authors.

**Competing interests**

The authors declare no competing interests.

**Data availability**

All data within this manuscript and other findings of this study are available from the corresponding author upon reasonable request.

## Methods

### R6G/hBN device fabrication

Extended Data Fig. 1 illustrates the dry-transfer methods we employed to fabricate R6G/hBN devices for STM/STS measurements in this study. Two different methods were used. R6G/hBN D1 & D7 were fabricated by a modified flip-chip method. Multilayer graphene and hBN flakes were exfoliated onto Si substrate with a 285 nm thick $SiO_2$ dielectric. The thicknesses of hBN flakes were measured by atomic force microscopy (AFM) and reported in Extended Data Table 1. The layer number of graphene flakes was identified via optical contrast and AFM topography, and its rhombohedral nature was identified via a rapid infrared imaging technique[81] and Raman spectroscopy using a laser wavelength of 532 nm (Extended Data Fig. 1). To avoid structural relaxation, rhombohedral graphene flakes were delineated into isolated pieces from Bernal graphene flakes using a femtosecond laser. To assemble aligned R6G/hBN devices, we used a polypropylene carbonate (PPC) film covered polydimethylsiloxane (PDMS) stamp to first pick up an hBN flake and then an R6G flake with their straight edges in alignment. A possible 30$^{\circ}$ misalignment could occur, but such devices could be identified in later STM measurements and excluded. Care needs to be taken when picking up the R6G flakes to minimize the possibility of strain-induced relaxation. The PPC film was then peeled off and flipped onto $SiO_2$/Si substrate to expose the rhombohedral graphene surface. The as-fabricated structure was then annealed in an Ar/$H_2$ atmosphere at 350 $^{\circ}$C to remove the PPC film. Electrical contacts were made by evaporating Cr/Au (5nm/50nm) onto the heterostructure through a SiN stencil mask. The rhombohedral graphene surface was cleaned gently by contact-mode AFM. Raman spectroscopy was performed throughout the fabrication process to verify the rhombohedral nature of the sample. The twist angles of the fabricated devices were later directly visualized through STM imaging.

The other R6G/hBN devices were fabricated using a more direct method. The exfoliation, identification, and delineation processes of R6G flakes were identical to the previous fabrication method. Instead of using PPC flip, R6G flakes were directly picked up by a poly (bisphenol A carbonate) (PC) film covered PDMS stamp, and then the PC film supporting the R6G flake was released onto a hBN flake with their straight edges in alignment. The remaining processes involving evaporating electrical contacts and AFM cleaning were identical to that in the previous method. This second method was less time consuming than the first one, and no noticeable surface quality difference was found after careful AFM cleaning.

| Sample | Twist angle | Corresponding moiré period | Data sets | Strain | hBN substrate thickness |
|---|---|---|---|---|---|
| R6G D1 | 0.28° | 14.1 nm | Figs. 1,3,4; EFigs.[a] 3,8,10; Fig.[b] S4 | 0.06% | 49 nm |
| R6G D2 | 0.42° | 13.5 nm | Fig. 1b; EFig. 6; Fig. S11 | 0.20% | 25 nm |
| R6G D3 | 0.52° | 13.0 nm | EFigs. 4,5, 7k-o; Figs. S10,S12 | 0.09% | 31 nm |
| R6G D4 | 1.22° | 9.2 nm | Fig. 2g | 0.06% | 6 nm |
| R6G D5 | 1.31° | 8.7 nm | Fig. 2f | 0.18% | 28 nm |
| R6G D6 | 1.40° | 8.4 nm | Fig. 2a-d; EFig. 7a-j | 0.10% | 41 nm |
| R6G D7 | 1.78° | 7.0 nm | Fig. 2e; Fig. S13 | 0.30% | 42 nm |
| R6G D8 | 0.17° | 14.5 nm (dual gate) | Fig. 1f (transport) | —— | —— |

a. EFig. is short for Extended Data Figure (same below).
b. Fig. S is short for Supplementary Information Figure (same below).

**Extended Data Table 1. Summary of R6G/hBN devices studied in this work.**

**STM/STS measurements**

STM/STS measurements were performed using a customized CASAcme Joule-Thomson STM system operated under ultra-high vacuum conditions. The reported phenomena show robustness within the measured temperature range (Supplementary Information Section 9). All STM microtips used were calibrated against a noble metal surface (Cu(100), Cu(111), Au(111)) before STM/STS measurements, except in Extended Data Fig. 7 where the same microtip was used to subsequently measure D6 and D3 (without tip fixing in between). Only tips with good work-function matching to graphene were used, as evaluated by fits to entire gate-dependent data sets (next section). Several macroscopically different tungsten tip wires were used and yielded similar results. STM d$I$/d$V$ spectra were obtained using standard lock-in techniques with a small bias modulation $V_{\mathrm{mod}}$ at 421 Hz (with $V_{\mathrm{mod}}$ values reported in figure captions). All d$I$/d$V$ conductance maps were performed in the constant-current mode by sequentially acquiring d$I$/d$V$ spectra on individual sites of a real-space square-shaped grid with the same current and bias voltage setpoints. This mode can minimize topographic effects because the bias voltage can be chosen to be far from the flat bands or on the opposite bias polarity, so that topographic variations can be largely included in the constant-current setpoint, thus minimizing their effect in the d$I$/d$V$ maps. Very different setpoints were used, including positive/negative bias voltages and very high voltage setpoints located inside the R6G remote bands (i.e., energetically far from the flat bands), and they show similar results (Fig. 4, Extended Data Figs. 4-7, Supplementary Information Section 10). The nearly homogeneous d$I$/d$V$ maps at energies away from the flat band (e.g., Fig. 4d,h,l) also evidence the minimization of the topographic effects. Before obtaining each set of maps the STM tip was parked near the sample surface for at least 4 hours to minimize piezoelectric drift effects.

We located R6G samples in the STM based on an optical alignment procedure, which was sufficient for our sample size and *in-situ* optical resolution. The STM used in this study is equipped with a reentrant viewport and a homemade long-working-distance camera, which is able to acquire sufficiently magnified images of the devices. It also helps to carefully align such *in-situ* optical images with higher-resolution images acquired outside with a large-NA optical microscope. We could then locate the STM tip above the sample region and approach the tip to the R6G samples.

**Simulations for fitting to gate-dependent STM spectra**

To simulate the measured gate-dependent STM spectra, we found it useful to parametrize each microtip by its work function difference from graphene

$$\Delta\Phi = W_{\mathrm{g}} - W_{\mathrm{tip}}, \quad (1)$$

which induces an approximate local displacement field of

$$\frac{D_{\mathrm{tg}}}{\varepsilon_0} = -\frac{\Delta\Phi}{e d_{\mathrm{t-s}}}, \tag{2}$$

and an approximate local doping of electrons of

$$n_{\mathrm{tg}} = -\frac{D_{\mathrm{tg}}}{e} = \frac{\varepsilon_0 \Delta\Phi}{e^2 d_{\mathrm{t-s}}}. \tag{3}$$

The bottom gate was simulated in the conventional manner, where a bottom gate voltage $V_{\mathrm{bg}}$ applied to the bottom gate electrode leads to a displacement field $D_{\mathrm{bg}}/\varepsilon_0 = C_{\mathrm{bg}}V_{\mathrm{bg}}/\varepsilon_0$ and carrier density $n_{\mathrm{bg}} = C_{\mathrm{bg}}V_{\mathrm{bg}}/e$, with $C_{\mathrm{bg}}$ being the bottom gate capacitance per unit area. Combining both effects, multilayer graphene sample experiences a total carrier density

$$n_{\mathrm{total}} = n_{\mathrm{bg}} + n_{\mathrm{tg}} = \frac{C_{\mathrm{bg}}V_{\mathrm{bg}}}{e} + \frac{\varepsilon_0 \Delta\Phi}{e^2 d_{\mathrm{t-s}}}, \tag{4}$$

and an averaged displacement field of

$$\frac{D_{\mathrm{ave}}}{\varepsilon_0} = \left(\frac{D_{\mathrm{bg}}}{\varepsilon_0} + \frac{D_{\mathrm{tg}}}{\varepsilon_0}\right)/2 = \frac{C_{\mathrm{bg}}V_{\mathrm{bg}}}{2\varepsilon_0} - \frac{\Delta\Phi}{2e d_{\mathrm{t-s}}}. \tag{5}$$

We note that such displacement fields are local and may bear different implications from reported values in transport using global gates. Finally, this displacement field was converted to an electrostatic energy difference between two neighboring graphene layers by

$$V_{\mathrm{D}} = \frac{D_{\mathrm{ave}}}{\varepsilon_0}\frac{e d_{\mathrm{ABC}}}{\varepsilon_{\mathrm{ABC}}}. \tag{6}$$

Equations (4) and (6) entered the simulations of rhombohedral graphene (next section). Using experimental values of $d_{\mathrm{ABC}} = 0.335$ nm, $W_{\mathrm{g}} = 4.7$ eV and taking $\varepsilon_{\mathrm{ABC}} = 14$, we could fit all experimental data sets of $V_{\mathrm{bg}}$-dependent STM spectra obtained by different STM tips on R6G/hBN by fine tuning $C_{\mathrm{bg}}$ and $\Delta\Phi/d_{\mathrm{t-s}}$, where $d_{\mathrm{t-s}}$ was experimentally measured to be ~1 nm. The best results of such fits yielded the local doping and the work function of a particular tip, $W_{\mathrm{tip}}$, and only tips with good work-function matching to graphene are used in this manuscript. From Equations (5) and (6), we obtained the resultant parameters of $(n_{\mathrm{total}}, D_{\mathrm{ave}}/\varepsilon_0)$ for different data sets, and the filling factor ν can be determined by $\nu = n_{\mathrm{total}}/n_{\mathrm{moiré}}$, in which $n_{\mathrm{moiré}} = 1/A_{\mathrm{M}}$ with $A_{\mathrm{M}}$ being the area of each moiré supercell. These values were plotted in Extended Data Fig. 9 and reported throughout the text.

**Tight-binding modelling of R6G**

The Hamiltonian of R6G takes the standard form as detailed elsewhere[60]. Briefly, the intralayer Hamiltonian term for the $l$th-layer in bare R6G reads

$$H_l^{\text{intra}} = \begin{pmatrix} V_l & v_F(k_x - ik_y) \\ v_F(k_x - ik_y) & V_l \end{pmatrix},$$

where the Dirac velocity was taken as $v_F = 630.62\ \text{meV} \cdot \text{nm}$. $V_l = lV_D$ is a potential induced by the displacement field *D*, where the layer label $l = 0$ to 5 corresponds to bottom to top layers of R6G, respectively (same convention as in Fig. 1a).

The interlayer hopping term from the $(l+1)$-th layer (rows) to the $l$-th layer (columns) reads

$$H_{l+1,l}^{\text{inter}} = \begin{pmatrix} -v_4(k_x + ik_y) & t_1 \\ -v_3(k_x - ik_y) & -v_4(k_x + ik_y) \end{pmatrix},$$

where fitting to experimental remote-band positions (here and Ref. [60]) yielded $t_1 = 400\ \text{meV}$, $v_3 = v_4 = 0.04\ v_F$. In the following, we denote the two sublattices as A/B, corresponding to the first/second row in the above Hamiltonians. By comparing with experimental band-structure measurements, we found the intrinsic inversion-symmetric potential and farther-layer hoppings to be negligible and hence neglected them.

**Simulating the relaxed moiré interface and stacking configurations in R6G/hBN**

The moiré potential between R6G and hBN was included using standard methods[55] as summarized below. Suppose the two sublattices of each graphene layer $l$ are located at $\boldsymbol{\tau}^{\mathbf{A}} = \frac{a_0}{\sqrt{3}}(0, l+1)$ and $\boldsymbol{\tau}^{\mathbf{B}} = \frac{a_0}{\sqrt{3}}(0, l+2)$, respectively, where $a_0 = 0.246$ nm is the graphene lattice constant. As illustrated in Extended Data Fig. 2, we use stacking configuration $\xi = 0$ to denote the case where boron and nitrogen atoms are located at $\boldsymbol{\tau}^{\mathbf{Boron}} = R_{-\theta}(1+\epsilon)\frac{a_0}{\sqrt{3}}(0,1)$ and $\boldsymbol{\tau}^{\mathbf{Nitrogen}} = R_{-\theta}(1+\epsilon)\frac{a_0}{\sqrt{3}}(0,2)$, respectively, where $\epsilon = 0.0163$ is the lattice constant mismatch between hBN and graphene, and $\theta$ is the small twist angle between R6G and hBN. Exchanging boron and nitrogen atomic positions leads to the stacking configuration $\xi = 1$. With both R6G and hBN breaking the $C_{2z}$ symmetry, these two local stacking configurations are nonequal[58].

To simulate lattice relaxation in R6G/hBN, let us first consider the R6G/hBN moiré interface. There, the $C_{BN}$ and $C_N$ sites are expected to have an energy of 100 meV and 90 meV per microscopic unit cell higher than $C_B$, leading to significant (in-plane) lattice relaxation that expands the $C_B$ area[55]. Following Ref. [55], we obtained the relaxed local stacking pattern by solving the elastic system consisting of the $l = 0$-layer graphene and an hBN layer. In accordance with first-principle simulations[56], we assumed that upper graphene layers conform to this graphene/hBN interface, as their additional structural relaxations are expected to decay exponentially[56].

Such lattice relaxation significantly alters the moiré potential landscape felt by R6G electrons on the moiré-proximate (bottom) layer (Extended Data Fig. 2c). As a standard approximation, we assumed such potential to depend only on the local stacking configuration at position $\boldsymbol{r}$, which could be encoded in phases $\varphi_j(\boldsymbol{r}) = \left(\boldsymbol{G_j^G} - \boldsymbol{G_j^{hBN}}\right) \cdot \boldsymbol{r} - \frac{\boldsymbol{G_j^G} + \boldsymbol{G_j^{hBN}}}{2} \cdot \boldsymbol{u}(\boldsymbol{r}) \bmod 2\pi$. Here, $\boldsymbol{u}(\boldsymbol{r})$ is the relative atomic displacement due to relaxation, $\boldsymbol{G_j^G} = \frac{4\pi}{\sqrt{3}a_0}(\sin\frac{2\pi(j-1)}{3}, \cos\frac{2\pi(j-1)}{3})$ with $j = 1,2,3$ are three graphene reciprocal lattice vectors, and $\boldsymbol{G_j^{hBN}} = R_{-\theta}\frac{G_j^G}{1+\epsilon}$. For example, the $C_{BN}$ sites have all $\varphi_j = 0$; taking all $\varphi_j = \frac{2\pi}{3}$ yields $C_N$ at $\xi = 0$ and $C_B$ at $\xi = 1$. The contribution of co-moving atomic displacement to $\varphi_j$ is at least $O(\epsilon, \theta)$ smaller and hence neglected. Following Ref. [56], the moiré potential felt by the moiré-proximate R6G layer could be written as

$$V(\mathbf{r}) = V_0\begin{pmatrix}1 & 0\\0 & 1\end{pmatrix} + V_1\sum_j e^{i(\varphi_j+\psi_\xi)}\begin{pmatrix}1 & e^{-i\frac{2\pi}{3}j}\\ e^{i\frac{2\pi}{3}(j+1)} & e^{i\frac{2\pi}{3}}\end{pmatrix} + h.c.,$$

where the diagonal elements are scalar potentials on the A and B sublattices and off-diagonal elements are responsible for inter-sublattice transitions with $\psi_\xi = -136.55°$ for $\xi = 0$ and $\psi_\xi = 16.55°$ for $\xi = 1$. $h.c.$ represents the Hermitian conjugate. $V_0$ is a uniform potential that raises the energy of bottom R6G layer with respect to other layers, and it should be combined with the $G = 0$ component of $V_1$ to fit the relative energies of the two surfaces. $V_1$ determines the moiré potential modulation on the bottom (moiré-proximate) surface at the graphene/hBN interface. Its recent direct measurements[59] allow us to take an experiment-based value of $V_1 = 20$ meV, so that after self-consistent Hartree simulations, the moiré potential modulation on the bottom surface is 40~50 meV. Since in R6G the low-energy electron states are sublattice polarized on each layer, we neglect hopping variations induced by in-plane stretches.

**Self-consistent mean-field simulations of R6G/hBN**

With the relaxed moiré potential on the moiré-proximate layer modelled, we could simulate the entire system within self-consistent mean-field theory. Importantly, the moiré-distant surface experiences significant interactions from electron density modulations on the moiré-proximate interface (Fig. 5, Extended Data Fig. 2). We found that the Hartree approximation suffices to capture this effect and reproduce the experimental results. The Hartree term reads $U_H(r) = \int U(r-r')\,\rho(r')$, where $\rho(r)$ is the total electron density variation with respect to a uniform charge-neutral background, and $U(r-r') = \frac{e^2}{4\pi\varepsilon_r\varepsilon_0}\left(\frac{1}{|r-r'|} - \frac{1}{\sqrt{|r-r'|^2+\zeta^2}}\right)$ is the Coulomb potential screened by a metallic gate placed $\zeta$ = 30 nm away with the relative dielectric constant of hBN taken as $\varepsilon_r = 6$. The decay of Coulomb potential across R6G is neglected as the R6G thickness (~1.7 nm) is small. We did not include the Fock term in our model, as it is known to overestimate layer polarizations at small $D$, leading to a large charge-neutrality bandgap that does not agree with our experiments. Different subtraction schemes (also known as reference fields, reference states, or referenced densities) of the density matrix are known to yield very different physical consequences in self-consistent moiré calculations[27,28]. As discussed in more detail in Supplementary Information Section 6, our choice could be regarded as either the average scheme or a charge-neutrality scheme for moiréless R*n*G, as they both produce the same uniform charge-neutral total density background that we subtracted here. If we adopted a charge-neutral scheme for moiréful (hBN-aligned) R*n*G, on the other hand, the density redistributions caused by hBN alignment would be subtracted as a background at low filling ν, which was found to yield negligible moiré electronic modulations on the top surface at all twist angles.

Our simulations allow us to distinguish the influences of the twist angles, the stacking configurations, and lattice relaxation on the observed trans-moiré renormalization:

(1) The main control knob of moiré renormalization lies in the twist angles, as shown in Fig. 5 and extensively discussed in the main text.

(2) In contrast, according to post-relaxation simulations, the two different stacking configurations lead to similar moiré renormalization strengths on the moiré-distant surface in the transport-relevant $\nu \geqslant 0, D < 0$ regime (Extended Data Fig. 2, Supplementary Information Fig. 4). A previous theoretical puzzle is that while the two R*n*G/hBN stacking orientations ($\xi$ = 0 and 1) both exhibit moiré-distant QAHE[13,30,58] in experiment, the two R*n*G/hBN stacking orientations ($\xi$ = 0, 1) have very different consequences in the average scheme, with Chern insulators essentially only form under $\xi$ = 1[28].

Our imaging results, combined with theoretical simulations, find that moiré lattice relaxation is a key factor in the proper simulation of the Chern electronic states in rhombohedral moiré graphene. As shown in Extended Data Fig. 2, the three high-symmetry stacking sites ($C_N$, $C_{BN}$, $C_B$) of the moiré interface have different local charge densities due to R$n$G/hBN alignment, hence through a Hartree potential, leading to different on-site energies for electrons living on the moiré-distant surface. Without relaxation, such a trans-moiré potential has a unique minimum per moiré unit cell, $C_{BN}$ for $\xi$ = 1, and $C_N$ for $\xi$ = 0. Consequently, the lowest moiré conduction band is localized around these minima, leading to different implications between the two stacking orientations.

In the presence of lattice relaxation, however, the structurally most stable $C_B$ regions expands and 'absorbs' electrons from the other two sites, hence largely smearing out their differences. Consequently, the trans-moiré potential difference at $C_N$ and $C_{BN}$ decreases, leading to a hollow-cage-like 'network' of the trans-moiré potential minima (connecting $C_N$ and $C_{BN}$ sites) for both stackings. Only these post-relaxation electronic structures are consistent with our imaging results (Fig. 4). Importantly, in terms of topology, we have found relaxed trans-moiré potentials to favor the formation of topological non-trivial bands and can lead to Chern-band with $|C|$=1 in both stacking configurations ( $\xi$ = 1 and 0) under transport-accessible conditions within our approximations (see Supplementary Information Fig. 4 and discussions there), in agreement with experiment[13,30,58].

**Correspondence between real-space trans-moiré orbitals and reciprocal-space moiré minibands**

Supplementary Information Fig. 4 illustrates the connection between the experimentally observed trans-moiré orbitals and the theoretical moiré minibands in the $\nu \geq 0, D < 0$ ($V_D > 0$) regime. As shown there, the lowest-energy hollow-cage-like trans-moiré orbital corresponds to the lowest-lying moiré minibands; the match is better for $\xi = 1$ under small $D$ but becomes increasingly similar for both stackings at larger $D$. Higher-energy trans-moiré orbitals correspond to the real-space LDOS distributions from higher constant-energy cuts of moiré minibands—the spatial distributions are more parameter-sensitive and exhibit a range of different spectral shapes (Supplementary Information Figs. 4, 5). At even higher energies that correspond to the third lowest-lying moiré minibands, the main LDOS contributions are located near $C_B$ sites, hence consistently showing pancake-like spectral shape (Supplementary Information Figs. 4, 5).

We found that the lowest-lying moiré minibands, corresponding to the observed hollow-cage-like orbitals, hosts a Chern number $|C| = 1$ in the parameter regime of interest ($\nu = 1$ at large negative $D$), which, upon breaking valley and spin symmetries, can produce a Chern insulator with quantum anomalous Hall effects. Importantly, both the observed orbital shapes and the Chern miniband formation persist to the large-$D$ limit that is relevant to transport measurements, as shown in Supplementary Information Fig. 4 and discussed in detail in Supplementary Information Section 5. This allows us to extrapolate our conclusion into the QAH transport parameter regime, as detailed there.

**Testing tip-induced charging via charging peaks and tip-height-dependent maps**

As mentioned in the main text, tip-induced charging could generate a variety of real-space patterns in related moiré systems[38-40] that might carry visual similarities to our imaging results. We designed various experimental tests to rule out tip-induced charging as the origin of our observations.

Tip-induced charging occurs when the tip's local gating, tunable by tip's $x$, $y$, $z$ position and bias voltage $V_b$, becomes strong enough to trigger charging of a proximate local object. These dependences of tip-induced charging thus provide us with various means to test whether the observation originates from charging.

We shall start with the $V_b$-dependence as it is widely established that charging is tied to a pronounced charging d$I$/d$V$ feature (a 'charging peak') in STM d$I$/d$V$ spectra that marks the onset of charging. For example, in conventional charging of single atoms/molecules[62,82], the charging rings in STM imaging are associated with such charging-peak features in STM d$I$/d$V$ spectra. This is also the case of charging in moiré 'quantum-dot arrays', where the sharp charging d$I$/d$V$ peaks are visible in the gate-dependent d$I$/d$V$ spectra in addition to smoothly varying electronic structure features[38-40]. In contrast, after carefully examining all data sets of our acquired gate-dependent d$I$/d$V$ spectra (acquired on seven different samples with various different set-points), we found none of them to show signatures of charging peaks. Extended Data Fig. 10, for example, presents gate-dependent d$I$/d$V$ spectra acquired at all high-symmetry stacking sites ($C_{BN}$, $C_B$, and $C_N$) in R6G/hBN D1 (that accompany main Fig. 3h)—where no signatures of charging peaks were observed beyond the anticipated electronic structure features from simulations (cf. Extended Data Fig. 8), inconsistent with a charging origin.

A second test of tip-induced charging is by changing the tip heights, $z$, as the tip-sample distance should influence the local tip-gating capacitance, thus affecting both the real-space patterns of charging rings and spectroscopic positions of d$I$/d$V$ charging peaks[61-63]. To test whether such effects are present in our data, on the same area of R6G/hBN D3 with the same microtip, we measured three sets of d$I$/d$V$ maps and spectra at three different tip heights and found very little tip-height dependence (Extended Data Fig. 5). First, by taking tip-height dependent d$I$/d$V$ spectra (right panels in Extended Data Fig. 5), we found the d$I$/d$V$ spectroscopic features to be nearly identical at varying tip heights, which is hence inconsistent with their interpretation as charging peaks or other charging-induced d$I$/d$V$ features, but consistent with interpretation as electronic structure features (which agrees with material simulations in Extended Data Fig. 8). Second, by taking constant-current d$I$/d$V$ maps at different current setpoints using the same microtip, we found nearly identical d$I$/d$V$ real-space patterns at different tip-sample distances at all bias voltages within the measurement range (Extended Data Fig. 5; the complete data sets are shown in Supplementary Movie 2, where the weak stripes were caused by a very slow temperature oscillation occurring during this data acquisition). These observations highlighted the robustness of the trans-moiré orbitals at different tip-sample distances, inconsistent with their interpretations as tip-induced charging features.

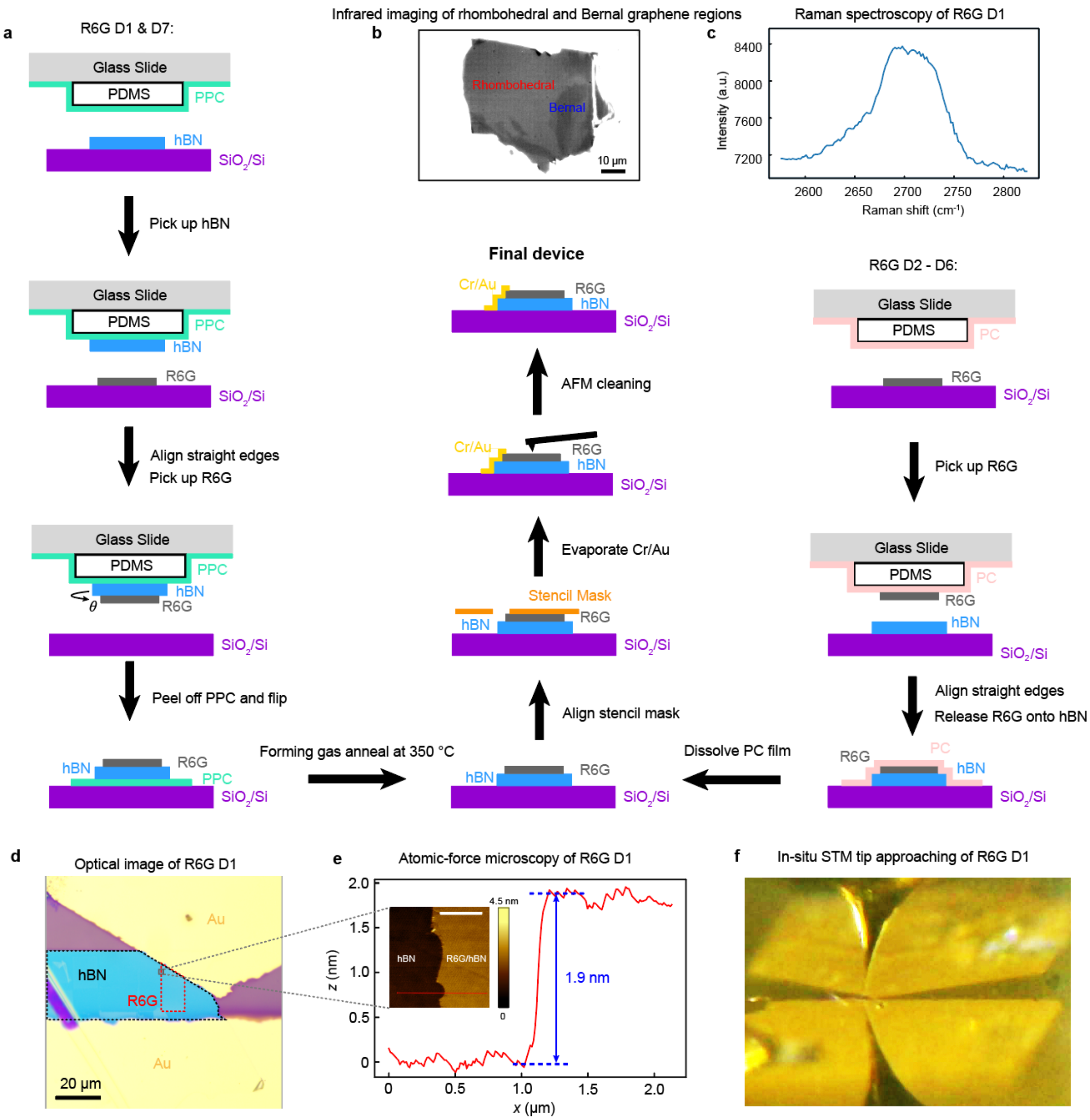

**Extended Data Figure 1. R*n*G/hBN device fabrication. a**, Fabrication procedures of STM-grade R6G/hBN devices (Methods). **b,** Infrared imaging showing rhombohedral and Bernal regions of a hexalayer graphene flake (which are later separated by laser cutting to avoid collateral relaxation). **c,** Raman spectrum of R6G D1 showing its rhombohedral nature; other devices show similar results. **d,** Optical image of R6G D1, where the dashed red box marks the R6G flake position. **e,** AFM image across the R6G/hBN step of R6G D1 showing the hexalayer flake thickness; other devices show similar results. Scale bar in inset: 1 μm. **f,** Optical image taken inside the STM upon approaching the STM tip to R6G/hBN D1 (the bottom 'tip' is the reflection).

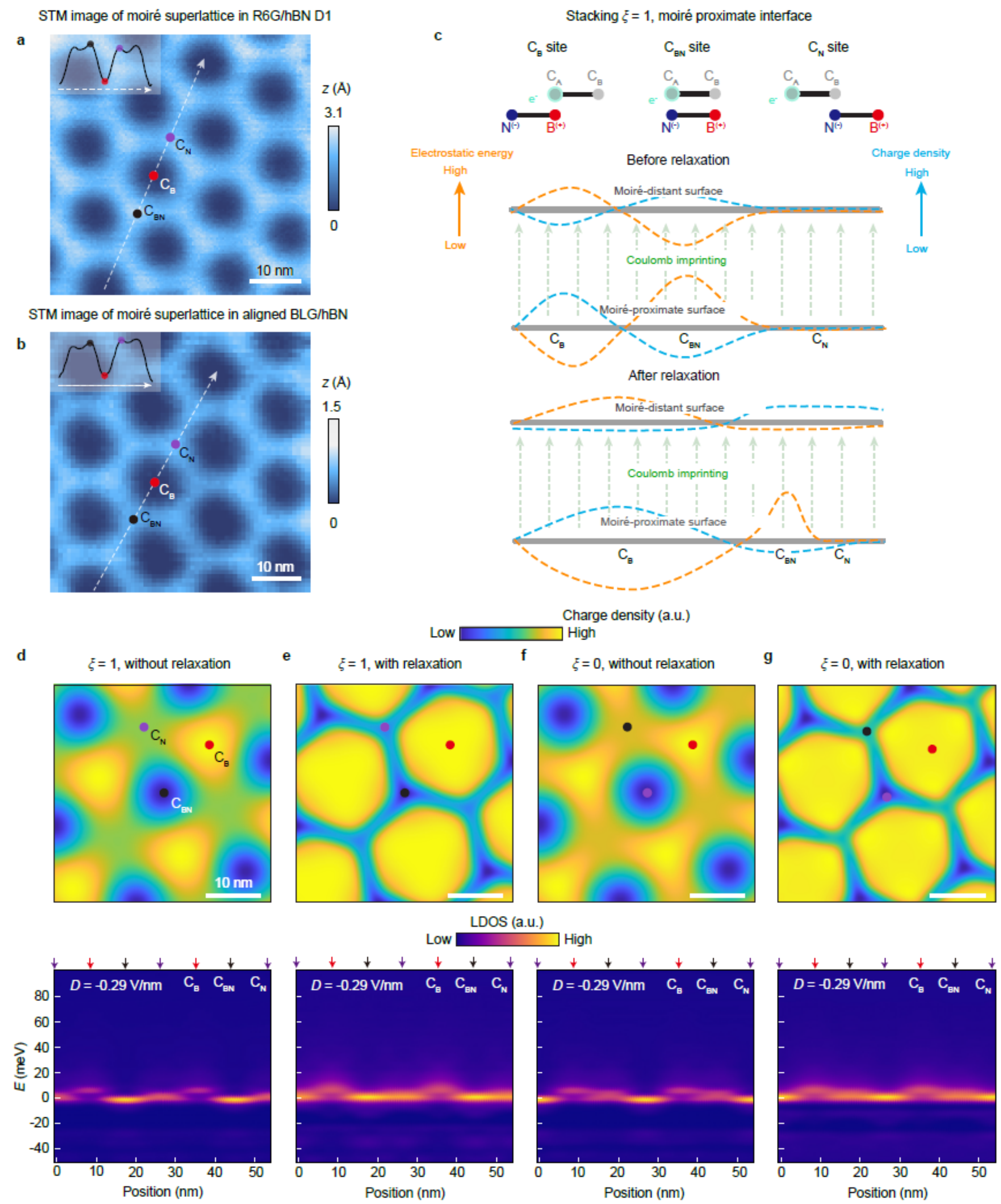

**Extended Data Figure 2. Lattice relaxation and its influences on electronic structures of R6G/hBN. a**, STM topograph of aligned R6G/hBN (D1, similar to Fig. 1b) ($V_b$ = -0.2 V, $I_t$ = 5 pA). **b**, STM topograph of bilayer graphene/hBN showing overall similar topographic features to R6G/hBN ($V_b$ = 0.2 V, $I_t$ = 10 pA). **c**, Top: Schematics of the $\xi = 1$ R6G/hBN stacking configuration. Middle: Schematics of the electrostatic energies & charge density distributions at the moiré-proximate and moiré-distant layers before lattice relaxation. Bottom: Schematics after lattice relaxation. **d–g,** Top: Simulated charge density distributions on the moiré-proximate surface under $\xi = 0$ and 1, without and with lattice relaxation. The two stacking configurations produce similar charge density distributions only after relaxation. Bottom: Simulated moiré-distant flat-band renormalization, showing similar behavior under $\xi = 0$ and 1.

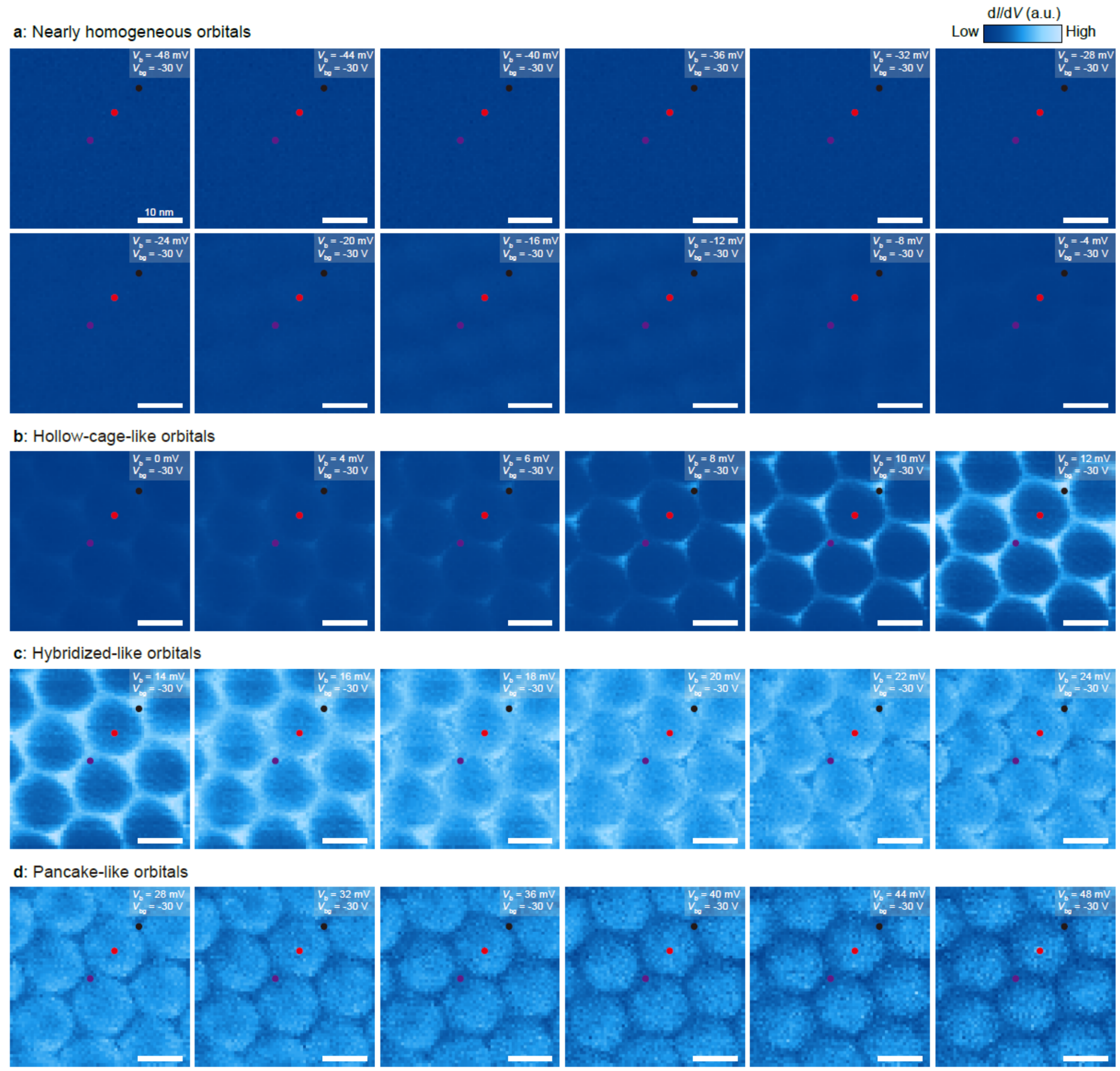

**Extended Data Figure 3. Complete set of d$I$/d$V$ maps acquired at different bias voltages on R6G/hBN D1 ($\theta$ = 0.28°). a,** At energies below the flat band, the moiré-distant surface exhibits nearly spatially homogeneous d$I$/d$V$ signals. **b**, Inside the flat band, the lowest-energy orbitals feature a distinct hollow-cage like shape. **c**, At intermediate energies, the moiré-distant flat band shows spectral shapes that resemble more strongly linked or hybridized orbitals. **d**, The upper half of the moiré-distant flat band is dominated by pancake-like orbitals that stay very similar in shape with enhanced spectral weights in the $C_B$ stacking regions ($V_b$ = -0.1 V, $I_t$ = 30 pA, $V_{mod}$ = 3 mV, $V_{bg}$ = -30 V, $T$ = 3.4 K).

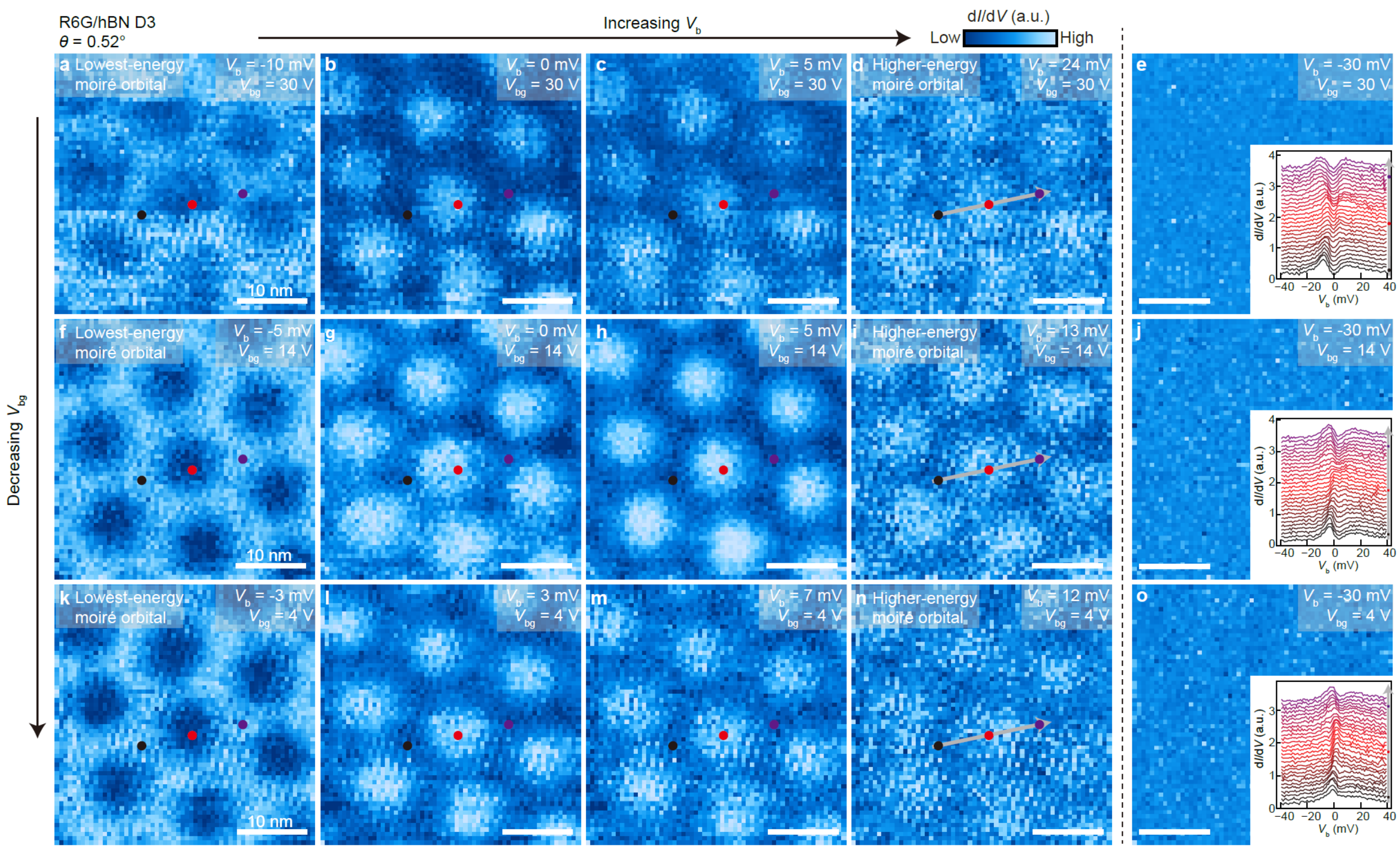

**Extended Data Figure 4. Trans-moiré orbitals measured in R6G/hBN D3 ($\theta$ = 0.52°), akin to Fig. 4. a-e,** STM d$I$/d$V$ maps at $V_{bg}$ = 30 V, corresponding to $v$ = 2.4, $D$ = -0.19 V/nm, showing electrons filled into the hollow-cage-like trans-moiré orbitals (**a**) ($V_b$ = -0.3 V, $I_t$ = 20 pA, $V_{mod}$ = 3 mV). Inset in **e**: Spatially dependent d$I$/d$V$ spectra acquired along a high-symmetry direction (arrow in **d**). The overall features observed in R6G/hBN D3 resembles the results of R6G/hBN D1 (Fig.4). **f-j,** Same as **a-e,** but at $V_{bg}$ = 14 V, corresponding to $v$ = 1.0, $D$ = -0.10 V/nm, showing electrons filled into the hollow-cage-like trans-moiré orbitals (**f**) ($V_b$ = -0.3 V, $I_t$ = 20 pA, $V_{mod}$ = 3 mV). **k-o,** Same as **a-e,** but at $V_{bg}$ = 4 V, corresponding to $v$ = 0.1, $D$ = -0.04 V/nm ($V_b$ = -0.3 V, $I_t$ = 20 pA, $V_{mod}$ = 3 mV). ($v$, $D$) parameters are summarized in Extended Data Fig. 9. All measurements performed at $T$= 2.3 K.

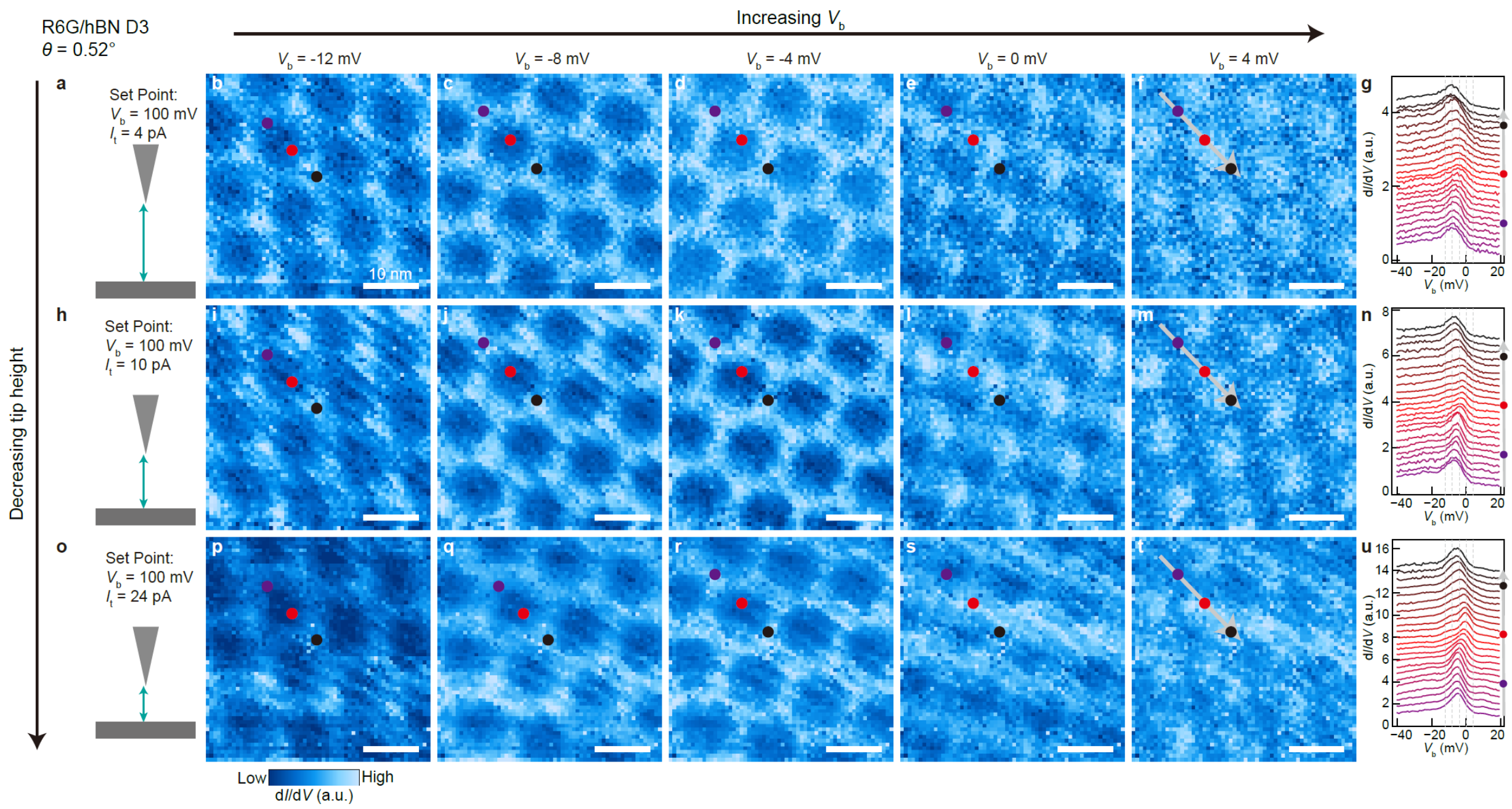

**Extended Data Figure 5. Tip-height-dependent d$I$/d$V$ maps of R6G/hBN D3 ($\theta$ = 0.52°) showing insensitivity of the measurement results to tip heights. a,** Schematic showing the large tip-sample separation under the labelled setpoints. **b-f,** Constant-current d$I$/d$V$ maps taken when the tip is far from the R6G surface, showing overall similarities to Fig. 4 ($I_t$ = 5 pA, $V_b$ = 0.1 V). **g,** Spatially dependent d$I$/d$V$ spectra acquired along a high-symmetry direction (grey arrow in **f**). **h-n,** Same as **a**-**f**, but with $I_t$ = 10 pA, $V_b$ = 0.1 V, so that the tip is at an intermediate separation from the sample, but nevertheless showing similar results. **o-u,** Same as **h**-**n**, but with $I_t$ = 24 pA, $V_b$ = 0.1 V, so that the tip is close to the sample, also showing similar results. All measurements with $V_{mod}$ = 3 mV and $V_{bg}$ = 0 V. The complete data sets are shown in Supplementary Movie 2. All measurements performed at $T$= 2.1 K.

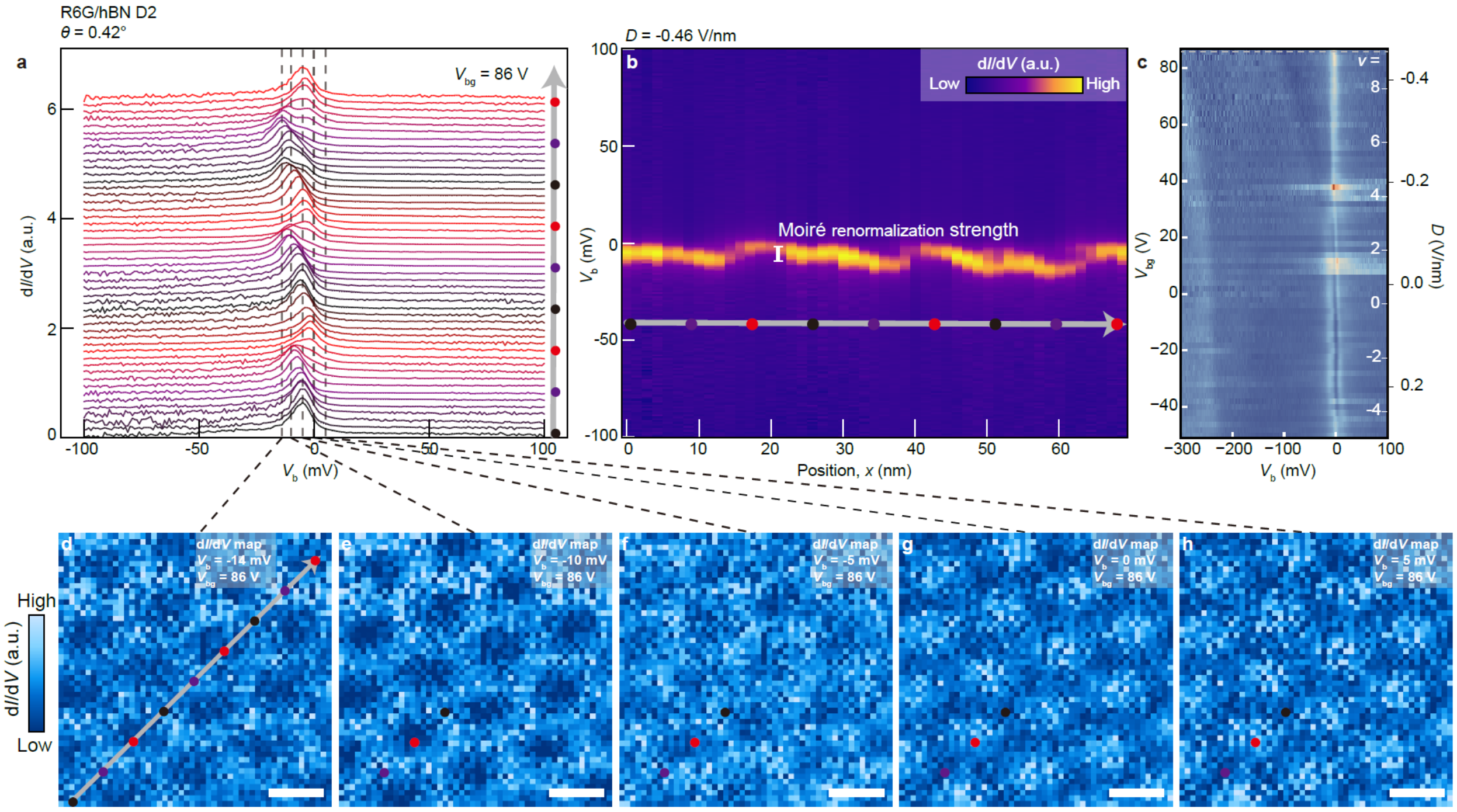

**Extended Data Figure 6. Robustness of trans-moiré orbitals under QAH-relevant displacement fields ($D$ = -0.46 V/nm) in R6G/hBN D2 ($\theta$ = 0.42°). a,** Spatially dependent d$I$/d$V$ spectra (along the grey single arrow in **d**) showing strong moiré-periodic flat-band modulations even under this large-negative $D$ field (taken at a large $V_{bg}$ of 86 V; corresponding to $D$ = -0.46 V/nm, $v$ = 9.5). **b**, Same data as in **a**, respectively, but rotated by 90° and plotted in color. **c**, Gate-dependent d$I$/d$V$ spectra acquired at a $C_B$ site of R6G/hBN D2. **d-h,** Constant-current d$I$/d$V$ maps at $V_{bg}$ = 86 V ($v$ = 9.5, $D$ = -0.46 V/nm) showing emergent trans-moiré orbitals with overall similarities to Fig. 4. In particular, the low-energy edge of the flat band (filled below the Fermi level) still exhibits hollow-cage-like orbitals (panels **d**, **e**) with low LDOS near $C_B$ sites (red dots) and high LDOS forming a hexagonal network connecting $C_N$ and $C_{BN}$ sites. Setpoints: $V_b$ = 0.3 V, $I_t$ = 60 pA, $V_{mod}$ = 3 mV(**a-b** & **d-h**); $V_b$ = -0.3 V, $I_t$ = 80 pA, $V_{mod}$ = 3 mV(**c**). All measurements performed at $T$ = 6.0 K.

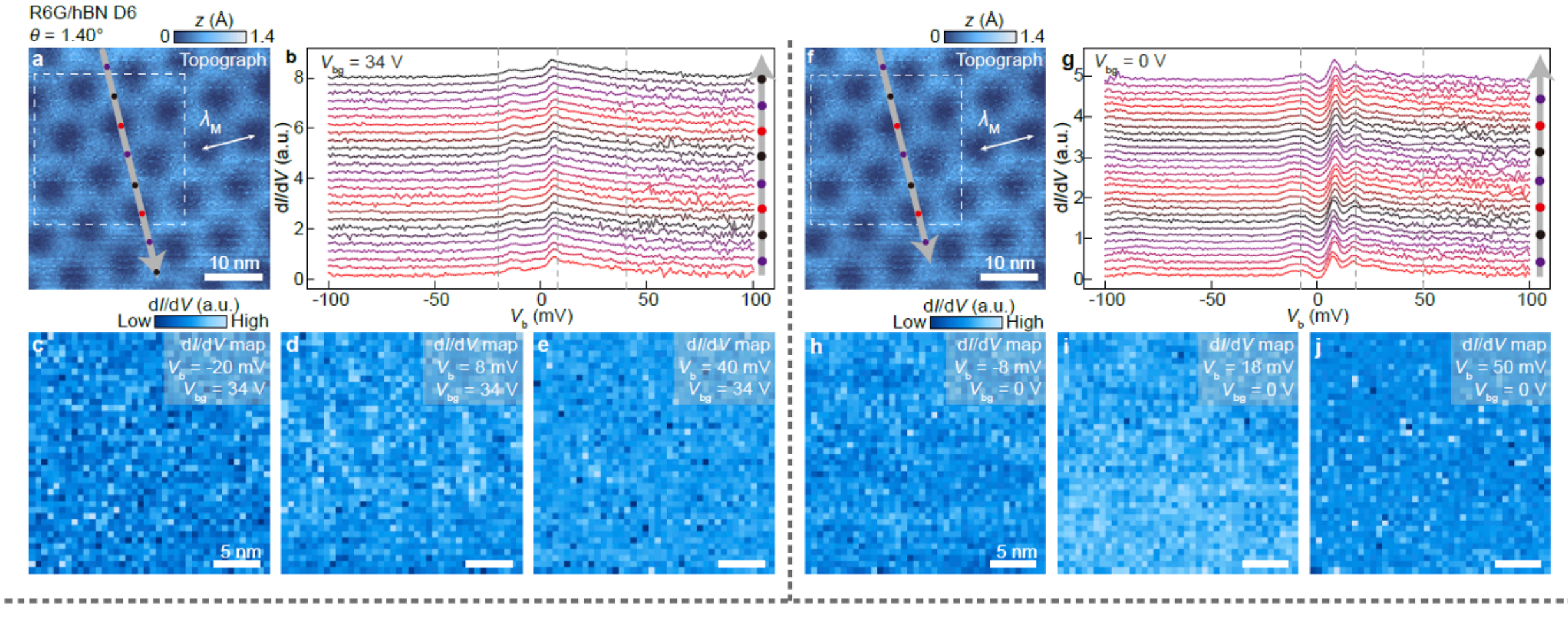

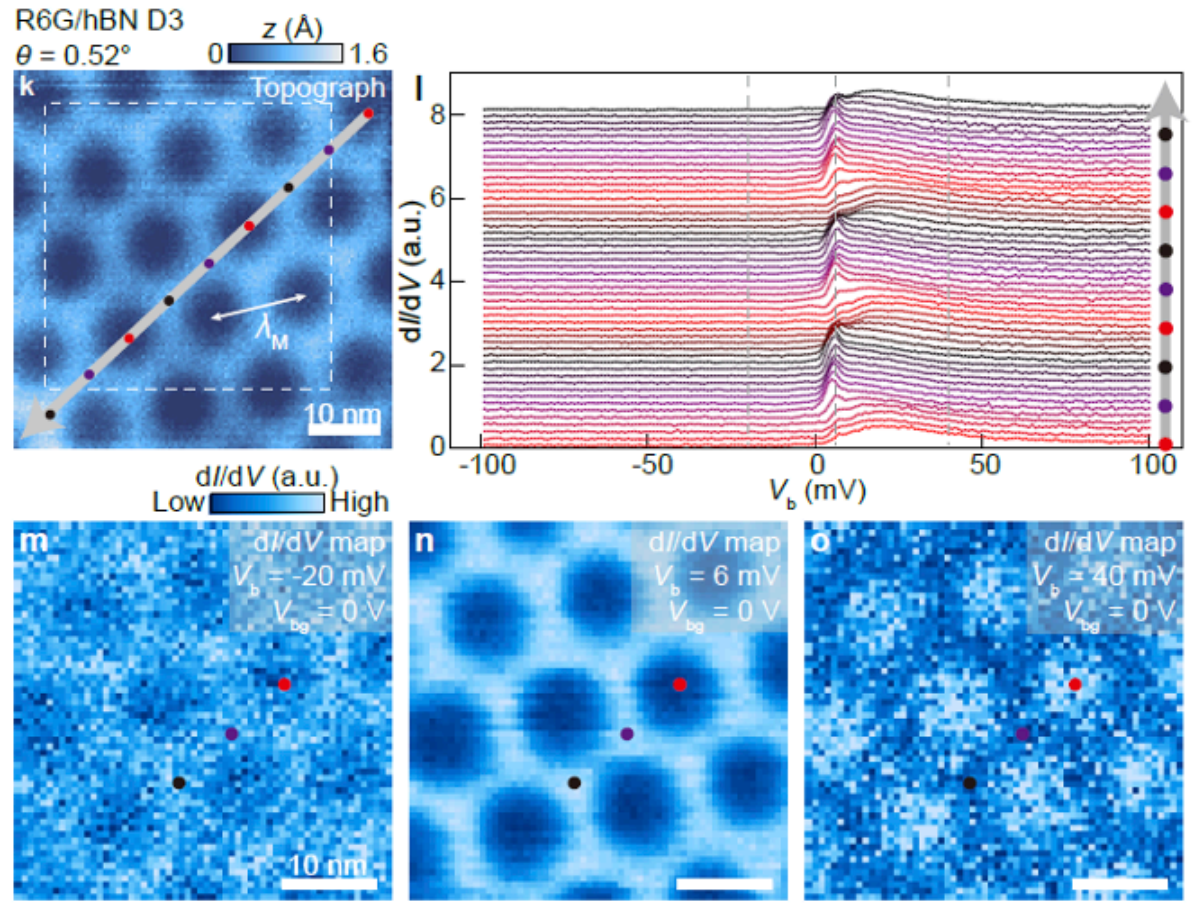

**Extended Data Figure 7. STM/STS measurement results of R6G/hBN D6 ($\theta$ = 1.40°) and D3 ($\theta$ = 0.52°) using the same microtip. a-j,** Measurement results of R6G/hBN D6 at $V_{bg}$ = 34 V & 0 V, showing spatially homogeneous spectral density, similar to Fig. 2. **k-o**, Measurement results of R6G/hBN D3 with the same microtip, showing the emergence of trans-moiré orbitals and moiré flat-band renormalization, similar to Figs. 3, 4. This experiment explicitly invalidates tip dependence as a possible cause of our observations. Setpoints: **a**, $V_b$ = 0.3 V, $I_t$ = 3 pA; **b**, $V_b$ = 0.1 V, $I_t$ = 20 pA, $V_{mod}$ = 2 mV; **c-e**, $V_b$ = 0.3 V, $I_t$ = 40 pA, $V_{mod}$ = 2 mV; **f**, $V_b$ = 0.3 V, $I_t$ = 3 pA; **g**, $V_b$ = 0.1 V, $I_t$ = 20 pA, $V_{mod}$ = 2 mV; **h-j**, $V_b$ = 0.1 V, $I_t$ = 20 pA, $V_{mod}$ = 2 mV; **k**, $V_b$ = -0.3 V, $I_t$ = 3 pA; **l**, $V_b$ = 0.2 V, $I_t$ = 20 pA, $V_{mod}$ = 2 mV; **m-o**, $V_b$ = 0.3 V, $I_t$ = 40 pA, $V_{mod}$ = 2 mV. All measurements performed at $T$ = 2.3 K.

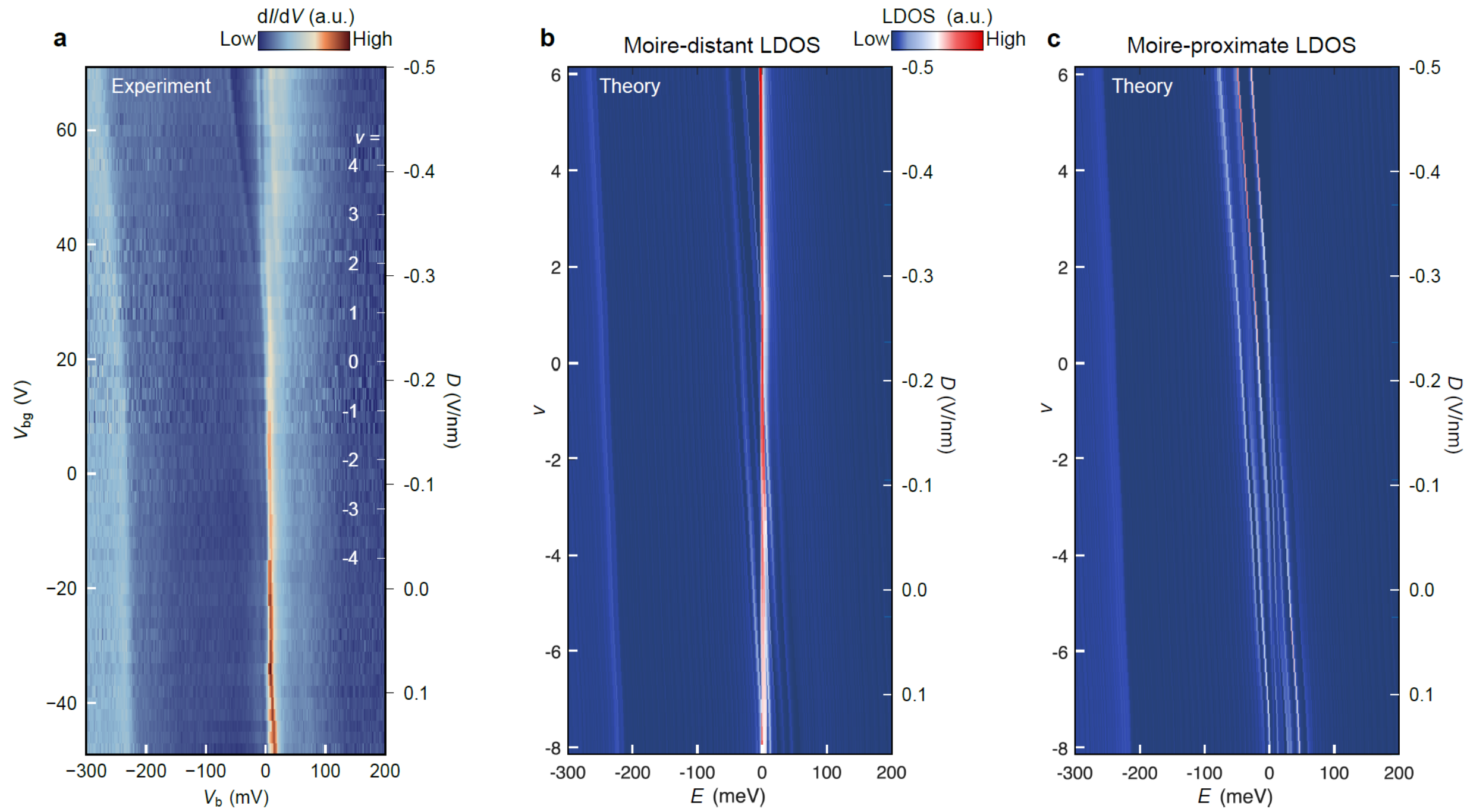

**Extended Data Figure 8. Simulations of experimental gate-dependent d$I$/d$V$ spectra in Fig. 3h. a**, Gate-dependent d$I$/d$V$ spectrum of R6G D1, same as Fig. 3h ($V_b$ = -0.4 V, $I_t$ = 30 pA, $V_{mod}$ = 3 mV, $T$ = 3.4 K). **b**, Theoretical gate-dependent moiré-distant LDOS from best-fit simulations of R6G/hBN D1 showing similar features to **a**, including the gate-dependent evolutions of the low-energy moiré-distant flat band, the displacement-field-induced LDOS suppression around positive fillings, and moiré-related faint LDOS features that cross the moiré-distant flat band. **c**, Same as **b** for moiré-proximate LDOS. The distinct slope from the moiré-distant flat band arises from the displacement-field-induced energy separation of the two surface states (see more discussion in Supplementary Information Section 3).

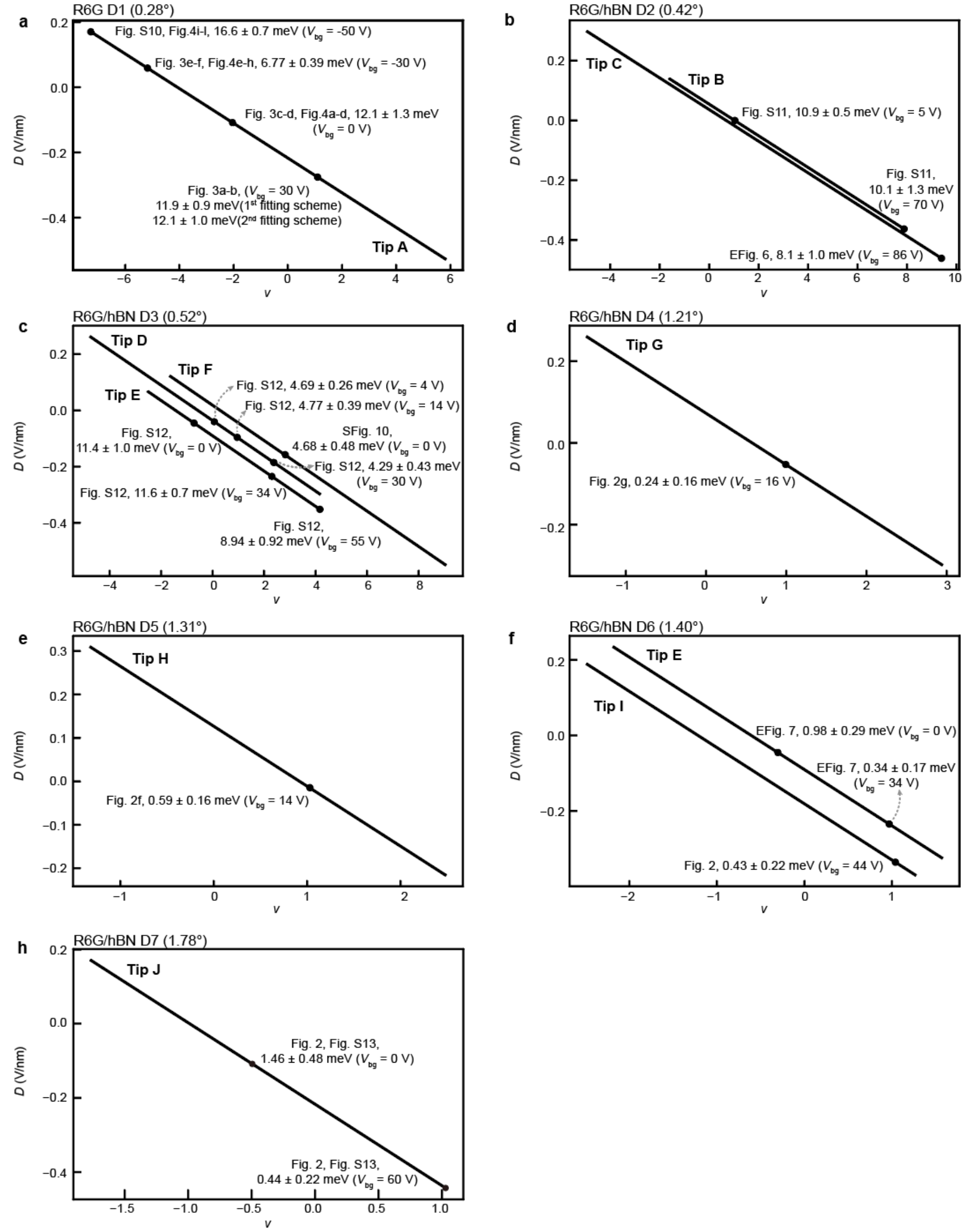

**Extended Data Figure 9. Summary of ($v$, $D$) parameters and fitted moiré renormalization strengths.**

Lines and dots representing ($v$, $D$) parameters obtained from best-fit simulations to gate-dependent d$I$/d$V$ spectra, as reported in corresponding Main, Extended Data, and Supplementary Information Figures. Labels showing moiré renormalization strengths from two-step fits to spatially resolved d$I$/d$V$ spectra (as detailed Supplementary Information Section 7). Each panel corresponds to a different device and each line represents a different tip; as only tips with good work-function matching to graphene is used, different lines nearly align with each other.

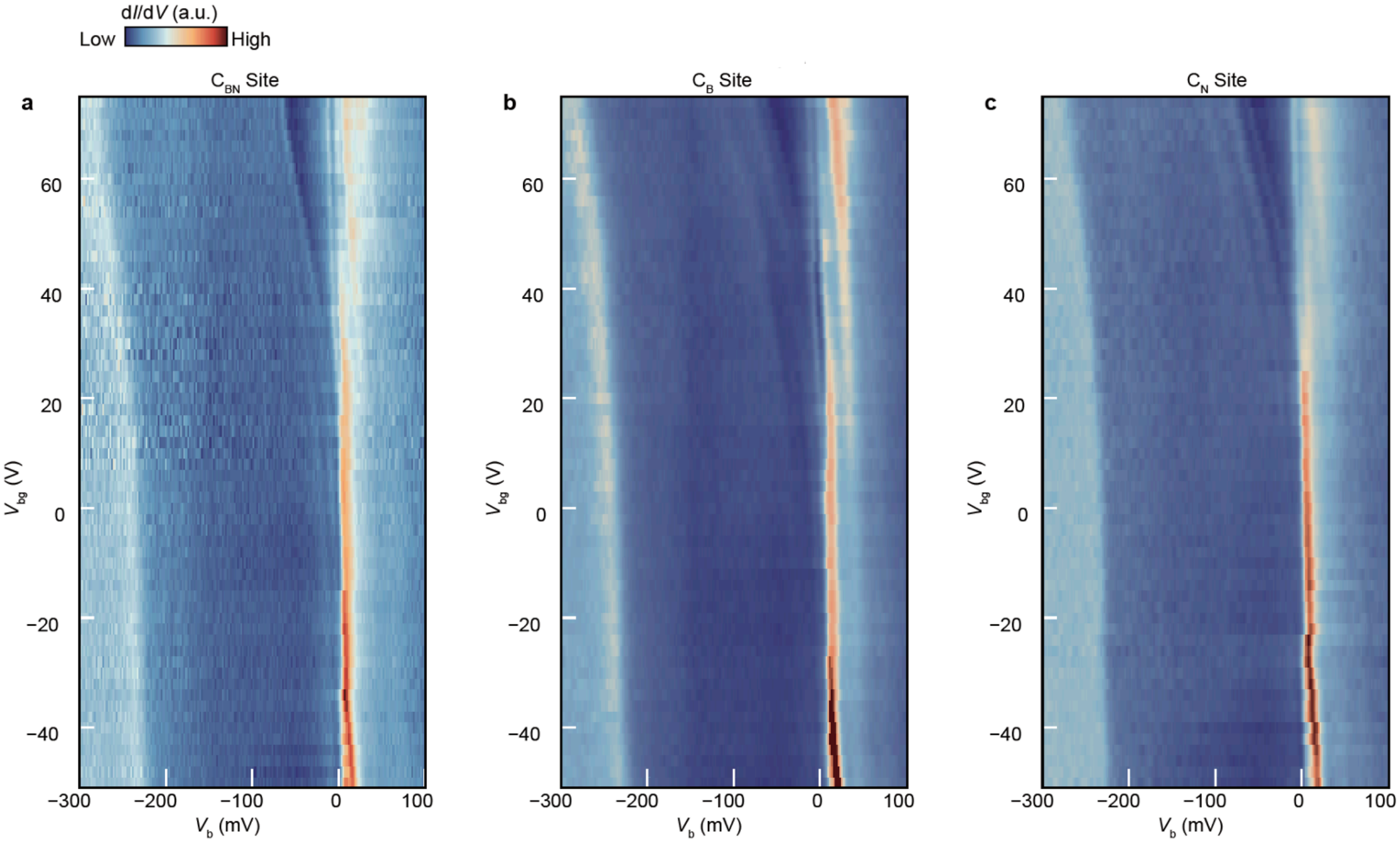

**Extended Data Figure 10. Gate-dependent d*I*/d*V* spectra acquired at different high-symmetry stacking sites in R6G/hBN D1. a,** Gate-dependent d*I*/d*V* spectra acquired at a $C_{BN}$ site in R6G/hBN D1, same as Fig. 3h. **b,** Gate-dependent d*I*/d*V* spectra acquired at a $C_B$ site in R6G/hBN D1. **c,** Gate-dependent d*I*/d*V* spectra acquired at a $C_N$ site in R6G/hBN D1. The spectra taken at different sites show overall similar features, with an energy shift due to the observed flat-band renormalization. All spectra were measured at $V_b$ = -0.4 V, $I_t$ = 30 pA, $V_{mod}$ = 3 mV, $T$ = 3.4 K.

**References:**

1 Regnault, N. & Bernevig, B. A. Fractional chern insulator. *Physical Review X* **1**, 021014 (2011).

2 Neupert, T., Santos, L., Chamon, C. & Mudry, C. Fractional quantum Hall states at zero magnetic field. *Physical Review Letters* **106**, 236804 (2011).

3 Sheng, D., Gu, Z.-C., Sun, K. & Sheng, L. Fractional quantum Hall effect in the absence of Landau levels. *Nature Communications* **2**, 389 (2011).

4 Sun, K., Gu, Z., Katsura, H. & Das Sarma, S. Nearly flatbands with nontrivial topology. *Physical Review Letters* **106**, 236803 (2011).

5 Tang, E., Mei, J.-W. & Wen, X.-G. High-temperature fractional quantum Hall states. *Physical Review Letters* **106**, 236802 (2011).

6 Cai, J. *et al.* Signatures of fractional quantum anomalous Hall states in twisted MoTe2. *Nature* **622**, 63-68 (2023). https://doi.org/10.1038/s41586-023-06289-w

7 Zeng, Y. *et al.* Thermodynamic evidence of fractional Chern insulator in moiré MoTe2. *Nature* **622**, 69-73 (2023). https://doi.org/10.1038/s41586-023-06452-3

8 Park, H. *et al.* Observation of fractionally quantized anomalous Hall effect. *Nature* **622**, 74-79 (2023). https://doi.org/10.1038/s41586-023-06536-0

9 Xu, F. *et al.* Observation of Integer and Fractional Quantum Anomalous Hall Effects in Twisted Bilayer MoTe2. *Physical Review X* **13**, 031037 (2023). https://doi.org/10.1103/PhysRevX.13.031037

10 Lu, Z. *et al.* Fractional quantum anomalous Hall effect in multilayer graphene. *Nature* **626**, 759-764 (2024). https://doi.org/10.1038/s41586-023-07010-7

11 Xie, J. *et al.* Tunable fractional Chern insulators in rhombohedral graphene superlattices. *Nature Materials* **24**, 1042-1048 (2025). https://doi.org/10.1038/s41563-025-02225-7

12 Lu, Z. *et al.* Extended quantum anomalous Hall states in graphene/hBN moiré superlattices. *Nature* **637**, 1090-1095 (2025). https://doi.org/10.1038/s41586-024-08470-1

13 Li, C. *et al.* Stacking-Orientation and Twist-Angle Control on Integer and Fractional Chern Insulators in Moiré Rhombohedral Graphene. *arXiv preprint arXiv:2505.01767* (2025).

14 Choi, Y. *et al.* Superconductivity and quantized anomalous Hall effect in rhombohedral graphene. *Nature* **639**, 342-347 (2025). https://doi.org/10.1038/s41586-025-08621-y

15 Bernevig, B. A. *et al.* Fractional quantization in insulators from Hall to Chern. *Nature Physics* **21**, 1702-1713 (2025). https://doi.org/10.1038/s41567-025-03072-8

16 Cao, T., Fu, L., Ju, L., Xiao, D. & Xu, X. Fractional Quantum Anomalous Hall Effect. *Annual Review of Condensed Matter Physics* (2025). https://doi.org/10.1146/annurev-conmatphys-031524-071133

17 Dong, J. *et al.* Anomalous Hall Crystals in Rhombohedral Multilayer Graphene. I. Interaction-Driven Chern Bands and Fractional Quantum Hall States at Zero Magnetic Field. *Physical Review Letters* **133**, 206503 (2024). https://doi.org/10.1103/PhysRevLett.133.206503

18 Soejima, T. *et al.* Anomalous Hall crystals in rhombohedral multilayer graphene. II. General mechanism and a minimal model. *Physical Review B* **110** (2024). https://doi.org/10.1103/PhysRevB.110.205124

19 Zhou, B., Yang, H. & Zhang, Y.-H. Fractional Quantum Anomalous Hall Effect in Rhombohedral Multilayer Graphene in the Moiréless Limit. *Physical Review Letters* **133**, 206504 (2024). https://doi.org/10.1103/PhysRevLett.133.206504

20 Dong, Z., Patri, A. S. & Senthil, T. Stability of anomalous Hall crystals in multilayer rhombohedral graphene. *Physical Review B* **110** (2024). https://doi.org/10.1103/PhysRevB.110.205130

21 Dong, Z., Patri, A. S. & Senthil, T. Theory of Quantum Anomalous Hall Phases in Pentalayer Rhombohedral Graphene Moiré Structures. *Physical Review Letters* **133** (2024). https://doi.org/10.1103/PhysRevLett.133.206502

22 Tan, T. & Devakul, T. Parent Berry Curvature and the Ideal Anomalous Hall Crystal. *Physical Review X* **14** (2024). https://doi.org/10.1103/PhysRevX.14.041040

23 Zeng, Y., Guerci, D., Crépel, V., Millis, A. J. & Cano, J. Sublattice structure and topology in spontaneously crystallized electronic states. *Physical Review Letters* **132**, 236601 (2024).

24 Shen, X. *et al.* Stabilizing fractional Chern insulators via exchange interaction in moiré systems. *arXiv preprint arXiv:2405.12294* (2024).

25 Guo, Z., Lu, X., Xie, B. & Liu, J. Fractional Chern insulator states in multilayer graphene moiré superlattices. *Physical Review B* **110**, 075109 (2024). https://doi.org/10.1103/PhysRevB.110.075109

26 Huang, K., Das Sarma, S. & Li, X. Fractional quantum anomalous Hall effect in rhombohedral multilayer graphene with a strong displacement field. *Physical Review B* **111** (2025). https://doi.org/10.1103/PhysRevB.111.075130

27 Huang, K., Li, X., Das Sarma, S. & Zhang, F. Self-consistent theory of fractional quantum anomalous Hall states in rhombohedral graphene. *Physical Review B* **110** (2024).

https://doi.org/10.1103/PhysRevB.110.115146

28 Kwan, Y. H. *et al.* Moiré fractional Chern insulators. III. Hartree-Fock phase diagram, magic angle regime for Chern insulator states, role of moiré potential, and Goldstone gaps in rhombohedral graphene superlattices. *Physical Review B* **112**, 075109 (2025). https://doi.org/10.1103/PhysRevB.112.075109

29 Lu, X., Yang, Y., Guo, Z. & Liu, J. General Many-Body Perturbation Framework for Moiré Systems. *arXiv preprint arXiv:2509.19764* (2025).

30 Huo, Z. *et al.* Does Moire Matter? Critical Moire Dependence with Quantum Fluctuations in Graphene Based Integer and Fractional Chern Insulators. *arXiv preprint arXiv:2510.15309* (2025).

31 Andrei, E. Y. *et al.* The marvels of moiré materials. *Nature Reviews Materials* **6**, 201-206 (2021). https://doi.org/10.1038/s41578-021-00284-1

32 Nuckolls, K. P. & Yazdani, A. A microscopic perspective on moiré materials. *Nature Reviews Materials* **9**, 460-480 (2024). https://doi.org/10.1038/s41578-024-00682-1

33 Oh, M. *et al.* Evidence for unconventional superconductivity in twisted bilayer graphene. *Nature* **600**, 240-245 (2021).

34 Kim, H. *et al.* Evidence for unconventional superconductivity in twisted trilayer graphene. *Nature* **606**, 494-500 (2022).

35 Kim, H. *et al.* Resolving intervalley gaps and many-body resonances in moiré superconductors. *Nature* **650**, 592-598 (2026). https://doi.org/10.1038/s41586-025-10067-1

36 Nuckolls, K. P. *et al.* Strongly correlated Chern insulators in magic-angle twisted bilayer graphene. *Nature* **588**, 610-615 (2020).

37 Choi, Y. *et al.* Correlation-driven topological phases in magic-angle twisted bilayer graphene. *Nature* **589**, 536-541 (2021).

38 Li, H. *et al.* Imaging two-dimensional generalized Wigner crystals. *Nature* **597**, 650-654 (2021).

39 Li, H. *et al.* Mapping charge excitations in generalized Wigner crystals. *Nature nanotechnology* **19**, 618-623 (2024).

40 Li, H. *et al.* Imaging local discharge cascades for correlated electrons in WS2/WSe2 moiré superlattices. *Nature Physics* **17**, 1114-1119 (2021).

41 Han, T. *et al.* Signatures of chiral superconductivity in rhombohedral graphene. *Nature* **643**, 654-661 (2025). https://doi.org/10.1038/s41586-025-09169-7

42 Chen, G. *et al.* Evidence of a gate-tunable Mott insulator in a trilayer graphene moiré superlattice. *Nature Physics* **15**, 237-241 (2019). https://doi.org/10.1038/s41567-018-0387-2

43 Yang, J. *et al.* Spectroscopy signatures of electron correlations in a trilayer graphene/hBN moiré superlattice. *Science* **375**, 1295-1299 (2022). https://doi.org/doi:10.1126/science.abg3036

44 Chen, G. *et al.* Signatures of tunable superconductivity in a trilayer graphene moiré superlattice. *Nature* **572**, 215-219 (2019). https://doi.org/10.1038/s41586-019-1393-y

45 Zhou, H., Xie, T., Taniguchi, T., Watanabe, K. & Young, A. F. Superconductivity in rhombohedral trilayer graphene. *Nature* **598**, 434-438 (2021). https://doi.org/10.1038/s41586-021-03926-0

46 Zhou, H. *et al.* Half- and quarter-metals in rhombohedral trilayer graphene. *Nature* **598**, 429-433 (2021). https://doi.org/10.1038/s41586-021-03938-w

47 Chen, G. *et al.* Tunable correlated Chern insulator and ferromagnetism in a moiré superlattice. *Nature* **579**, 56-61 (2020). https://doi.org/10.1038/s41586-020-2049-7

48 Zhang, F., Sahu, B., Min, H. & MacDonald, A. H. Band structure of ABC-stacked graphene trilayers. *Physical Review B* **82** (2010). https://doi.org/10.1103/PhysRevB.82.035409

49 Koshino, M. & McCann, E. Trigonal warping and Berry's phase Nπ in ABC-stacked multilayer graphene. *Physical Review B* **80** (2009). https://doi.org/10.1103/PhysRevB.80.165409

50 Zhang, Y. *et al.* Layer-dependent evolution of electronic structures and correlations in rhombohedral multilayer graphene. *Nature Nanotechnology* **20**, 222-228 (2025).

51 Liu, Y. *et al.* Visualizing incommensurate inter-valley coherent states in rhombohedral trilayer graphene. *arXiv preprint arXiv:2411.11163* (2024).

52 Liu, Y. *et al.* Electronic Correlations in Rhombohedral Graphene at Atomic Scale. *Physical Review Letters* **135**, 156401 (2025).

53 Kerelsky, A. *et al.* Moiréless correlations in ABCA graphene. *Proceedings of the National Academy of Sciences* **118** (2021). https://doi.org/10.1073/pnas.2017366118

54 Jung, J., Raoux, A., Qiao, Z. & MacDonald, A. H. Ab initio theory of moiré superlattice bands in layered two-dimensional materials. *Physical Review B* **89**, 205414 (2014).

55 Krisna, L. P. A. & Koshino, M. Moiré phonons in graphene/hexagonal boron nitride moiré superlattice. *Physical Review B* **107**, 115301 (2023). https://doi.org/10.1103/PhysRevB.107.115301

56 Herzog-Arbeitman, J. *et al.* Moiré fractional Chern insulators. II. First-principles calculations and continuum models of rhombohedral graphene superlattices. *Physical Review B* **109** (2024). https://doi.org/10.1103/PhysRevB.109.205122

57 Woods, C. *et al.* Commensurate–incommensurate transition in graphene on hexagonal boron nitride. *Nature Physics* **10**, 451-456 (2014).

58 Uzan, M. *et al.* hBN alignment orientation controls moiré strength in rhombohedral graphene. *arXiv preprint arXiv:2507.20647* (2025).

59 Klein, D. R. *et al.* Imaging the sub-moire potential using an atomic single electron transistor. *Nature* (2026). https://doi.org/10.1038/s41586-025-10085-z

60 Zhang, H. *et al.* Moiré enhanced flat band in rhombohedral graphene. *Nature Materials* (2025). https://doi.org/10.1038/s41563-025-02416-2

61 Song, C.-L. *et al.* Gating the charge state of single Fe dopants in the topological insulator Bi${}_{2}$Se${}_{3}$ with a scanning tunneling microscope. *Physical Review B* **86**, 045441 (2012). https://doi.org/10.1103/PhysRevB.86.045441

62 Brar, V. W. *et al.* Gate-controlled ionization and screening of cobalt adatoms on a graphene surface. *Nature Physics* **7**, 43-47 (2011). https://doi.org/10.1038/nphys1807

63 Schuler, B. *et al.* Large Spin-Orbit Splitting of Deep In-Gap Defect States of Engineered Sulfur Vacancies in Monolayer ${\mathrm{WS}}_{2}$. *Physical Review Letters* **123**, 076801 (2019). https://doi.org/10.1103/PhysRevLett.123.076801

64 Wang, Y. & Song, Z. In preparation.

65 Tsui, Y.-C. *et al.* Direct observation of a magnetic-field-induced Wigner crystal. *Nature* **628**, 287-292 (2024). https://doi.org/10.1038/s41586-024-07212-7

66 Xiang, Z. *et al.* Imaging quantum melting in a disordered 2D Wigner solid. *Science* **388**, 736-740 (2025). https://doi.org/10.1126/science.ado7136

67 Li, H. *et al.* Wigner molecular crystals from multielectron moiré artificial atoms. *Science* **385**, 86-91 (2024). https://doi.org/10.1126/science.adk1348

68 Li, H. *et al.* Competing Chern states revealed by quasiparticle charging in moir\'e rhombohedral graphene. *arXiv preprint arXiv:2607.08710* (2026).

69 Seewald, E. *et al.* Mapping the moiré potential in multi-layer rhombohedral graphene. *arXiv preprint arXiv:2510.09548* (2025).

70 Regnault, N. *et al.* Catalogue of flat-band stoichiometric materials. *Nature* **603**, 824-828 (2022).

71 Călugăru, D. *et al.* General construction and topological classification of crystalline flat

bands. *Nature Physics* **18**, 185-189 (2022).

72 Zhang, Z. *et al.* Engineering correlated insulators in bilayer graphene with a remote Coulomb superlattice. *Nature Materials* **23**, 189-195 (2024).

73 Gu, J. *et al.* Remote imprinting of moiré lattices. *Nature Materials* **23**, 219-223 (2024).

74 He, M. *et al.* Dynamically tunable moiré exciton Rydberg states in a monolayer semiconductor on twisted bilayer graphene. *Nature Materials* **23**, 224-229 (2024).

75 Mullan, C. *et al.* Mixing of moiré-surface and bulk states in graphite. *Nature* **620**, 756-761 (2023).

76 Wang, X. *et al.* Moiré band structure engineering using a twisted boron nitride substrate. *Nature Communications* **16**, 178 (2025).

77 Qiao, L., Lu, X., Zhang, F.-C. & Liu, J. Charge imprinting biases topology of correlated insulator in hBN-aligned rhombohedral multilayer graphene. *arXiv preprint arXiv:2606.20377* (2026).

78 Regnault, N., Li, H., Kwan, Y. H., Bernevig, B. A. & Herzog-Arbeitman, J. The "Moiré Capacitor Effect" and Stabilization of Fractional Chern Insulators in Rhombohedral Graphene Superlattices. *arXiv preprint arXiv:2608.12452* (2026).

79 Xie, J. *et al.* Unconventional Orbital Magnetism in Graphene-based Fractional Chern Insulators. *arXiv preprint arXiv:2506.01485* (2025).

80 Zan, X. *et al.* Chern number reversal and emergent superconductivity in rhombohedral graphene induced by in-plane magnetic fields. *arXiv preprint arXiv:2604.27788* (2026).

81 Feng, Z. *et al.* Rapid infrared imaging of rhombohedral graphene. *Physical Review Applied* **23** (2025). https://doi.org/10.1103/PhysRevApplied.23.034012

82 Wickenburg, S. *et al.* Tuning charge and correlation effects for a single molecule on a graphene device. *Nature Communications* **7**, 13553 (2016). https://doi.org/10.1038/ncomms13553

## Supplementary Information:

## Emergent trans-moiré orbitals and topology in rhombohedral graphene

Yuqin Wang*, Jian Xie*, Yi-Jie Wang*, Jiajun Zhang*, Yiting Gao, Zaizhe Zhang, Da Yi, Yan Xie, Jingjing Shi, Guanqin Zhao, Chengyu Xiong, Kenji Watanabe, Takashi Taniguchi, Zhi-Da Song[†], Xiaobo Lu[†], and Yi Chen[†]

*These authors contributed equally to this work.

[†]E-mail: songzd@pku.edu.cn; xiaobolu@pku.edu.cn; yichen@pku.edu.cn

## Contents

## 1. Strain estimation and twist angle determination

In general, van der Waals moiré materials involving transfer processes inevitably host residual strain to some extent. Such strain can cause structural distortions to the moiré superlattice and may affect the electronic properties, as extensively investigated in magic angle twisted bilayer/trilayer graphene[1-3]. To quantitatively estimate the strain amplitudes in our R6G/hBN devices, here we use the heterostrain model presented in Ref. [1] to extract strain values from real-space STM images, a procedure that simultaneously determines the twist angle.

Suppose the twist angle between R6G and hBN is $\theta$ and strain is applied to R6G with an angle $\alpha$ to the R6G's $x$ axis (while maintaining hBN unstrained). The effect of strain is then to deform the reciprocal vectors $\boldsymbol{k}_{\mathrm{G1}}, \boldsymbol{k}_{\mathrm{G2}}, \boldsymbol{k}_{\mathrm{G3}}$ of R6G according to the strain tensor $\mathbf{E}(\epsilon)$:

$$\boldsymbol{k}'_{\mathrm{G}i} = \mathbf{R}(-\alpha)\mathbf{E}(\epsilon)\mathbf{R}(\alpha)\boldsymbol{k}_{\mathrm{G}i}$$

$$= \begin{pmatrix} \cos(\alpha) & \sin(\alpha) \\ -\sin(\alpha) & \cos(\alpha) \end{pmatrix} \begin{pmatrix} \frac{1}{1+\epsilon} & 0 \\ 0 & \frac{1}{1-\delta\epsilon} \end{pmatrix} \begin{pmatrix} \cos(\alpha) & -\sin(\alpha) \\ \sin(\alpha) & \cos(\alpha) \end{pmatrix} \boldsymbol{k}_{\mathrm{G}i},$$

in which $\mathbf{R}(\alpha)$ is the rotation matrix, $\epsilon$ is the strain amplitude, $\delta = 0.16$ is the Poisson ratio of graphene[1], and $i = 1, 2, 3$ labels the three directions. The reciprocal vectors $\boldsymbol{k}_{\mathrm{BN}1}, \boldsymbol{k}_{\mathrm{BN}2}, \boldsymbol{k}_{\mathrm{BN}3}$ of hBN are given by

$$\boldsymbol{k}'_{\mathrm{BN}i} = \mathbf{R}(\theta)\boldsymbol{k}_{\mathrm{BN}i}\,.$$

The resulting moiré wavelengths along the three $C_3$-connected directions are given by

$$\lambda_i = \frac{4\pi}{\sqrt{3}|\boldsymbol{k}'_{\mathrm{BN}i} - \boldsymbol{k}'_{\mathrm{G}i}|}\,.$$

By fitting the experimentally measured moiré wavelengths $\lambda_i$ to the above formula, both the strain amplitude $\epsilon$ and the twist angle $\theta$ can be extracted.

In STM studies of twisted graphene, the input parameters above ($\lambda_i$'s) are conventionally acquired in the real space by measuring three moiré periods in a single moiré supercell[1]. Here, owing to the weak moiré inhomogeneity of our samples (Fig. 1b, Supplementary Information Fig. 1), we use a reciprocal-space method to determine the moiré wavevectors by taking fast Fourier transform (FFT) of a large STM topograph covering multiple moiré supercells (as shown in Supplementary Information Fig. 1). We can then use these moiré wavevectors as inputs and extract the strain value and twist angle of each sample as mentioned above and reported in Extended Data Table 1. The extracted strain values for our samples, typically between 0.06% to 0.2%, are somewhat smaller than typical strain values on the order of 0.2~0.7% in early twist-graphene devices[1] and already comparable with later optimized samples of ~0.1%[2,3], thus highlighting the quality of our R*n*G/hBN devices. Due to such low strain values, we expect weak strain effects in the observed electronic structures and have indeed observed homogeneous behavior across different real-space regions on the same sample surface.

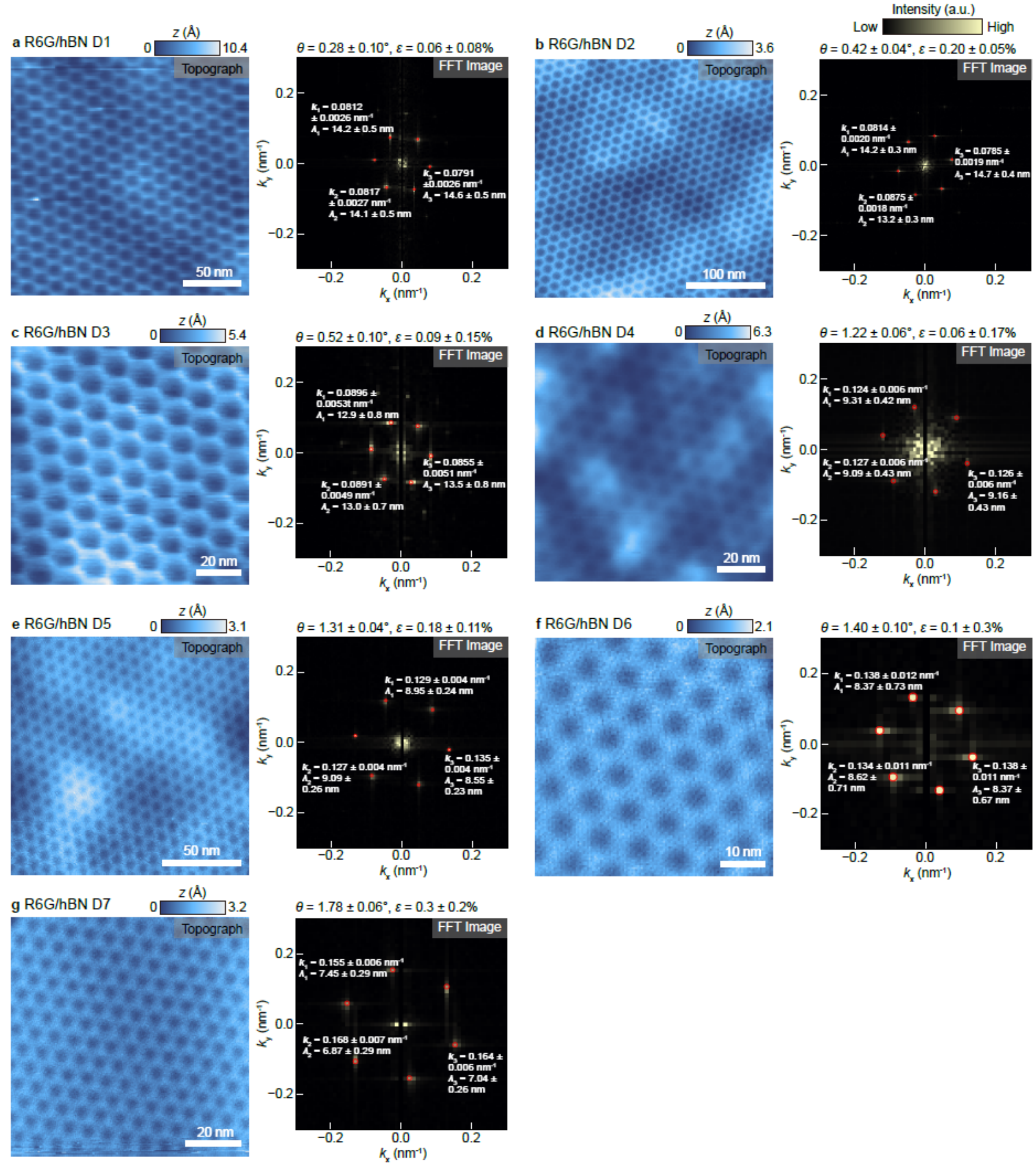

**Supplementary Information Figure 1. Summary of real-space STM topographs and correspondent FFT spectra of seven R6G/hBN devices studied in this work; values reported in Extended Data Table 1.**

## 2. Comparison of electronic structures between R6G and R3G

In this section and the next we provide a comparison between the electronic structures between R6G and the more familiar case of R3G[4-7], especially in terms of the STM d$I$/d$V$ spectra. In this section we discuss single STM d$I$/d$V$ spectra of R6G and R3G. While their d$I$/d$V$ spectra show overall similarity and exhibit pronounced spectral features corresponding to the flat-band electronic structure[4-7], one noticeable distinction appears under zero displacement field. In R3G, the d$I$/d$V$ spectrum exhibits two apparent spectral peaks under zero displacement field[4-6], while it appears that only one spectral peak dominates the d$I$/d$V$ spectrum in the case of R6G. This arises from differences in the flat-band electronic structures of R3G and R6G as illustrated in the single-particle band structures of R3G and R6G in Supplementary Information Fig. 2. In Supplementary Information Fig. 2a, R3G under zero displacement field indeed exhibits two peaks in the local-density-of-states (LDOS) plots, each corresponding to a van Hove singularity of the flat-band electronic structure. Since the two flat bands are layer unpolarized under zero displacement field, two apparent d$I$/d$V$ peaks are expected in STM d$I$/d$V$ spectroscopy of R3G[4-6]. In contrast, in R6G, the band structure is much flatter and the two van Hove singularities largely overlap in the LDOS plot of R6G, leading to a single-peak-like LDOS feature with much enhanced density of states. This accounts for the apparent single d$I$/d$V$ peak in our measurements.

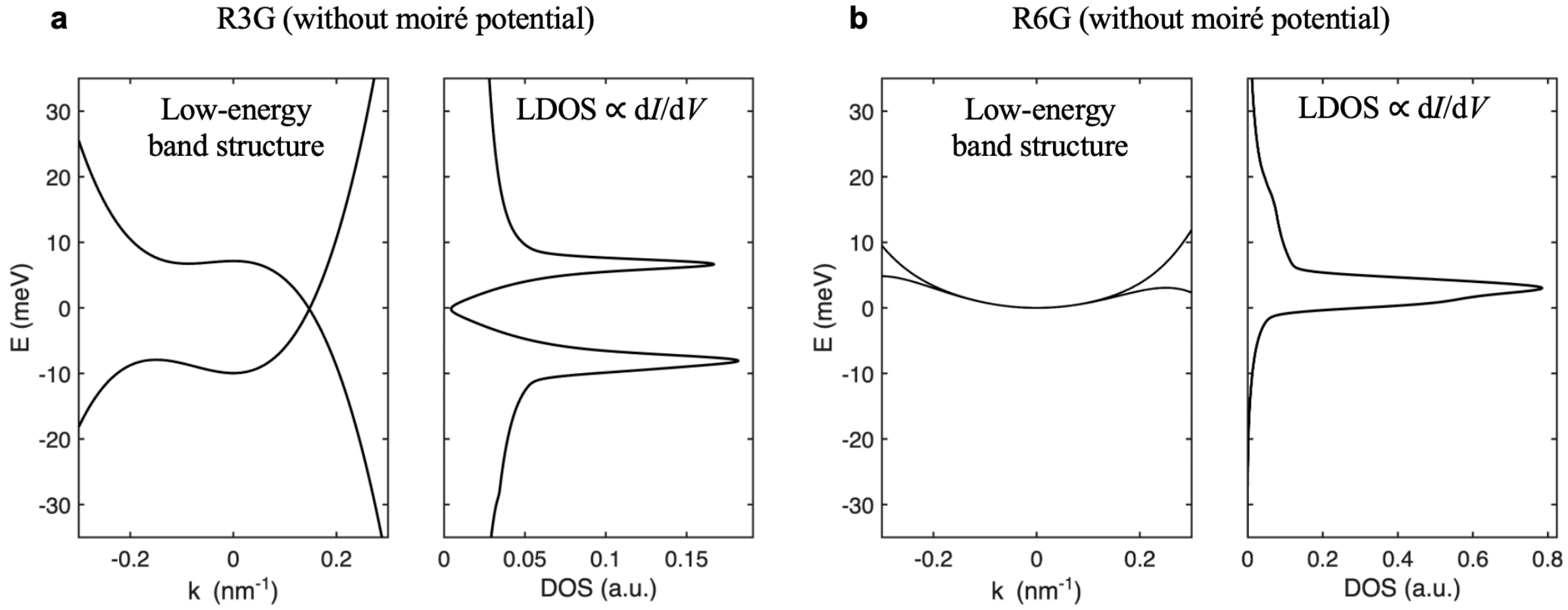

**Supplementary Information Figure 2. Simulated electronic structures of R3G and R6G. a,** Left panel: Single-particle low-energy band structure of R3G. Right panel: Corresponding LDOS plot showing two apparent peaks around the Fermi level. **b,** Same as **a** but for R6G, showing only one apparent LDOS peak.

## 3. Discussion of gate-dependent d*I*/d*V* spectra of R6G/hBN

In this section we discuss gate-dependent d*I*/d*V* spectra of R6G/hBN and their certain differences from R3G.

Overall, such set of measured spectra contains two groups of features: low-energy flat-band-related features near the Fermi level, as well as higher-energy remote-band features located at least 200 meV away from the Fermi level (Supplementary Information Fig. 3a). Both groups of features are affected by the back-gate voltage, but in different ways. Generally speaking, due to the electronic properties of R*n*G, the effects of a back-gate voltage on R*n*G are two-fold: first, the back gate voltage induces $V_{bg}$-dependent doping into the system, leading to a shift in $E_F$ (which affects all electronic d*I*/d*V* features); second, the back gate voltage induces a $V_{bg}$-dependent displacement field that acts to separate the two surface states of R*n*G (with the local probe only measuring the moiré-distant flat band spatially close to the probe).

To experimentally disentangle these dual effects, we have found it helpful to acquire a wide-bias-range of d*I*/d*V* spectra covering both flat and remote bands, as shown in Supplementary Information Fig. 3a. The remote bands, exhibiting shifts from changes in the Fermi level ($-E_F$, the green curve in Supplementary Information Fig. 3b), provide a basis for estimating the doping effect. On the other hand, the pair of low-energy flat bands experiences additional *D*-induced energy separation, schematically representable by the energetic evolutions of $+5V_D/2 - E_F$ and $-5V_D/2 - E_F$ (blue and red curve in Supplementary Information Fig. 3b). This provides a basis for an estimate of $V_D$, the *D*-induced layer potential difference.

In order to compare to the R3G results[4-7] we should note that these effects are layer-number sensitive. In particular, the energy shifts due to $-E_F$ is related to the magnitude of low-energy density of states; as mentioned above, because R6G has higher flat-band LDOS than R3G, the resultant $E_F$ shift is much less, making the $V_{bg}$ dependence less drastic than R3G.

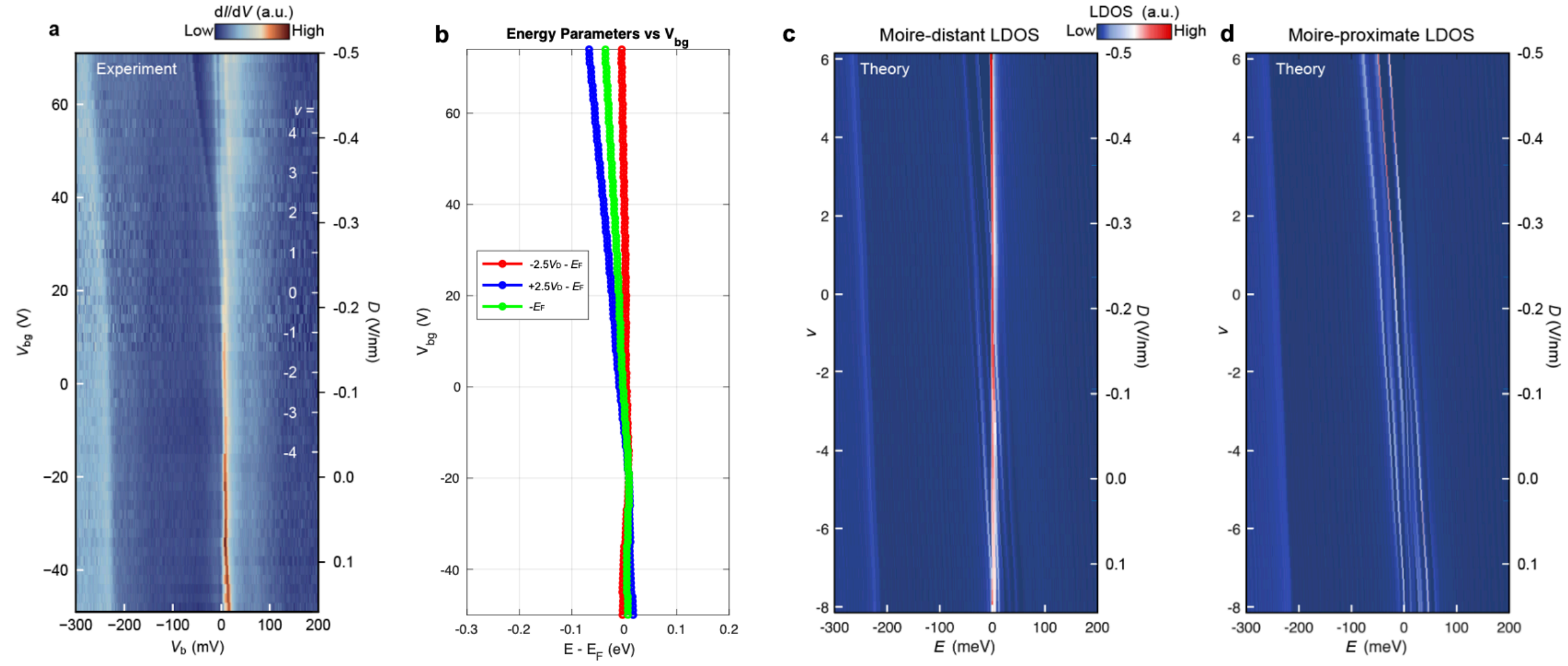

**Supplementary Information Figure 3. Comparison between experimental gate-dependent d*I*/d*V* spectrum and LDOS calculations. a,** Gate-dependent d$I$/d$V$ spectrum acquired at a $C_{BN}$ site in R6G/hBN, same as Fig.3h. **b,** Theoretically calculated $V_{bg}$ dependence of $\pm 2.5V_D$-$E_F$ & -$E_F$. **c,** Theoretically simulated gate-dependent LDOS spectra on the moiré-distant surface (at a $C_{BN}$ site) that carries overall similarities to **a**. **d,** Same as **c** but for the moiré-proximate surface.

In addition to these general R*n*G characteristics, the inclusion of a small-twist-angle moiré interface in R*n*G/hBN leads to various moiré-related effects, e.g., inducing moiré-related splittings. Capturing these effects requires more than the above handwaving arguments but full-fledged moiré R*n*G modelling as shown in Supplementary Information Fig. 3c and detailed in Methods. The moiré interface first and foremost affects the moiré-proximate electronic structure (Supplementary Information Fig. 3d), which is now strongly moiré-modulated into a group of three main bands that span a wide energy range of around 50 meV. This in turn leads to changes of the moiré-distant electronic structure detectable in the STM spectroscopy, for example, the faint d$I$/d$V$ line-like features near the main flat-band feature in Supplementary Information Fig. 3a. The moiré interface also affects the apparent electron-hole asymmetry in the spectra. For example, a $D$-induced flat-band separation (a low-energy LDOS suppression) is only visible at $V_{bg} \gtrsim 35$ V because under negative $V_{bg}$, the moiré-induced broadening causes strong overlap (and hence no separation) between the two surface flat bands, as shown in Supplementary Information Fig. 3c, d; the negative-$V_{bg}$ separation region only appears at even more negative $V_{bg}$.

Correlation may also play a certain role in the measured gate-dependent d$I$/d$V$ spectra, especially on the spectral splittings near the Fermi level in the $\nu > 4$ region in Supplementary Information Fig. 3a. A plausible explanation of such splittings is a Stoner-type transition, widely observed in rhombohedral multilayer graphene[5,8], but the precise nature requires additional investigations.

## 4. A complete data set of bias-dependent d*I*/d*V* maps

Here, we present a complete data set of d$I$/d$V$ maps of R6G/hBN D1 taken at $V_{bg}$ = -30 V, which better visualizes the energy-dependent evolution of trans-moiré orbitals (Extended Data Fig. 3 & Supplementary Video 1).

The general evolution is as follows. At energies outside the flat band, the moiré-distant surface exhibits nearly spatially homogeneous d$I$/d$V$ signals (Extended Data Fig. 3a**)**. Inside the flat band, there are primarily three groups of d$I$/d$V$ mapping patterns, and a representative image within each group is plotted in Fig. 4.

These three groups of orbitals are:

(1) Hollow-cage-like orbitals: the low-energy edge of the moiré-distant flat band is dominated by a hollow-cage-like orbital, featuring enhanced spectral weights on the outermost rims of the moiré unit cells (Extended Data Fig. 3b).

(2) Hybridized-like orbitals: at intermediate energies, the moiré-distant flat band shows spectral shapes that resemble more strongly linked or hybridized orbitals.

(3) Pancake-like orbitals: the upper half of the moiré-distant flat band is dominated by pancake-like orbitals that stay very similar in shape with enhanced spectral weights in $C_B$ stacking regions and appear like the negative of (1).

We have also plotted this data set of d$I$/d$V$ maps sequentially in Supplementary Video 1.

## 5. Persistence of trans-moiré renormalization and orbitals under large-negative *D* fields

Given that STM imaging requires an open-surface condition that is not compatible a tunable top gate, the Gauss's law dictates that the reachable $\nu$-$D$ parameters are in a much narrower range than dual-gated transport devices. As a result, the most natural way to test out the proposed mechanism, simultaneously performing STM orbital imaging (which requires open surfaces) and QAHE transport measurements (which requires dual gates to reach $v = 1$ and large negative $D$), cannot be performed on the same devices. Facing this technological challenge, we have designed three alternative tests to evaluate whether our conclusion holds in the large-$D$, transport-relevant regime, and conclude in the positive, as detailed below.

The first test: It involves selecting devices with excellent dielectric performance to directly go into the large-$D$ regime. One reason why we have chosen to study R6G/hBN (rather than thinner R5G/hBN or R4G/hBN) is that the critical $D$ field for QAH states in R6G/hBN is lower. As shown in Refs.[9,10], in well-aligned R6G/hBN samples, the onset of QAH states appears at a critical displacement field of -0.45 V/nm (with the FQAHE appearing at a higher field), lower than that in R5G/hBN[11] and R4G/hBN[12].

This forms the basis of the following experimental test, which involves two steps. First, we find that for certain devices with a high breakdown field, it is possible to increase the back-gate voltage to 80 V ~ 100 V without dielectric breakdown. By this means, a large $D$ field of -0.46 V/nm can be obtained, already reaching the QAH displacement field requirement of R6G/hBN (Extended Data Figure 6). Second, STM imaging can be performed under this large $D$ field. Due to the absence of a top gate, the Gauss's law dictates that the filling number $v$ now is inevitably much larger than 1, which means that the "$v = 1$" electronic states relevant to the QAHE, or more precisely, the lowest-energy Chern conduction miniband on the moiré-distant surface, are now well below the Fermi level and not detectable by Fermi-level probes like transport; but these electronic states can nevertheless be detected by the STM, because a finite STM bias voltage allows tunneling into these selected filled states. More specifically, one can now test whether these relevant electronic states exhibit real-space moiré renormalization and moiré orbital patterns under large-*D* conditions. Imaging of such sort is presented in Extended Data Figure 6, which shows that under a *D* field of -0.46 V/nm, the electronic states still experience significant moiré-periodic modulations, similar to the situation under small-*D* fields presented in the main text. Such insensitivity to *D* within experimentally accessible ranges can be understood within our interaction-driven picture as detailed later.

While strictly speaking this first test is not the same as direct imaging of the large-*D* and $\nu$ = 1 states (because of different $\nu$'s), the following theoretical observation add to the validity of this approach. By keeping a large *D* field while gradually decreasing $\nu$ from 8 to 1, we find the trans-moiré renormalization only to slightly increase (Supplementary Information Fig. 4h), suggesting that significant moiré renormalization exists in the QAHE regime. This increase is due to fewer electrons occupy the low-energy $C_{BN}/C_N$ orbitals on the moiré-distant surface, hence lowering their potential energy relative to that of $C_B$.

The second test: While the first test accesses a large-$D$, large-$\nu$ state experimentally and extrapolates the behavior to the QAHE regime via theoretical simulations with decreasing $\nu$, in the second test, we reach a $\nu = 1$ but small-$D$ state experimentally (Extended Data Fig. 4) and evaluate the effects upon increasing $D$. As shown in Supplementary Information Fig. 4, large-$D$ simulations show generally similar results to small-$D$ results across a wide range of $D$ fields. To find the real-space modulation of moiré renormalization for $D$ fields ranging from -0.1 V/nm to beyond -1.0 V/nm, in Supplementary Information Fig. 4g we simulate $D$-dependent LDOS spectra at the three stacking sites, $C_B$, $C_{BN}$, $C_N$, on the moiré-distant surface, so that the difference between the $C_B$ LDOS peak energy (red lines) and the $C_{BN}$/$C_N$ LDOS peak energy (black and purple lines) gives us an estimate of the moiré renormalization strength. We find the main effect of increasing $D$ is to flatten the conduction band and amplify the overall density of states. The moiré renormalization strength shows a weak diminishing trend with increasing $D$ but remains strong (~3.5 meV) even under FQAHE-relevant $D$ = -0.8 V/nm. Therefore, we conclude that moiré renormalization still defines a significant energy scale in the FQAHE-relevant parameter regime, shaping the moiré-distant electronic states.

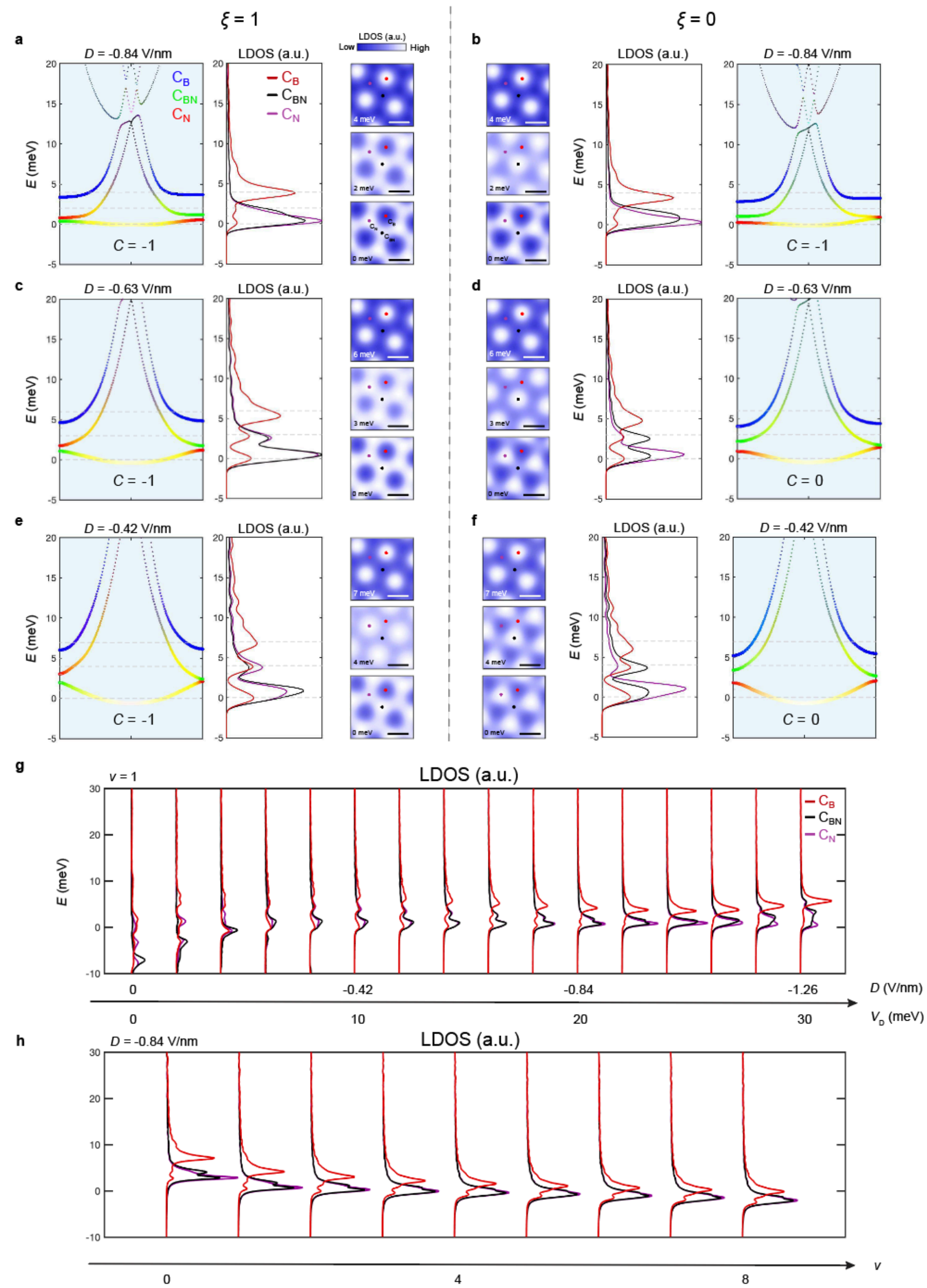

**Supplementary Information Figure 4. Correspondence between trans-moiré orbitals and moiré minibands under different displacement fields. a, b,** Moiré conduction bands in + valley at filling $\nu$=1 and $D$ = -0.84 V/nm (corresponding to $V_D$ = 20 meV) under stacking configuration $\xi = 1$ **(a)** and $\xi = 0$ **(b)**. The dot size represents the moiré-distant spectral weight. The Bloch state amplitudes at real-space locations $w = \mathrm{C_N},\ \mathrm{C_{BN}},\ \mathrm{C_B},\ \sum_{\alpha=A,B}\left|\langle r_w|\psi_{top,\alpha}(k)\rangle\right|^2$, are further mapped into the R-G-B triplet of the dot colors, respectively. The corresponding LDOS spectra and real-

space distributions are plotted on the side. **c, d,** Same as **a, b,** but at $D$ = -0.63 V/nm (corresponding to $V_D$ = 15 meV). **e, f,** Same as **a, b,** but at $D$ = -0.42 V/nm (corresponding to $V_D$ = 10 meV). **g**, Evolution of moiré-site-dependent LDOS with increasing displacement field at filling $\nu = 1$, crossing the QAH region. On increasing *D*, the lowest-energy moiré minibands flattens more, and the LDOS distributions at stacking configurations $\xi = 1$ and 0 become increasingly similar. As a reference, in typical transport experiments of R6G/hBN, quantum anomalous Hall plateaus appear at $D$ = -0.45 V/nm. **h**, Evolution of moiré-site-dependent LDOS with increasing filling $\nu$ at $D$ = -0.84 V/nm.

The third test: While the previous two tests strive to adapt STM measurements to approximate the transport QAHE measurement regime as closely as possible, one perhaps should acknowledge that without a top gate, STM R*n*G/hBN devices are not going to be the same as transport dual-gated devices. In other words, it is beyond current experimental capabilities to simultaneously image moiré orbitals and measure the QAHE in one set of samples. Facing this challenge, we have designed a powerful albeit painstaking third test—to prepare *two* sets of R*n*G/hBN samples, one set with exposed surfaces for STM imaging, and the other with dual gates to reach large *D* for transport QAHE measurements. Importantly, to cross-compare them, each set of samples should have a range of R*n*G-hBN twist angles $\theta$'s, because twist angles are known to be a main control knob of the QAHE in R*n*G/hBN (it shows up in small-twist-angle R6G/hBN samples at $\theta \lesssim 0.92^\circ$)[9,10].

It is thus clear how to perform this test: if trans-moiré orbitals were indeed the underlying mechanism of Chern band formation and the QAHE in R6G/hBN, STM-observed trans-moiré orbitals should be seen at whatever twist angles the QAHE appears in transport samples, and vice versa. If, on the other hand, STM-observed trans-moiré orbitals had any different twist-angle dependence from the QAHE (e.g., not vanishing at any angles, or only showing up at large angles, or even having the right trend but with a different critical angle), we would conclude that our proposed mechanism had failed this test.

This test is performed with a total of seven STM samples (together with seven transport samples[9]) as discussed in the main text. As shown in Fig. 5b, strikingly, it is found that the "on" and "off" of transport QAHE phenomenon are concurrent with the "on" and "off" of STM-observed moiré potentials. We thus conclude that our proposed trans-moiré mechanism has passed this test, and the similar critical angles of the two sets of samples (both $\sim 1^{\circ}$) highlight the intimate linkage between STM-observed moiré potentials and transport-observed QAHE.

General comment on the insensitivity of the trans-moiré potential landscape to a wide range of $\nu$ and $D$: Lastly we discuss the general insensitivity of the trans-moiré potential landscape to experimentally accessible $\nu$ and $D$ values. The main underlying reason of this insensitivity is that this potential is derived from strong valence-band charge accumulations trapped at the moiré interface. Using experimentally observed strength of trans-moiré renormalization, we conclude that a $V_1$ of 20~30 meV can reproduce the observed phenomenology in our self-consistent Hartree calculations (Fig. 3, for example). This is akin to the SET-measured moiré potential at the monolayer graphene/hBN interface[13], which is estimated to be on the order of 40~50 meV. Due to this strong potential, the charge accumulated at the interface is at least composed of three highest moiré valence minibands, interlaced with additional three dispersive bands; depleting them would require $\nu \sim -24$ to -28. Therefore, this rather rigid density background induces a rather rigid trans-moiré potential, insensitive to the tunable range of experimentally accessible $\nu$'s, as shown in Supplementary Information Fig. 4g.

Next we discuss the reason for the insensitivity of moiré renormalization to the experimentally accessible *D* fields. If *D* were infinitely larger than the moiré folding strength from hBN, $V_1$, the charge-neutrality (CN) gap set by *D* would prevent significant folding and moiré renormalization—so the key is to obtain a proper estimate of the CN gap size and $V_1$. The latter, as mentioned above, is determined by our experiment to be of 20~30 meV. In comparison, consider the large-*D* FQAHE regime (e.g., *D* = -0.8 V/nm, corresponding to an energy of $V_D \sim 20$ meV); at the moiré Brillouin zone corners (where the moiré folding is the most obvious), the generated CN gap for bare moiré-less R6G is about 75 ~ 80 meV, and that of hBN-aligned R6G is reduced to ~60 meV. As a result, we conclude that even in the transport large-*D* FQAH regime, a $V_1$ of 20~30 meV should not be regarded as negligible compared with the CN gap. For example, even at *D* = -0.8 V/nm, we have found an emergent moiré modulation amplitude of ~3.5 meV in our simulations (Supplementary Information Fig. 4), hence still defining a significant energy scale for moiré-distant electrons.

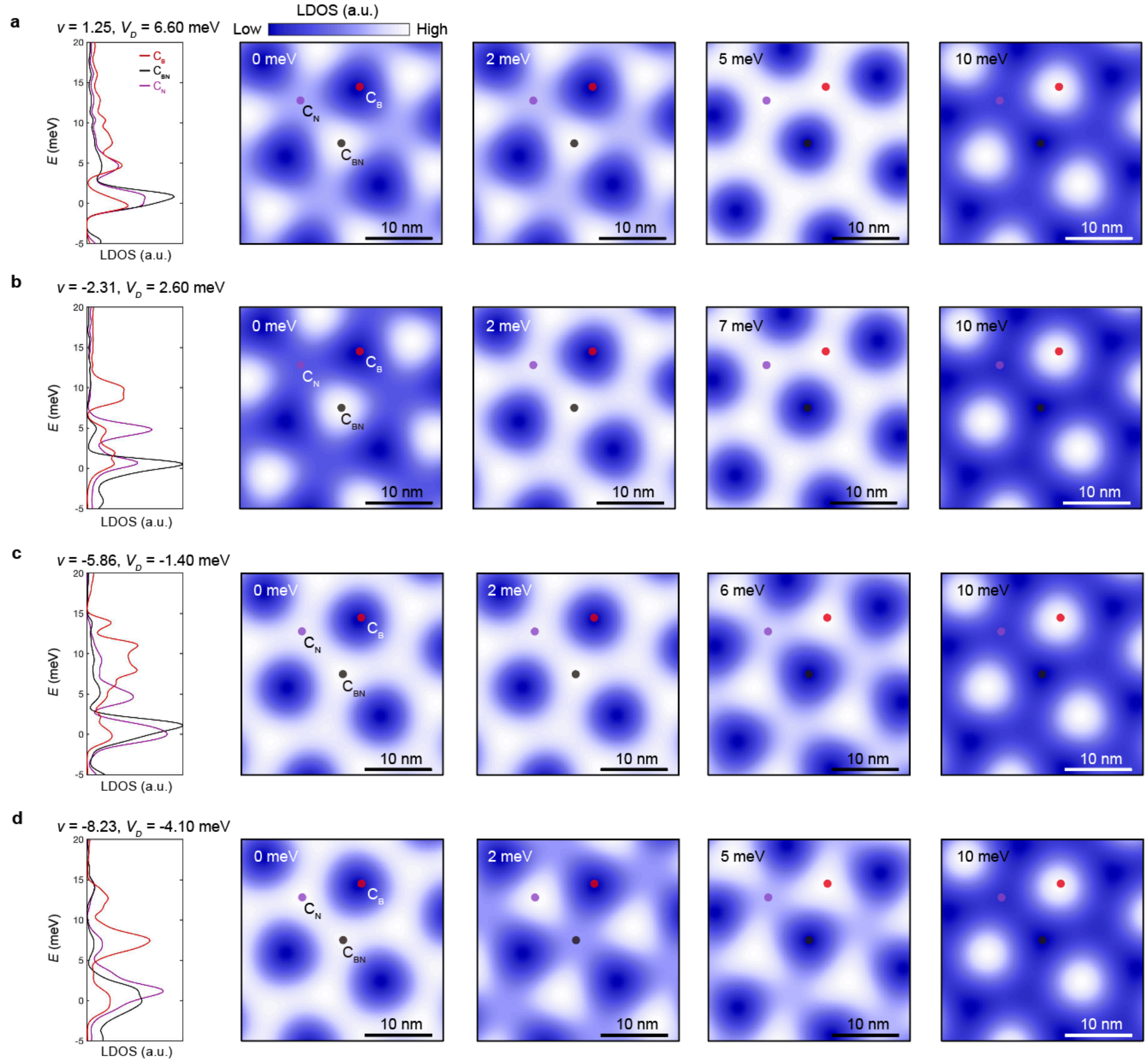

**Supplementary Information Figure 5. Theoretical LDOS maps of renormalized moiré-distant flat band. a,** Leftmost column: simulated moiré-site-resolved local density of states (LDOS) at the moiré-distant surface at $\nu = 1.25$, $V_D = 6.60$ meV, same parameters as in Fig. 3a,b,i. Right columns: representative LDOS distribution maps at different energies, showing correspondence to the observed trans-moiré orbitals. **b,** Same as **a**, but at $\nu = -2.31$, $V_D = 2.60$ meV, same parameters as in Fig. 3c,d,j and Fig. 4a-d. **c,** Same as **a**, but at $\nu = -5.86$, $V_D = -1.40$ meV, same parameters as in Fig. 3e,f,k and Fig. 4e-h. **d,** Same as **a**, but at $\nu = -8.23$, $V_D = -4.10$ meV, same parameters as in Fig. 4i-l.

## 6. Implications of observed trans-moiré orbitals for theoretical modelling of R*n*G/hBN

A crucial yet controversial question in this field is how to most faithfully describe the physics of rhombohedral moiré graphene. In this section we discuss how our imaging results may contribute in this regard.

On the choice of the reference field: One of the central debates on modelling of rhombohedral moiré graphene is the choice of various schemes of the reference field $P^{\mathrm{ref}}$ (also known as the reference density, the reference state, or the subtraction scheme), which is a largely unsettled theoretical issue (see, e.g., Refs.[14-16]).This choice is, on the other hand, very influential, yielding drastically different physical consequences. Physically this is because for an electron state with density matrix $P$, only the deviation from the reference field, $P - P^{\mathrm{ref}}$, enters the mean-field electron Hamiltonian and affects electronic behaviors.

There have been several different choices of $P^{\mathrm{ref}}$. A leading example in R*n*G studies is the moiréful charge neutrality (CN) scheme, where $P^{\mathrm{ref}}$ is calculated from the filled moiré valence bands[17-19]. In such a scheme, the valence electron density, modulated by hBN alignment, is largely subtracted as a background; as a result, the conduction electrons (on the moiré-distant surface under large negative $D$) are distributed homogeneously in space unless additional spontaneous translational symmetry breaking occurs (which leads to Wigner crystallisation or anomalous Hall crystallisation depending on the topology). In comparison, in the so-called average scheme where $P^{ref}$ is taken as the infinite-temperature density matrix[14,15], or in the moiréless CN scheme where $P^{ref}$ is the valence density of bare R6G (without hBN)[14], $P^{ref}$ does not contain moiré valence-band modulations, hence $P$ - $P^{ref}$ can still generate a 'moiréful' potential to the conduction band electrons. This choice is crucial for understanding exotic physics of rhombohedral moiré graphene, as discussed in several publications (e.g., Refs. [14,15]); but even these two theoretical works, after carefully comparing various schemes, conclude differently on the choice of the reference field, and both with compelling theoretical arguments.

This theoretical dilemma highlights the need for diagnostic experiments. In our work, through high-resolution STM imaging of the moiré-distant electronic states, we visualize a surprisingly large, 10-meV-scale moiré-periodic flat-band modulation across a very wide range of $v$ and $D$ values (reaching -0.46 V/nm, for example), showing that the low-energy physics on the far side of the moiré interface is still strongly influenced by an emergent moiré potential. This originates from the crucial influence of valence-band electrons located at the moiré-proximate interface, hence providing direct experimental evidence for the use of reference fields that do not suppress valence-band contributions. We expect that our imaging results of the trans-moiré potentials, taken over a wide range of twist angles, filling numbers, and displacement fields, can help determine the most faithful scheme to describe rhombohedral moiré graphene.

On a faithful atomistic description of the moiré superlattice: A key aspect in understanding the FQAHE in R*n*G/hBN lies in a faithful atomistic description of its moiré structures at relevant tiny twist angles, which to our knowledge has not been visualized. Such lattice relaxation can significantly affect the electronic structures and almost all energetics involved in rhombohedral moiré graphene, as discussed extensively in Ref. [20].

Our imaging results, combined with theoretical simulations, find that moiré lattice relaxation is a key factor in the simulations of topology in rhombohedral moiré graphene. As discussed in Methods, a previous theoretical puzzle is that while the two R*n*G/hBN stacking orientations ($\xi = 0$ and 1) both exhibit moiré-distant QAHE[9,21] in experiment, the two RnG/hBN stacking orientations ($\xi = 0$, 1) have different consequences in the average scheme, with Chern insulators essentially only form under $\xi = 1$[15]. This, as shown in Extended Data Fig. 2, is related to the fact that without relaxation, the lowest moiré conduction band is localized at $C_{BN}$ for $\xi = 1$ and $C_N$ for $\xi = 0$, hence leading to different implications between the two stacking orientations.

In the presence of lattice relaxation, however, we have found that the structurally most stable $C_B$ regions expands and 'absorbs' electrons from the other two sites, hence largely smearing out their differences. Consequently, the trans-moiré potential difference at $C_N$ and $C_{BN}$ decreases, leading to a hollow-cage-like 'network' of the trans-moiré potential minima (connecting $C_N$ and $C_{BN}$ sites) for both stackings. Only these post-relaxation electronic structures are consistent with our imaging results. Importantly, in terms of topology, we have found relaxed trans-moiré potentials to favor the formation of topological non-trivial bands and can lead to Chern-band with $|C|=1$ in both stacking configurations ($\xi = 1$ and 0) under transport-accessible conditions within our approximations (Supplementary Information Fig. 4). The final trans-moiré landscape, arising from a nontrivial interplay between the moiré interface, electron-electron interactions, and lattice relaxation, is directly visualized in this work.

On the physical origin of the FQAHE in rhombohedral moiré graphene: Related to the aforementioned debate on the reference field, there are different proposed physical origins of the FQAHE/QAHE in rhombohedral moiré graphene. They are expected to carry different real-space information that can be distinguished based on STM imaging.

The anomalous Hall crystals, for example, can be seen as the topological counterpart of Wigner crystals and should appear in STM imaging as either a triangular array of electronic-density modulations in the clean limit or a distorted array when pinned by disorder[22,23]. Importantly, these electron crystals should have an inter-electron periodicity that is very sensitive to the filling number $v$ and scales at least roughly as $1/\sqrt{v}$. Even in the presence of a strong moiré potential, upon changing $v$, an electron crystal can be identified by a varying number of carriers filling into individual sites of the triangular moiré lattice[24].

Our STM imaging, on the other hand, shows electronic orbitals have similar patterns with fixed periodicities over a wide range covering at least $-4 \leq v \leq 4$, inconsistent with the occurrence of an electron crystal. On the other hand, this electronic periodicity is well matched with the moiré periodicity (the topographic periodicity set by the moiré interface), hence strongly suggesting their origin as our proposed emergent trans-moiré orbitals. Importantly, as discussed above, our systematic twist-angle study shows that the 'on' and 'off' conditions of STM-observed moiré orbitals are concurrent with those of the transport-observed QAHE with Chern band formation, highlighting the intimate linkage between the STM-observed trans-moiré orbitals and the transport-observed QAHE. While anomalous Hall crystallisation may still dominate in parameter regimes not covered in our experiments, especially at millikelvin temperatures where the extended quantum anomalous Hall effect is observed, our observations of trans-moiré renormalization provides the electronic-structure basis for proper theoretical descriptions of this material system, and its fine interplay with electron crystallisation and fractionalization invites future theoretical investigations.

## 7. Fitting of moiré renormalization strengths

In the following, we describe two fitting schemes for the extraction of moiré renormalization strengths from our data. Each to-be-fitted experimental data set is composed of spatially dependent d$I$/d$V$ spectra acquired along a real-space high-symmetry direction that crosses three high-symmetry sites ($C_N$, $C_{BN}$ and $C_B$) multiple times (e.g., Fig. 3a).

The first fitting scheme, working well for d$I$/d$V$ spectra that do not contain significant spectral splitting, comprises of two steps (Supplementary Information Fig. 6):

1. Fitting each individual STM d$I$/d$V$ spectrum with a preset spectral shape that contains a single spectral peak to find the flat-band peak voltage and its uncertainty (Supplementary Information Fig. 6a). Due to possible asymmetry of the peak, we employ a Fano lineshape for the fitting:

$$\frac{\mathrm{d}I}{\mathrm{d}V}=f\left(V_{\mathrm{peak}}^{\mathrm{fit}}\right)=A\frac{(q+\varepsilon^2)}{1+\varepsilon^2}+C,\ \varepsilon=\frac{V\text{-}V_{\mathrm{peak}}^{\mathrm{fit}}}{\Gamma/2}$$

The flat-band peak voltage $V_{\mathrm{peak}}^{\mathrm{fit}}$ and its uncertainty $\Delta_{\mathrm{peak}}^{\mathrm{fit}}$ are obtained by minimizing the sum of squared residuals $\sum\left(\frac{dI}{dV}\text{-}f\left(V_{\mathrm{peak}}^{\mathrm{fit}}\right)\right)^2$.

2. Feeding this series of flat-band peak voltages and spatial positions into a statistical model to find the moiré-periodic oscillation amplitude. Along the high-symmetry direction (e.g., Fig. 3), the flat-band peak voltages are expected to exhibit a periodicity of $\sqrt{3}\lambda_{\mathrm{M}}$, with $\lambda_{\mathrm{M}}$ being the moiré periodicity. We can thus model the spatial dependence of flat-band peak voltages using a Fourier series (with experimentally determined $\lambda_{\mathrm{M}}$) as in

$$V_{\mathrm{peak}}=\sum_0^n A_i\cos(ikx+\psi_i),\ k=2\pi/\sqrt{3}\lambda_{\mathrm{M}}.$$

To account for the uncertainties of peak voltages, the fitting is performed using a standard statistical model via minimizing the sum of squared residuals weighted by the corresponding uncertainties (both obtained in the first step):

$$\sum\left(V_{\mathrm{peak}}^{\mathrm{fit}}-V_{\mathrm{peak}}\right)^2/\left(\Delta V_{\mathrm{peak}}^{\mathrm{fit}}\right)^2$$

The moiré renormalization strength is then taken as the difference between the maximum and minimum of the fitted curve: $\max\left(V_{\mathrm{peak}}\right)$ -$\min(V_{\mathrm{peak}})$.

For most of measured spectra, the doping and displacement fields are such that the measured flat band is not located right on the Fermi level; in such cases, the main effect of moiré renormalization is to create a shift in the peak energy with spectral reshaping being the secondary effect, so that the first fitting model works well. We have found that for R6G D1 to D3, a third-order Fourier series fit the data well (Supplementary Information Fig. 6b-d). R6G D4 to D7 have very weak moiré renormalization strengths with peak-voltage variations largely buried in fitting uncertainties in the first step; to mitigate the problem of overfitting, we reduce the model to use a Fourier series of the second order (Supplementary Information Fig. 6e-h), but even so overfitting and overestimation of the moiré renormalization strengths still exist in D4 to D7.

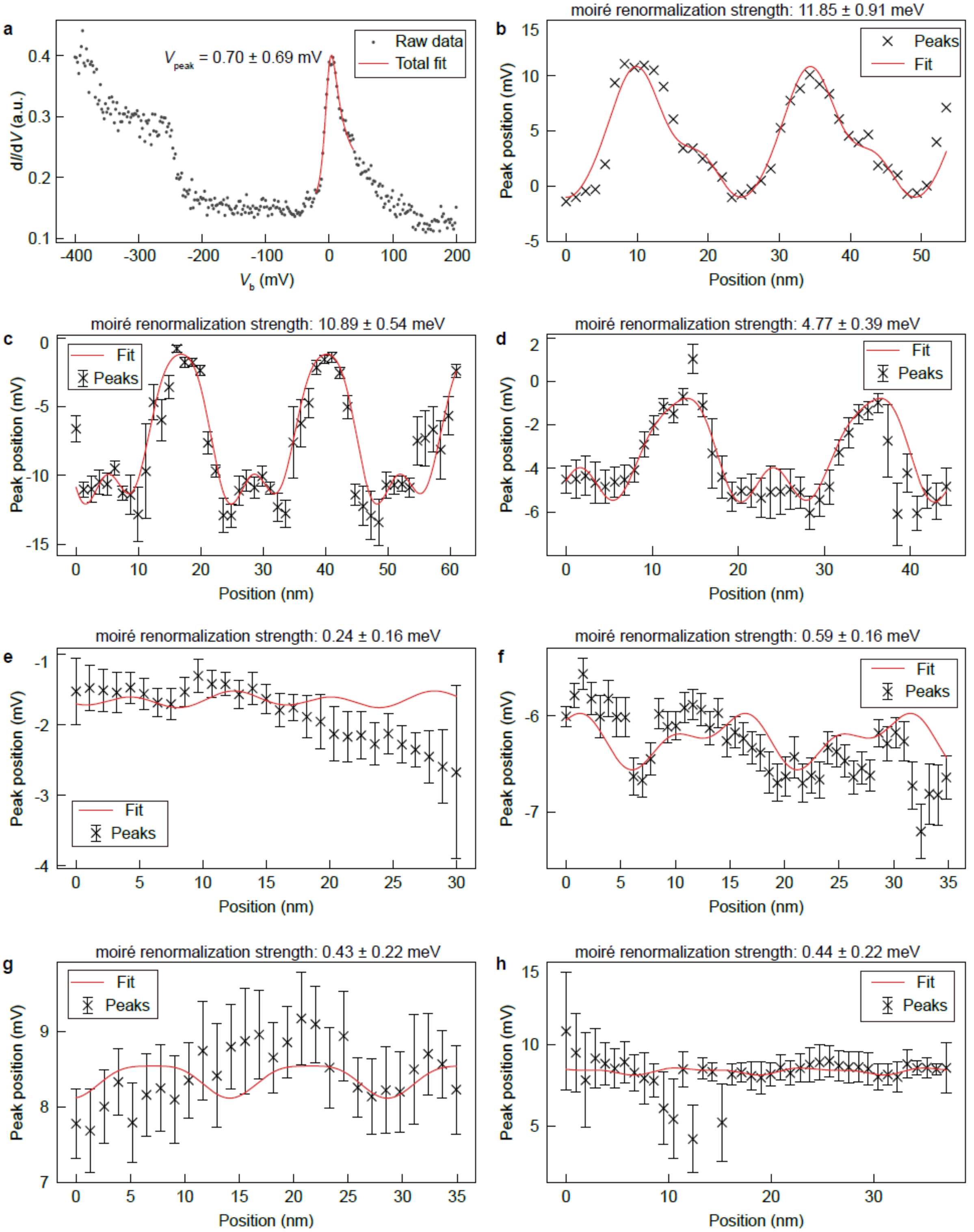

**Supplementary Information Figure 6. Fitting procedure and results of moiré renormalization strengths.** **a**, Example of flat-band peak fitting using a Fano lineshape. **b**, Fitting results for R6G D1 in Fig. 5b ($\nu$ = 1.1, $D$ = -0.28 V/nm). **c**, Fitting results for R6G D2 in Fig. 5b ($v$ = 1.0, $D$ = 0.0 V/nm). **d**, Fitting results for R6G D3 in Fig. 5b ($v$ = 1.0, $D$ = -0.10 V/nm). **e**, Fitting results for R6G D4 in Fig. 5b ($v$ = 1.1, $D$ = -0.07 V/nm). **f**, Fitting results for R6G D5 in Fig. 5b ($v$ = 1.0, $D$ = -0.02 V/nm). **g**, Fitting results for R6G D6 in Fig. 5b ($v$ = 1.0, $D$ = -0.34 V/nm). **h**, Fitting results for R6G D7 in Fig. 5b ($v$ = 1.1, $D$ = -0.44 V/nm).

There are cases such as in Fig. 3a where the flat band can exhibit additional spectral reshaping or splitting. In these cases, we can perform a more in-depth study composed of cross comparison between two different fitting schemes for the moiré renormalization strength. In the first scheme, we still attempt to fit the moiré renormalization strength using the method above, i.e., by performing a single-peak fitting hence ignoring all spectral reshaping effects. As a comparison, we also perform a second, more involved fitting scheme that accounts for the spectral reshaping/splitting, which contains:

1. Peak fitting using a more flexible spectral shape that takes possible spectral splitting into account, i.e., using a spectral shape containing two spectral peaks with their independent fitting parameters: $\frac{\mathrm{d}I}{\mathrm{d}V}=f(V_{\mathrm{peak}})=A_1\frac{(q_1+\varepsilon_1^2)}{1+\varepsilon_1^2}+A_2\frac{(q_2+\varepsilon_2^2)}{1+\varepsilon_2^2}+C$, $\varepsilon_{1/2}=\frac{V\text{-}V_{\mathrm{peak},\,1/2}}{\Gamma_{1/2}/2}$. We then take the average of the two positions as the fitted center position of this two-peak spectral shape and an overall uncertainty to cover both peaks. We have found that this procedure also works well for curves with a single spectra peak because the two fitted peaks tend to add up to form this single peak.

2. Similar to the first method by minimizing a statistical residual function obtained from fitted parameters from the first step.

An example is shown in Supplementary Information Fig. 7, where the first scheme gives a fitted moiré renormalization strength of 11.9 ± 0.9 meV (Supplementary Information Fig. 7b) and the second fitting scheme gives a fitted moiré renormalization strength of 12.1 ± 1.0 meV (Supplementary Information Fig. 7d), hence statistically agree with each other. As a result, we conclude that our fitting schemes are valid even in the presence of certain spectral reshaping.

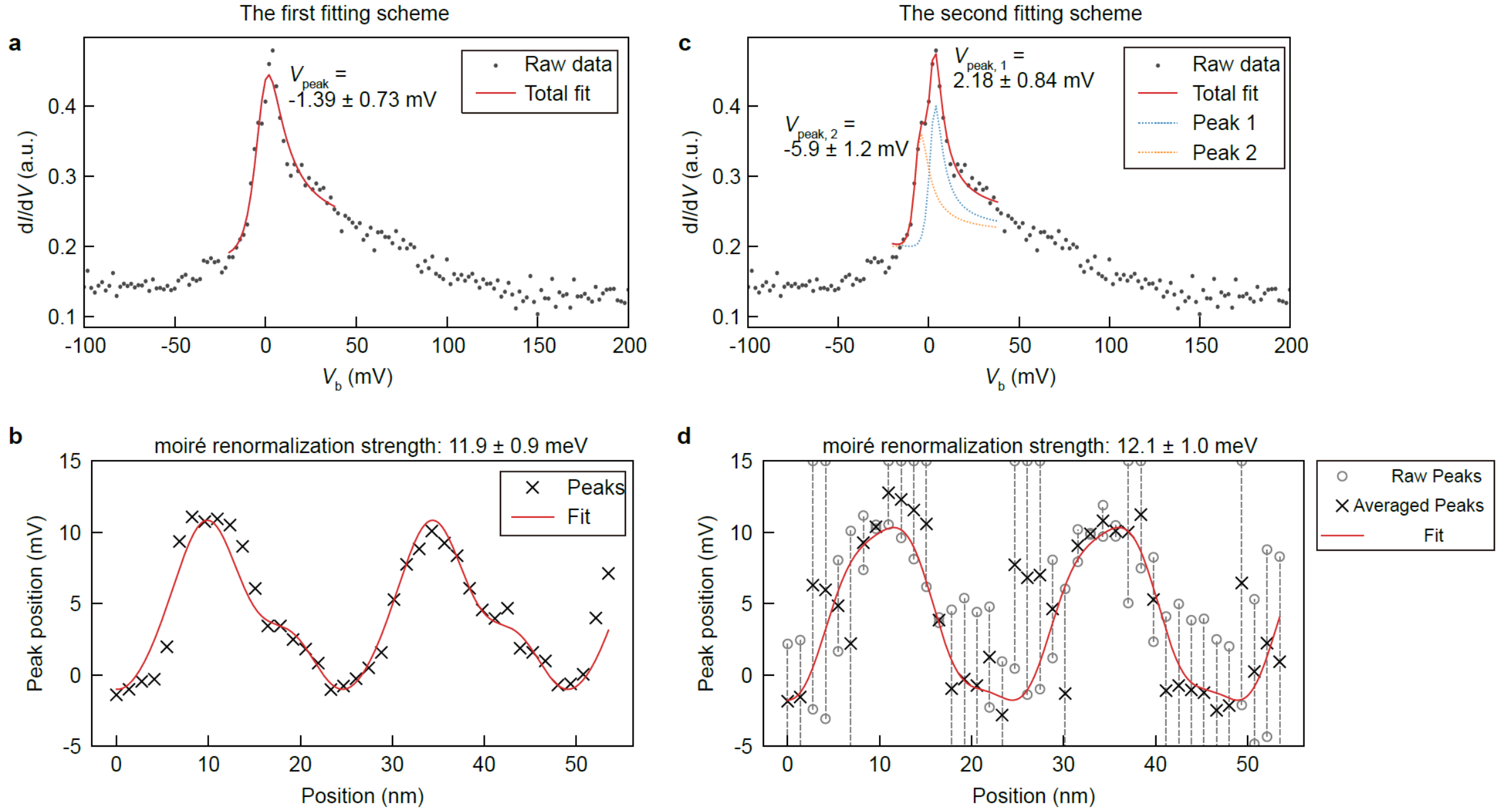

**Supplementary Information Figure 7. Comparison of two fitting schemes using the same data set (data corresponding to Fig. 3a acquired in R6G D1 at *ν* = 1.1, *D* = -0.28 V/nm). a**, Example of the first step (single-peak fitting) in the first fitting scheme. **b**, Example of the second step in the first fitting scheme to determine the moiré renormalization strengths. **c**, **d**, same as **a**, **b** but for the second fitting scheme.

## 8. **Spatial homogeneity of flat-band peaks at large vs. small twist angles**

By performing such fitting procedures, we can further quantitatively characterize the spatial distributions of the flat-band peaks and compare results obtained for R6G/hBN devices with large and small twist angles.

Take R6G/hBN D6 with $\theta = 1.40^{\circ}$ as an example of large-twist-angle R6G/hBN, whose d$I$/d$V$ grid measurements at $V_{\mathrm{bg}} = 44$ V, shown in Fig. 2b-d, exhibit negligible moiré-periodic electronic modulations. By taking the d$I$/d$V$ spectrum at every spatial pixel of this grid measurement and fitting the flat-band peak energies of each individual spectrum, we obtain a histogram of peak energies as shown in Supplementary Information Fig. 8. The extracted peak energies can be well described by a Gaussian distribution centered at $E = 8.68$ meV with a standard deviation of $\sigma = 0.5$ meV. As a comparison, the typical full width at half maximum (FWHM) of the flat-band peak is on the order of 10 meV. Therefore, the spatial variation of the peak energy is much smaller than the intrinsic spectral width of the flat band. This result indicates that the flat-band peak remains uniform in energy across the entire field of view covering multiple moiré unit cells, consistent with the absence of observable moiré-periodic electronic modulations in large-twist-angle R6G/hBN.

In comparison, the histogram of flat-band peak positions of R6G/hBN D1, a small-twist-angle device with $\theta = 0.28^{\circ}$, shows a complicate multi-peak structure that corresponds to the spatially dependent moiré renormalization and spans a much wider energy window. This stark contrast provides additional evidence for the observed trans-moiré renormalization only in small-twist-angle R6G/hBN devices.

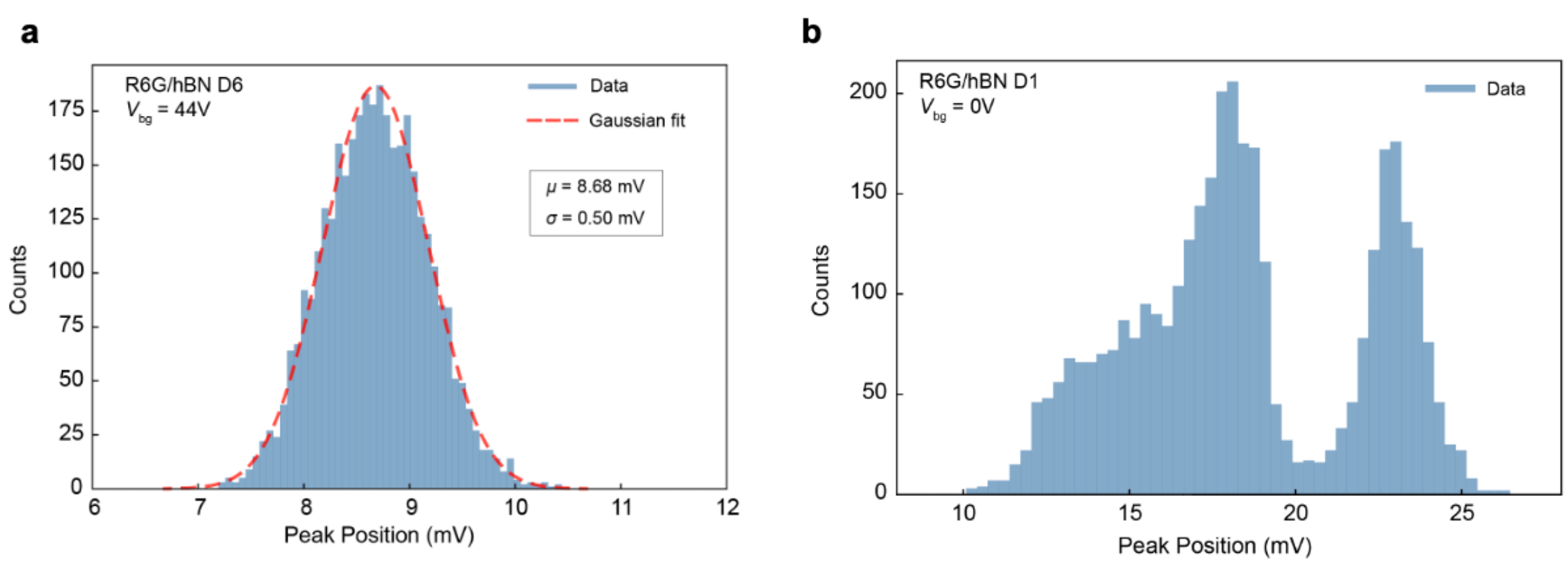

**Supplementary Information Figure 8. Spatial distribution of flat-band peak energies in R6G/hBN D6 & D1. a,** Histogram of flat-band peak positions in R6G/hBN D6 with $\theta = 1.40^{\circ}$ showing a single, narrow Gaussian distribution with a mean energy of 8.68 meV and a standard deviation of 0.50 meV. These peak positions are extracted from individual $dI/dV$ spectra taken at each pixel of a grid measurement in Fig. 2b-d. **b,** Same as **a** but for R6G/hBN D1 with $\theta = 0.28^{\circ}$ showing a complicate multi-peak structure that corresponds to the spatially dependent moiré renormalization and spans a much wider energy window.

## 9. Temperature dependence of moiré renormalization

As shown in Supplementary Information Fig. 9, temperature-dependent measurements of moiré renormalization strengths at $T = 2.5$ K, 10.5 K & 15.5 K show similar results, suggesting that the observed moiré renormalization is robust across our measured temperature range.

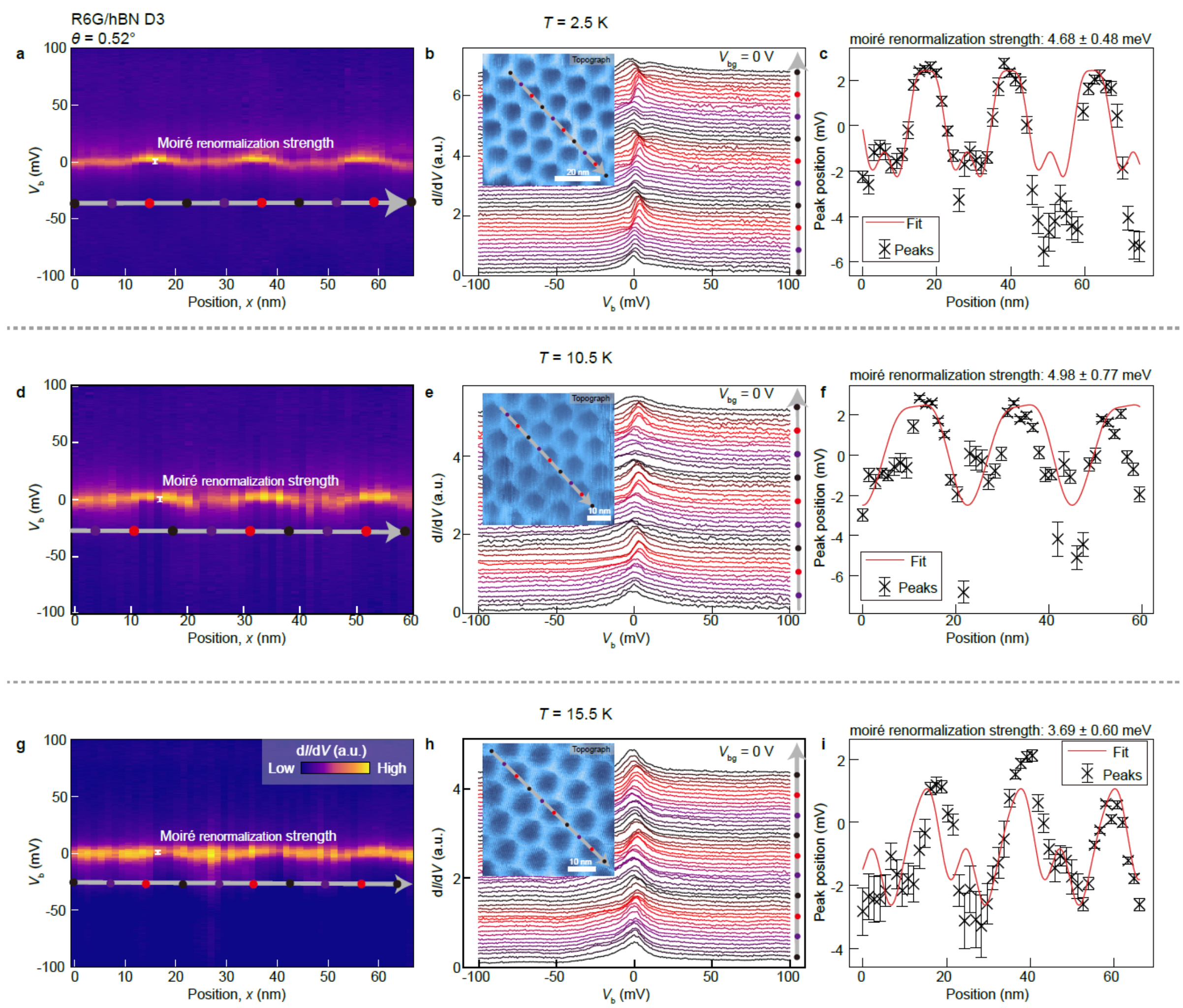

**Supplementary Information Figure 9. Flat-band renormalization in R6G/hBN D3 measured at different temperature ($\theta = 0.52^{\circ}$). a,** Colormap plot of spatially dependent d$I$/d$V$ spectra acquired at $T = 2.5$ K (along the grey arrow shown in the inset of **b**) ($V_b$ = -0.3 V, $I_t$ = 20 pA, $V_{bg}$ = 0 V, $V_{mod}$ = 3 mV). **b,** Same data as in **a** but rotated by 90° into a waterfall plot. Inset: STM topograph of R6G/hBN D3 ($V_b$ = -0.3 V, $I_t$ = 3 pA). **c,** Fitting results of moiré renormalization strength for **a**. **d-f,** Same as **a-c**, but at $T = 10.5$ K ($V_b$ = 0.1 V, $I_t$ = 9 pA, $V_{mod}$ = 3 mV, $V_{bg}$ = 0 V).

**g-i,** Same as **a-c**, but at $T$ = 15.5 K ($V_b$ = 0.1 V, $I_t$ = 9 pA, $V_{mod}$ = 3 mV, $V_{bg}$ = 55 V). Setpoint of inset in **e:** $V_b$ = -0.3 V, $I_t$ = 5 pA; **h:** $V_b$ = -0.3 V, $I_t$ = 3 pA.

## 10. **Additional or supporting data sets**

In this section we present additional data sets of STM/STS measurements of small- and large-twist-angle R6G/hBN devices. They show consistent results with main figures where emergent flat-band renormalization and trans-moiré orbitals are observed only on the moiré-distant surface of small-twist-angle R6G/hBN devices with $\theta \lesssim 1^{\circ}$, as shown in Supplementary Information Figs. 10 and 11 below. On the other hand, large-twist-angle R6G/hBN devices with $\theta \gtrsim 1^{\circ}$ show no perceptible moiré-periodic modulations, as shown in Supplementary Information Fig. 12.

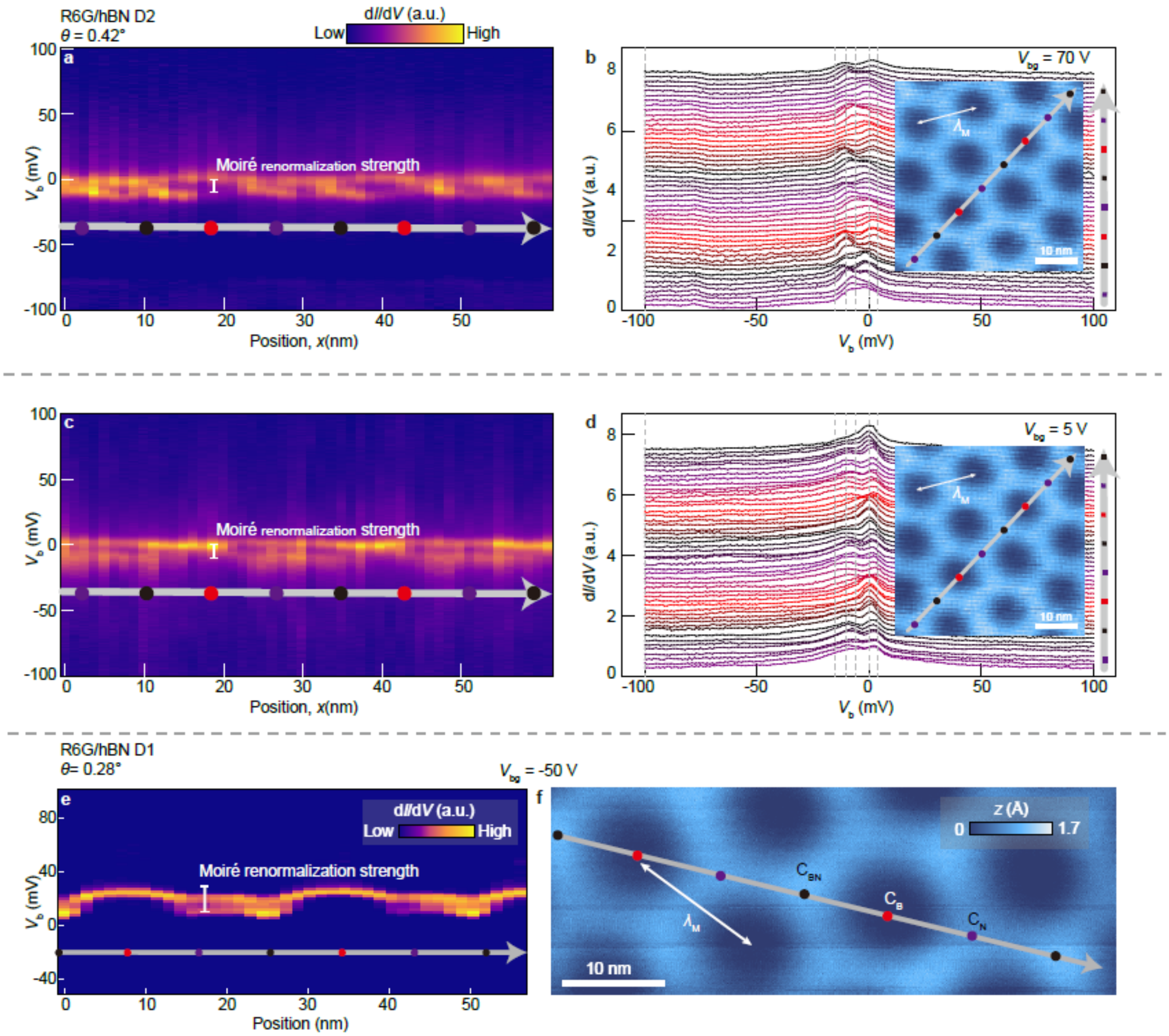

**Supplementary Information Figure 10. STM/STS measurement results of R6G/hBN D2 ($\theta$ = 0.42°) & R6G/hBN D1 ($\theta$ = 0.28°). a,** Colormap plot of spatially dependent d$I$/d$V$ spectra acquired at $V_{bg}$ = 70 V (along the grey arrow shown in the inset of **b**) ($V_b$ = -0.3 V, $I_t$ = 40 pA, $V_{mod}$ = 3 mV). **b,** Same data as in **a** but rotated by 90° into a waterfall plot. Inset: STM topograph of R6G/hBN D2 ($V_b$ = -0.3 V, $I_t$ = 3 pA). **c & d,** Same as **a** & **b**, but at $V_{bg}$ = 5 V ($V_b$ = -0.3 V, $I_t$ = 40 pA, $V_{mod}$ = 3 mV). **e,** Colormap plot of spatially dependent d$I$/d$V$ spectra (along the grey arrow in **b**) taken at $V_{bg}$ = -50 V with tip A ($V_b$ = -0.4 V, $I_t$ = 30 pA, $V_{mod}$ = 3 mV, $T$ = 3.7 K). **f,** STM topograph of R6G D1 ($V_b$ = -0.4 V, $I_t$ = 3 pA). ($\nu$, $D$) parameters are summarized in Extended Data Fig. 9. Measurements performed at $T$ = 6.0 K (**a-d**) & $T$ = 3.7 K (**e-f**).

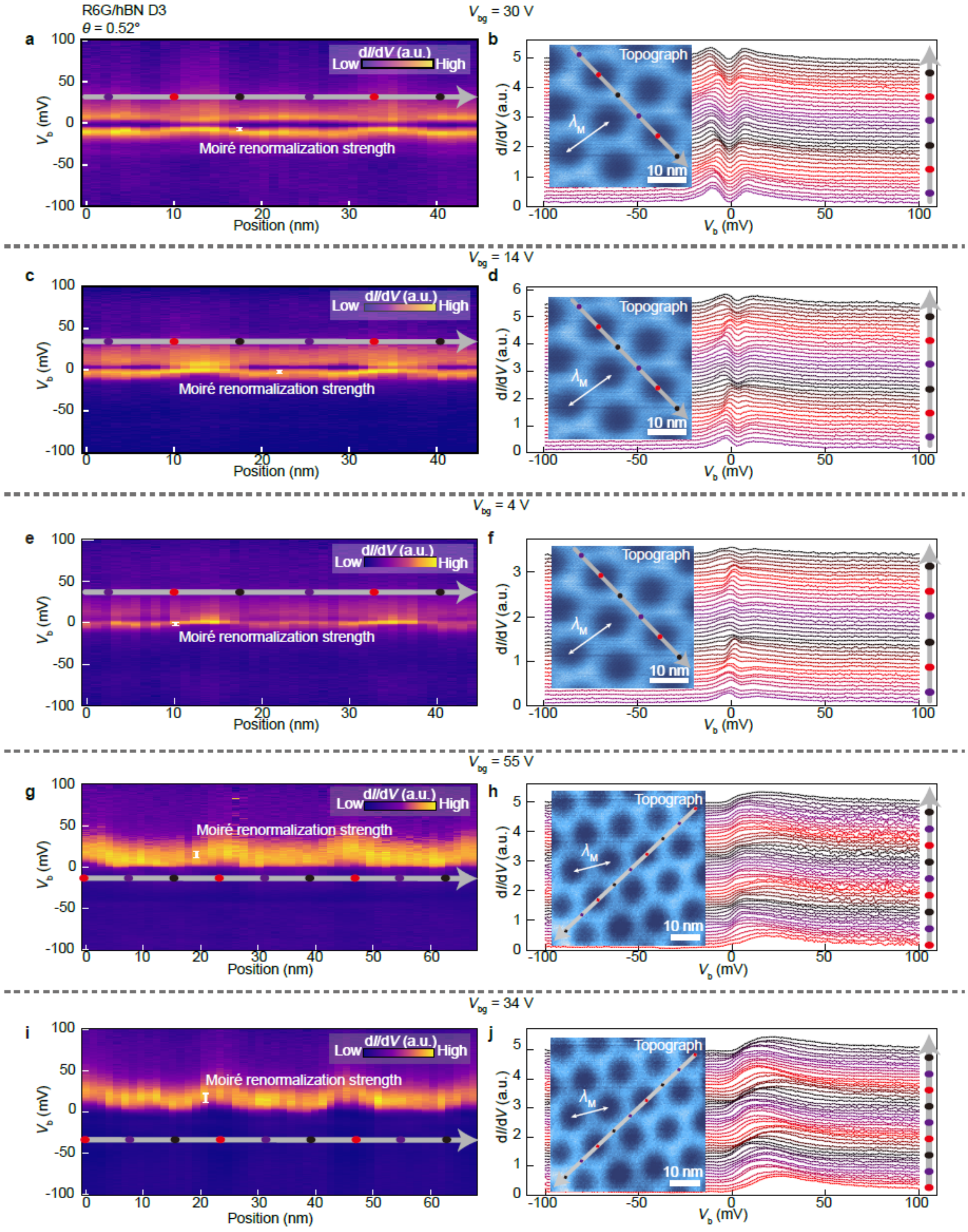

**Supplementary Information Figure 11. Emergent flat-band renormalization in R6G/hBN D3 ($\theta$ = 0.52°). a,** Colormap plot of spatially dependent d$I$/d$V$ spectra acquired at $V_{bg}$ = 30 V (along the grey arrow shown in the inset of **b**) ($V_b$ = -0.3 V, $I_t$ = 20 pA, $V_{mod}$ = 3 mV). **b,** Same data as in **a** but rotated by 90° into a waterfall plot. Inset: STM topograph of R6G/hBN D3 ($V_b$ = 0.3 V, $I_t$ = 5 pA). **c & d,** Same as **a** & **b**, but at $V_{bg}$ = 14 V ($V_b$ = -0.3 V, $I_t$ = 20 pA, $V_{mod}$ = 3 mV). **e & f,** Same as **a** & **b**, but at $V_{bg}$ = 4 V ($V_b$ = -0.3 V, $I_t$ = 20 pA, $V_{mod}$ = 3 mV). **g & h,** Same as **a** & **b**, but at $V_{bg}$ = 55 V ($V_b$ = -0.3 V, $I_t$ = 40 pA, $V_{mod}$ = 3 mV). **i & j,** Same as **a** & **b**, but at $V_{bg}$ = 34 V ($V_b$ = -0.3 V, $I_t$ = 40 pA, $V_{mod}$ = 3 mV). Setpoint of inset in **h** & **j:** $V_b$ = -0.3 V, $I_t$ = 3 pA. Fitted ($\nu$, $D$) parameters are **a-b**: $\nu$ = 2.4, $D$ = -0.19V/nm, **c-d**: $\nu$ =1.0, $D$ = -0.10V/nm, **e-f**: $\nu$ = 0.07, $D$ = -0.04 V/nm, **g-h**: $\nu$ = 4.2, $D$ = -0.35 V/nm, **i-j**: $\nu$ = 2.3, $D$ = -0.23 V/nm (summarized in Extended Data Fig. 9). All measurements performed at $T$ = 2.3 K.

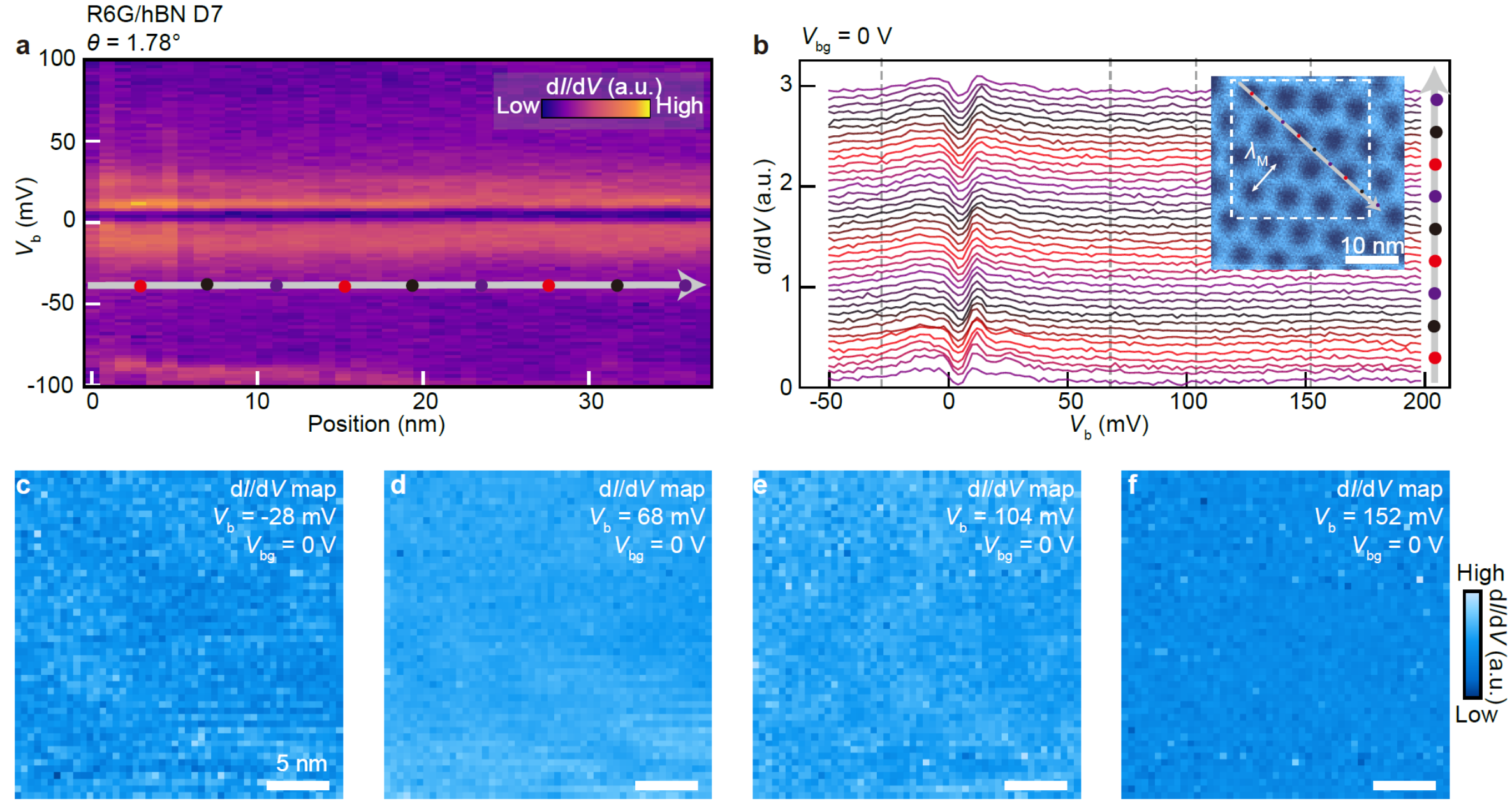

**Supplementary Information Figure 12. STM/STS measurement results of R6G/hBN D7 ($\theta$ = 1.78°). a,** Colormap plot of spatially dependent d$I$/d$V$ spectra (along the grey arrow in the inset of **b**) showing no noticeable moiré-periodic electronic modulations, taken at $V_{bg}$ = 0 V ($V_b$ = -0.5 V, $I_t$ = 30 pA, $V_{mod}$ = 3 mV). **b,** Same data as **a** by 90° into a waterfall plot. Inset: STM topograph of R6G D7 ($V_b$ = 0.5 V, $I_t$ = 3 pA). **c-f,** Constant-current d$I$/d$V$ maps showing spatially homogeneous spectral density ($V_b$ = -0.1 V, $I_t$ = 30 pA, $V_{mod}$ = 3mV). Fitted ($\nu$, $D$) parameters are: $\nu$ = -0.51, $D$ = -0.11 V/nm (Extended Data Fig. 9). All measurements performed at $T$ = 2.3 K.

# 11. **Reference**

1 Kerelsky, A. *et al.* Maximized electron interactions at the magic angle in twisted bilayer graphene. *Nature* **572**, 95-100 (2019). https://doi.org/10.1038/s41586-019-1431-9

2 Nuckolls, K. P. *et al.* Quantum textures of the many-body wavefunctions in magic-angle graphene. *Nature* **620**, 525-532 (2023). https://doi.org/10.1038/s41586-023-06226-x

3 Kim, H. *et al.* Imaging inter-valley coherent order in magic-angle twisted trilayer graphene. *Nature* **623**, 942-948 (2023). https://doi.org/10.1038/s41586-023-06663-8

4 Seewald, E. *et al.* Mapping the moiré potential in multi-layer rhombohedral graphene. *arXiv preprint arXiv:2510.09548* (2025).

5 Liu, Y. *et al.* Electronic Correlations in Rhombohedral Graphene at Atomic Scale. *Physical Review Letters* **135**, 156401 (2025).

6 Liu, Y. *et al.* Visualizing incommensurate inter-valley coherent states in rhombohedral trilayer graphene. *arXiv preprint arXiv:2411.11163* (2024).

7 Zhang, Y. *et al.* Layer-dependent evolution of electronic structures and correlations in rhombohedral multilayer graphene. *Nature Nanotechnology* **20**, 222-228 (2025).

8 Zhou, H. *et al.* Half- and quarter-metals in rhombohedral trilayer graphene. *Nature* **598**, 429-433 (2021). https://doi.org/10.1038/s41586-021-03938-w

9 Huo, Z. *et al.* Does Moire Matter? Critical Moire Dependence with Quantum Fluctuations in Graphene Based Integer and Fractional Chern Insulators. *arXiv preprint arXiv:2510.15309* (2025).

10 Xie, J. *et al.* Tunable fractional Chern insulators in rhombohedral graphene superlattices. *Nature Materials* **24**, 1042-1048 (2025). https://doi.org/10.1038/s41563-025-02225-7

11 Lu, Z. *et al.* Fractional quantum anomalous Hall effect in multilayer graphene. *Nature* **626**, 759-764 (2024). https://doi.org/10.1038/s41586-023-07010-7

12 Li, C. *et al.* Stacking-Orientation and Twist-Angle Control on Integer and Fractional Chern Insulators in Moiré Rhombohedral Graphene. *arXiv preprint arXiv:2505.01767* (2025).

13 Klein, D. R. *et al.* Imaging the sub-moire potential using an atomic single electron transistor. *Nature* (2026). https://doi.org/10.1038/s41586-025-10085-z

14 Huang, K., Li, X., Das Sarma, S. & Zhang, F. Self-consistent theory of fractional quantum anomalous Hall states in rhombohedral graphene. *Physical Review B* **110** (2024). https://doi.org/10.1103/PhysRevB.110.115146

15 Kwan, Y. H. *et al.* Moiré fractional Chern insulators. III. Hartree-Fock phase diagram, magic angle regime for Chern insulator states, role of moiré potential, and Goldstone gaps in rhombohedral graphene superlattices. *Physical Review B* **112**, 075109 (2025). https://doi.org/10.1103/PhysRevB.112.075109

16 Bernevig, B. A. *et al.* Fractional quantization in insulators from Hall to Chern. *Nature Physics* **21**, 1702-1713 (2025). https://doi.org/10.1038/s41567-025-03072-8

17 Dong, Z., Patri, A. S. & Senthil, T. Theory of Quantum Anomalous Hall Phases in Pentalayer Rhombohedral Graphene Moiré Structures. *Physical Review Letters* **133** (2024). https://doi.org/10.1103/PhysRevLett.133.206502

18 Zhou, B., Yang, H. & Zhang, Y.-H. Fractional Quantum Anomalous Hall Effect in Rhombohedral Multilayer Graphene in the Moiréless Limit. *Physical Review Letters* **133**, 206504 (2024). https://doi.org/10.1103/PhysRevLett.133.206504

19 Dong, J. *et al.* Anomalous Hall Crystals in Rhombohedral Multilayer Graphene. I. Interaction-Driven Chern Bands and Fractional Quantum Hall States at Zero Magnetic Field. *Physical Review Letters* **133**, 206503 (2024). https://doi.org/10.1103/PhysRevLett.133.206503

20 Herzog-Arbeitman, J. *et al.* Moiré fractional Chern insulators. II. First-principles calculations and continuum models of rhombohedral graphene superlattices. *Physical Review B* **109** (2024). https://doi.org/10.1103/PhysRevB.109.205122

21 Li, C. *et al.* Tunable Chern Insulators in Moiré-Distant and Moiré-Proximal Rhombohedral Pentalayer Graphene. *arXiv preprint arXiv:2505.01767* (2025).

22 Tsui, Y.-C. *et al.* Direct observation of a magnetic-field-induced Wigner crystal. *Nature* **628**, 287-292 (2024). https://doi.org/10.1038/s41586-024-07212-7

23 Xiang, Z. *et al.* Imaging quantum melting in a disordered 2D Wigner solid. *Science* **388**, 736-740 (2025). https://doi.org/10.1126/science.ado7136

24 Li, H. *et al.* Wigner molecular crystals from multielectron moiré artificial atoms. *Science* **385**, 86-91 (2024). https://doi.org/10.1126/science.adk1348